\documentclass[12pt,english,american]{article}
\usepackage[T1]{fontenc}
\usepackage[latin9]{inputenc}
\usepackage{geometry}
\usepackage{color}
\usepackage{babel}
\usepackage{amsthm}
\usepackage{amstext}
\usepackage{amssymb}
\usepackage{amsmath}
\usepackage{cancel}   
\usepackage[unicode=true,pdfusetitle,
 bookmarks=true,bookmarksnumbered=false,bookmarksopen=false,
 breaklinks=true,pdfborder={0 0 0},backref=false,colorlinks=true]
 {hyperref}
\hypersetup{
 linkcolor=blue,citecolor=blue, urlcolor=blue}

\makeatletter

\newcommand{\hde}{\hat\de}
\newcommand{\ca}{\mathrm{\check{a}}}
\newcommand{\cb}{\mathrm{\check{b}}}
\newcommand{\hubns}{\boldsymbol{\hat b}_{\boldsymbol{\ns}}}
\newcommand{\huans}{\boldsymbol{\hat a}_{\boldsymbol{\ns}} }
\newcommand{\cubns}{\boldsymbol{\check b}_{\boldsymbol{\ns}}}
\newcommand{\cuans}{\boldsymbol{\check a}_{\boldsymbol{\ns}} }

\newcommand{\Span}{\mathrm{Span}}
\newcommand{\hpa}{\hat\pa}
\newcommand{\unb}{\boldsymbol{b}}
\newcommand{\una}{\boldsymbol{a}}
\newcommand{\hub}{\boldsymbol{\hat b}}
\newcommand{\hua}{\boldsymbol{\hat a}}
\newcommand{\hb}{\hat b}
\newcommand{\ha}{\hat a}
\newcommand{\ns}{\mathrm{ns}}

\newcommand{\Op}{\mathrm{Op}}
\newcommand{\Pol}{\mathrm{Pol}}

\newcommand{\mrm}{\mathrm}

\newcommand{\hal}{\al}
\newcommand{\ir}{\si^{-\de_{\la_0}}}

\newcommand{\hf}{\hat f_{P,\si}}
\newcommand{\cpsi}{\check\psi}

\newcommand{\cphi}{\check{\phi}}
\newcommand{\mm}{n}
\newcommand{\m}{m}

\newcommand{\tic}{c}
\newcommand{\sym}{\mathrm{sym} }
\newcommand{\PS}{S}

\newcommand{\fib}{\mathrm{fi}}
\newcommand{\free}{\mathrm{free}}

\newcommand{\mcA}{\mathcal A}
\newcommand{\mcB}{\mathcal B}

\newcommand{\kas}{\kappa_{*}}

\newcommand{\unk}{\un k}

\newcommand{\pho}{\mathrm{f}}
\newcommand{\alf}{\overline{\al}}

\newcommand{\vv}{v_{\alf}}

\newcommand{\g}{\la }

\newcommand{\wt}{\widetilde}

\newcommand{\ti}{\tilde}

\newcommand{\un}{\underline}

\newcommand{\vac}{\Omega}
\newcommand{\Om}{\Omega}

\newcommand{\La}{\Lambda}

\newcommand{\ka}{\kappa}

\newcommand{\be}{\beta}

\newcommand{\pa}{\partial}

\newcommand{\mfh}{\mathfrak{h}}

\newcommand{\eps}{\varepsilon}
\newcommand{\de}{\delta}

\newcommand{\De}{\Delta}

\newcommand{\nin}{\noindent}
\newcommand{\si}{\sigma}
\newcommand{\ph}{\phantom}
\newcommand{\h}{\fr{1}{2}}
\newcommand{\nat}{\mathbb{N}}
\newcommand{\hil}{\mathcal{H}}

\newcommand{\mco}{\mathcal{O}}

\newcommand{\fr}[2]{\frac{#1}{#2}}
\newcommand{\al}{\alpha}
\newcommand{\real}{\mathbb{R}}

\newcommand{\la}{\lambda}
\newcommand{\non}{\nonumber}
\newcommand{\Ga}{\Gamma}
\newcommand{\half}{\fr{1}{2}}
\newcommand{\lan}{\langle}
\newcommand{\ran}{\rangle}

\def\proof{\noindent{\bf Proof. }}
\def\qed{$\Box$\medskip}

\newcommand{\beq}{\begin{equation}}
\newcommand{\eeq}{\end{equation}}
\newcommand{\beqa}{\begin{eqnarray}}
\newcommand{\eeqa}{\end{eqnarray}}
\newcommand{\ben}{\begin{arabicenumerate}}
\newcommand{\een}{\end{arabicenumerate}}

\def\bel{\begin{lem} } 
\def\eel{\end{lem} }
\def\bet{\begin{thm}}
\def\eet{\end{thm}}
\def\bed{\begin{defn}}
\def\eed{\end{defn} }

\def\bec{\begin{cor}}
\def\eec{\end{cor}}
\def\ber{\begin{rem}}
\def\eer{\end{rem}}

\usepackage{enumitem}		
\theoremstyle{plain}
\newtheorem{thm}{\protect\theoremname}[section]
\theoremstyle{definition}
\newtheorem{defn}[thm]{\protect\definitionname}
\theoremstyle{plain}
\newtheorem{prop}[thm]{\protect\propositionname}
\theoremstyle{plain}
\theoremstyle{remark}
\newtheorem{rem}[thm]{\protect\remarkname}
\theoremstyle{plain}
\newtheorem{lem}[thm]{\protect\lemmaname}
\theoremstyle{plain}
\newtheorem{cor}[thm]{\protect\corollaryname}

\usepackage{txfonts,refstyle,xcolor}

\newref{con}{name = Conjecture\ }
\newref{prop}{name = Proposition\ }
\newref{def}{name = Definition\ }
\newref{sec}{name = Section\ }
\newref{sub}{name = Section\ }
\newref{thm}{name = Theorem\ }
\newref{lem}{name = Lemma\ }
\newref{cor}{name = Corollary\ }
\newref{fig}{name = Figure\ }
\newref{rem}{name =  Remark\ }

\usepackage{bbm}

\usepackage[all]{xy}

\newcommand{\xyR}[1]{%
     \makeatletter
     \xydef@\xymatrixrowsep@{#1}
     \makeatother
}

\newcommand{\xyC}[1]{%
     \makeatletter
     \xydef@\xymatrixcolsep@{#1}
     \makeatother
}

\newcommand{\ncol}[1]{\color{normalcolor}}

\makeatother

\addto\captionsamerican{\renewcommand{\corollaryname}{Corollary}}
\addto\captionsamerican{\renewcommand{\definitionname}{Definition}}
\addto\captionsamerican{\renewcommand{\lemmaname}{Lemma}}
\addto\captionsamerican{\renewcommand{\propositionname}{Proposition}}
\addto\captionsamerican{\renewcommand{\remarkname}{Remark}}
\addto\captionsamerican{\renewcommand{\theoremname}{Theorem}}
\addto\captionsenglish{\renewcommand{\corollaryname}{Corollary}}
\addto\captionsenglish{\renewcommand{\definitionname}{Definition}}
\addto\captionsenglish{\renewcommand{\lemmaname}{Lemma}}
\addto\captionsenglish{\renewcommand{\propositionname}{Proposition}}
\addto\captionsenglish{\renewcommand{\remarkname}{Remark}}
\addto\captionsenglish{\renewcommand{\theoremname}{Theorem}}
\providecommand{\corollaryname}{Corollary}
\providecommand{\definitionname}{Definition}
\providecommand{\lemmaname}{Lemma}
\providecommand{\propositionname}{Proposition}
\providecommand{\remarkname}{Remark}
\providecommand{\theoremname}{Theorem}
\newcommand{\y}{\!\!\!}
\newcommand{\fin}{\mathrm{fin}}

\begin{document}

\title{Coulomb scattering in the massless Nelson model III. \\  Ground state wave functions and non-commutative recurrence relations} 
 \author{
{\bf Wojciech Dybalski}\\
Zentrum Mathematik, Technische Universit\"at M\"unchen,\\
and\\
Fakult\"at f\"ur Mathematik, Ludwig-Maximilians-Universit\"at M\"unchen \\
E-mail: {\tt dybalski@ma.tum.de}
\and
{\bf Alessandro Pizzo}\\
Dipartimento di Matematica, Universit\`a di Roma ``Tor
Vergata''\\ 
E-mail: {\tt pizzo@axp.mat.uniroma2.it}}
 
 \date{}

\maketitle

\begin{abstract}
Let $H_{P,\sigma}$ be the single-electron fiber Hamiltonians of the massless Nelson model at total
momentum $P$ and infrared cut-off $\sigma>0$.  We establish detailed regularity properties of the 
corresponding $n$-particle  ground state  wave functions $f^n_{P,\sigma}$ as functions of $P$ and
$\sigma$. In particular, we show that
\[
|\partial_{P^j}f^{n}_{P,\sigma}(k_1,\ldots, k_n)|, \ \  |\partial_{P^j} \partial_{P^{j'}} f^{n}_{P,\sigma}(k_1,\ldots, k_n)|  \leq  \fr{1}{\sqrt{n!}} \fr{(c\lambda_0)^n}{\sigma^{\delta_{\lambda_0}}} 
\prod_{i=1}^n\frac{  \chi_{[\sigma,\kappa)}(k_i)}{|k_i|^{3/2}}, 
\]
where $c$ is a numerical constant, $\lambda_0\mapsto \delta_{\lambda_0}$ is a positive function of the maximal admissible coupling constant which satisfies $\lim_{\lambda_0\to 0}\de_{\lambda_0}=0$ and $\chi_{[\sigma,\kappa)}$ is the (approximate) characteristic function of the energy region between the infrared 
cut-off $\sigma$ and the
ultraviolet cut-off $\kappa$.
While the analysis of the first derivative is relatively
straightforward, the second derivative requires a new strategy. By solving a non-commutative recurrence relation we derive a novel formula for $f^n_{P,\sigma}$ with improved infrared properties. In this representation $\partial_{P^{j'}}\partial_{P^{j}}f^n_{P,\sigma}$ is amenable to sharp estimates 
obtained by iterative analytic perturbation theory in part II of this series of papers. The bounds 
 stated above are instrumental for scattering theory of two electrons in the
Nelson model, as explained in part I of this series. 
\end{abstract}

\maketitle


\section{Introduction}
\setcounter{equation}{0}

Recently there has been some revival of interest in infrared problems  within
the theoretical physics community, triggered by the work of Strominger \emph{et al.} on soft photon theorems (see \cite{St17} for a review). Apart from emphasizing the richness of infrared physics, this reference states clearly that
 the absence of an infrared regular $S$-matrix in QED `is a big elephant in the room of 
mathematical quantum field theory'.  And indeed,  the infrared  problem has been a  subject of continuing research  
in the mathematical physics community for many decades. Also the present paper, which advances our long term investigation of  
two-electron scattering in a simplified model of QED,  stems from this tradition and provides evidence for the richness of infrared \emph{mathematics}.
Namely, we find an unexpected connection between infrared problems and  the theory of non-commutative recurrence relations. This theory turns out to be a robust method to demonstrate
infrared regularity   of physical quantities  which suffer from superficial infrared divergencies even after implementation of multi-scale techniques.


To summarize briefly the theory of non-commutative recurrence relations,  we consider two, possibly non-commuting, linear operators $\ha$ and $\hb$ 
on a vector space $X$.  Suppose the following  relation holds
\beqa
x_n=\ha x_{n-1}+\hb x_{n-2}, \label{recurrence-intro}
\eeqa
for a sequence of vectors $\{x_n\}_{n\in\nat_0}$ from $X$ with given initial
conditions $x_0$ and $x_1=\ha x_0$. The first closed-form solution of such a non-commutative recurrence was given in \cite{JMNP07, JNM08} and then applied to a computation of scattering amplitudes in perturbative QFT in \cite{Pu15}. It was shown by induction in these references that it has the form (cf. formula (2.7) of \cite{Pu15})
\beqa
x_n=\sum_{i_1+2i_2=n}\{\ha^{(i_1)}, \hb^{(i_2)} \}x_0,
\eeqa
where the bracket $\{\ha^{(i_1)}, \hb^{(i_2)} \}$ denotes the sum over all possible distinct permutations of factors $\ha$, $\hb$ each one appearing $i_1$, $i_2$ times, respectively. 
In our work we establish a different representation for the solution of  (\ref{recurrence-intro}) which is more suitable for our applications. It has the form 
\beqa
x_n=\Op^{(n)}_{\hat a, \hat b}[\exp(\sum_{i=1}^{n-1}b_{i+1,i}\pa_{a_{i+1}}\pa_{a_{i}}   ) a_n a_{n-1}\ldots a_1]x_0,
\label{recurrence-solution-x-intro}
\eeqa
where $a_i$, $b_{i+1,i}$ are real variables so that $a_n a_{n-1}\ldots a_1$ is a monomial on which 
$b_{i+1,i}\pa_{a_{i+1}}\pa_{a_{i}}$ acts by replacing $a_{i+1}a_i$ with $b_{i+1,i}$. The `quantization map'
$\Op^{(n)}_{\hat a, \hat b}$ replaces every $a_{i'}$ in the resulting polynomial with the operator $\ha$ and every $b_{i'+1,i'}$ with the operator $\hb$ while keeping the order specified by the indices. For a more detailed discussion and a very simple proof that (\ref{recurrence-solution-x-intro}) solves (\ref{recurrence-intro}) we refer to Subsection~\ref{non-commutative-subsection}.

\newcommand{\red}{black}

  Let us now proceed to applications of the theory of non-commutative recurrence relations to infrared problems. {\color{\red} Let} $H_{P,\si}$ be the fiber Hamiltonian of the Nelson model at total momentum $P$ with the infrared cut-off $\si>0$. It is a self-adjoint operator on the symmetric Fock space over $L^2(\real^3)$   given by
\beqa
H_{P,\si}:=\h(P-P_{\pho})^2+H_{\pho}+\int d^3k\, \vv^{\si}(k)\, ( b(k)+b^*(k) ), \label{infrared-cut-off-Hamiltonian-intro}
\eeqa  
where $b^*(k), b(k)$ are the (improper) creation and annihilation operators, $P_{\pho}:=\int d^3k\, k\, b^*(k)b(k)$ is the photon momentum and the form-factor 
$\vv^{\si}(k):=\la \chi_{[\si,\ka)}(k)|k|^{\alf}/\sqrt{2|k|}$ is defined precisely by formula~(\ref{IR-cut-off-propagator}) below. It contains the coupling constant $\la$ whose absolute value will be kept sufficiently small in our discussion and the regularity parameter $\alf\geq 0$, which can be set to zero for the purpose of the present paper. (We keep it here for consistency with \cite{DP12.1} where $\alf>0$ was assumed). This Hamiltonian has a normalized ground state vector $\cpsi_{P,\si}$,  with {\color{\red}the phase fixed in (\ref{def-checkpsi}),} corresponding to the simple eigenvalue $E_{P,\si}$. In  \cite{DP17.1} we derived the following formulas for the first and second derivative of $\cpsi_{P,\si}$ w.r.t. to $P$ 
\beqa
\pa_{P^j} \cpsi_{P,\si} \y&=&\y R_{P,\si}(\La_{P,\si})^{j}\check{\psi}_{P,\si}, 
\label{first-der-intro-x} \\ [5pt] 
\pa_{P^j}\pa_{P^{j'}} \cpsi_{P,\si}\y&= &\y
\big( \bar Q_{P,\si}^{\perp} R_{P,\si}(\La_{P,\si})^{j'}R_{P,\si} 
 (\La_{P,\si})^{j}\check{\psi}_{P,\si}+(j\leftrightarrow j')\big)\non\\
& & -\check{\psi}_{P,\si}\langle \check{\psi}_{P,\si},(\La_{P,\si})^{j'}
 R_{P,\si}^2 (\La_{P,\si})^{j}\check{\psi}_{P,\si}\rangle,    \label{second-der-intro}\\ [5pt] 
 R_{P,\si}\y &:=&\y (H_{P,\si}-E_{P,\si})^{-1}, \quad \La_{P,\si}:=\nabla E_{P,\si}-(P-P_{\pho}), \quad \bar Q_{P,\si}:=|\cpsi_{P,\si}\ran \lan \cpsi_{P,\si}|.
\eeqa
We recall that $\La_{P,\si}$ has the property $\lan \cpsi_{P,\si}, \La_{P,\si}\cpsi_{P,\si}\ran=0$, so the above expressions are well defined. The control of the behaviour of 
$\pa_{P^j} \cpsi_{P,\si}$, $\pa_{P^j}\pa_{P^{j'}} \cpsi_{P,\si}$ as $\si\to 0$ is a difficult problem in spectral theory, concerning eigenvalues at the bottom of
the continuous spectrum. However, using  iterative analytic perturbation theory
 we  established the following estimates in  \cite{DP17.1}   \beqa
& &\| R_{P,\si}(\La_{P,\si})^{j}\check{\psi}_{P,\si}\|\leq  \fr{c}{\si^{\de_{\g_0}}}, \label{first-spectral-ingr}\\
& &\| \bar Q_{P,\sigma}^{\perp}R_{P,\si} (\La_{P,\sigma})^{j} R_{P,\si}
(\La_{P,\sigma})^{j'}\check{\psi}_{P,\sigma}\|\leq \frac{c}{\sigma^{\delta_{\g_0}}}, \label{eq:inequality-cont-1-intro-x}
\eeqa
where $c$ is a universal constant and $\g_0 \mapsto \de_{\g_0}$ is positive and  satisfies $\lim_{\g_0\to 0}\de_{\g_0}=0$. Thus the infrared singularity of this estimate is very mild in the weak coupling regime. Obviously, from (\ref{first-der-intro-x})-(\ref{eq:inequality-cont-1-intro-x}) we obtain
\beqa
\|\pa_{P^j} \cpsi_{P,\si}\|, \ \  \|\pa_{P^j}\pa_{P^{j'}} \cpsi_{P,\si}\|\leq \fr{c}{\sigma^{\delta_{\g_0}}}. \label{mild-vector-bound}
\eeqa

The purpose of the present paper is to obtain bounds similar to (\ref{mild-vector-bound}) also for the $\mm$-photon wave functions 
$\{f^{\mm}_{P,\si}\}_{\mm\in \nat_0}$ of $\cpsi_{P,\si}$ using the spectral ingredients (\ref{first-spectral-ingr}), (\ref{eq:inequality-cont-1-intro-x}) and their
slight generalizations stated in Theorem~\ref{spectr-theorem} below.  Our main results,  stated precisely in Theorem~\ref{main-theorem-spectral} below,  include the following estimates
\beqa
|\pa_{P^j}f^{\mm}_{P,\si}(k_1,\ldots, k_\mm)|, \ \  |\pa_{P^j} \pa_{P^{j'}} f^{\mm}_{P,\si}(k_1,\ldots, k_\mm)|  \leq  \fr{1}{\sqrt{\mm!}} \fr{c^n}{\si^{\de_{\g_0}}} 
\prod_{i=1}^n\fr{  \vv^{\si}(k_i)}{|k_i|}, \quad n\geq 0.
\label{infrared-spectral-bound-zero}
\eeqa
As explained in  detail in \cite{DP12.1}, these bounds constitute
 crucial technical {\color{\red}inputs} for the construction of two-electron scattering states.  The momentum derivatives {\color{\red}play a crucial role} in \cite{DP12.1}
via  non-stationary phase arguments. {\color{\red}The control up to the second derivative is} dictated by  the Cook's method. 

{\color{\red}The ground state wave function is expressed by the  following formula}
\beqa
f^n_{P,\si}(k_1,\ldots,k_n):=\fr{1}{\sqrt{n!}} \lan \Om,b(k_1)\ldots b(k_n)\cpsi_{P,\si}\ran. \label{wavefunction-formula}
\eeqa
We define $\check{f}^{n}_{P,\si}(k_1,\ldots, k_n):=b(k_1)\ldots b(k_n)\cpsi_{P,\si}$ and set $\unk_n:=k_1+\cdots+k_n$, $|\unk|_n:=|k_1|+\cdots+|k_n|$. 
Making use of the fact that $H_{P,\si}\cpsi_{P,\si}=E_{P,\si}\cpsi_{P,\si}$ and of the canonical commutation relations, one easily gets
\beqa
\check{f}_{P,\si}^{n}(k_1,\ldots,k_n)= (-)R_{P,\si; n}\sum_{i=1}^n \vv^{\si}(k_i)\check{f}_{P,\si}^{n-1}(k_1,\ldots\check{i}\ldots, k_n), \quad 
R_{P,\si;n}:=(H_{P-\unk_n,\si}-E_{P,\si}+|\unk|_n)^{-1},\quad
\eeqa 
with $\check{f}_{P,\si}^{0}=\cpsi_{P,\si}$. Suitably interpreted, this is a trivial example of the recurrence (\ref{recurrence-intro}) with $\hb=0$. The solution has the form
\beqa
\check{f}^{n}_{P,\si}=(-1)^nn!P_{\sym} (v_{\alf;1}^{\si}\ldots v_{\alf;n}^{\si}  )(R_{P,\si;n}\ldots R_{P,\si;1})\cpsi_{P,\si}, 
\label{check-intro}
\eeqa
where $P_{\sym}$ denotes symmetrization in $k_1,\ldots, k_n$ and 
$(v_{\alf;1}^{\si}\ldots v_{\alf;n}^{\si})(k_1,\ldots, k_n):= v_{\alf}^{\si}(k_1)\ldots
 v_{\alf}^{\si}(k_n)$.   By substituting (\ref{check-intro}) to (\ref{wavefunction-formula}), we reproduce the standard formula for the wave functions due to 
Fr\"ohlich \cite{Fr73, Fr74.1}.

The next task is to study the derivatives of (\ref{check-intro})
so as to obtain  estimates~(\ref{infrared-spectral-bound-zero}). Here the rules 
of the game are dictated by the following relations:
\beqa
\pa_{P^j} \cpsi_{P,\si}=R_{P,\si}(\La_{P,\si})^{j}\check{\psi}_{P,\si}, \quad \pa_{P^j}R_{P,\si;i}=R_{P,\si;i}(\La_{P,\si}^i)^jR_{P,\si;i}, \quad 
\|R_{P,\si; i}\|\leq c|\unk|^{-1}_i, \label{rules-of-the-game}
\eeqa
where $\La_{P,\si}^i:=\La_{P,\si}+\unk_{i}$ and the first identity simply recalls (\ref{first-der-intro-x}). Thus the action of $\pa_{P^j}$ typically increases the number of resolvents. If the additional resolvent comes from the second identity
in (\ref{rules-of-the-game}), we cannot apply the standard resolvent bound 
$\|R_{P,\si; i}\|\leq c|\unk|^{-1}_i$, as this would lead to more  factors
$|k_{i'}|^{-1}$ than {\color{\red} those appearing} in (\ref{infrared-spectral-bound-zero}). Furthermore,  if we exploited the support 
property of $v_{\alf}^{\si}$ to write $|k_{i'}|^{-1}\leq \si^{-1}$,
 this would destroy the mild infrared behaviour of our main result given by
$\ir$. Thus the additional resolvents must be absorbed by the spectral ingredients
(\ref{first-spectral-ingr}), (\ref{eq:inequality-cont-1-intro-x}). This is possible as long as the relevant expressions have the alternating structure of resolvents separated by $\La_{P,\si}$ factors.  As we demonstrate in Subsection~\ref{first-derivative-sub}, this strategy works for $\pa_{P^j} f^{n}_{P,\si}$. 
However, the analysis of the second derivative
based on (\ref{check-intro}) faces serious problems. In fact, already in the case of $n=1$ we encounter the following contribution
\beqa
\pa_{P^j}\pa_{P^{j'}} \check{f}^{1}_{P,\si}(k_1)
\ni-\vv^{\si}(k_1)  R_{P,\si;1}(\La_{P,\si}^1)^{j} R_{P,\si;1}R_{P,\si}
(\La_{P,\si})^{j'}\cpsi_{P,\si} \label{second-der-problem}
\eeqa
by the combined application of the first  two identities in (\ref{rules-of-the-game}).
As this term contains two resolvents not separated by $\La_{P,\si}$, it is not directly tractable
by the spectral ingredients (\ref{first-spectral-ingr}), 
(\ref{eq:inequality-cont-1-intro-x}). 
We leave it open here if these latter estimates could be generalized so as to control (\ref{second-der-problem}).
Given that the inductive proof of (\ref{eq:inequality-cont-1-intro-x})  in its present form extends over thirty pages,
such a generalization would surely  be a very formidable task. Fortunately, 
theory of non-commutative recurrence relations allows us to circumvent this problem, as we now explain.

We recall the dressing transformation $W_{P,\si}:=e^{b^*(f_{P,\si})-b(f_{P,\si}) }$ from {\color{\red}\cite{Fr73}}, where
\beqa
 f_{P,\si}(k):=\fr{  \vv^{\si}(k)    }{|k|\al_{P,\si}(\hat k) },
\quad \al_{P,\si}(\hat k):=(1-\hat k\cdot \nabla E_{P,\si}),\quad \hat k=k/|k|
 \label{f-P-functions}
\eeqa
and define the dressed operators $b_W(k):=W_{P,\si}^* b(k) W_{P,\si}=b(k)+f_{P,\si}(k)$. We also introduce
the objects $\hf^n(k_1,\ldots,k_n):=b_W(k_1)\ldots b_W(k_n)\cpsi_{P,\si}$ which now replace
$\check{f}^n_{P,\si}$ considered above. Since {\color{\red}the} functions $f_{P,\si}$ given by (\ref{f-P-functions}) are explicit
and their derivatives w.r.t. $P$ up to second order have mild dependence on $\si$, it is clear from (\ref{wavefunction-formula})
that it suffices to analyse $\hf^n$. We show in Subsection~\ref{novel-subsection}
that  the following recurrence relation holds true
\beqa
\hf^n (k_1,\ldots, k_n)\y&=&\y
(-)R_{P,\si; n}  \bigg(\sum_{i=1}^nF_{P,\si}(k_i,k_1+\cdots \check{i} \cdots +k_n)
\hf^{n-1}(k_1,\ldots \check{i} \ldots, k_n)\non\\
& &\ph{44444444444}+\sum_{1\leq i<i'\leq n} g_{P,\si}(k_i)\cdot g_{P,\si}(k_{i'})\hf^{n-2}(k_1,\ldots \check{i}\ldots \check{i}' \ldots, k_n)\bigg)
\label{recurrence-relation-W-intro}
\eeqa
with the initial conditions $\hf^0=\cpsi_{P,\si}$, $ \ \hf^1(k_1)=(-)R_{P,\si;1}F_{P,\si}(k_1,0)\cpsi_{P,\si}$ and the definitions
\beqa
F_{P,\si}(\ti k,k):=-g_{P,\si}(\ti k){\cdot}\bigg(\La_{P,\si}+k+\h \ti k\bigg), 
\quad g_{P,\si}(k):=f_{P,\si}(k)k. \label{g-definition}
\eeqa
Formula (\ref{recurrence-relation-W-intro}) can be seen as a recurrence relation
of the form (\ref{recurrence-intro}) with $\hat a, \hat b\neq 0$. From (\ref{recurrence-solution-x-intro}) we obtain the the following representation of the solution:
\beqa
\hf^n\y&=&\y n! P_{\sym}\sum_{\ell=0}^{[n/2]}\sum_{2\leq i_1 \ll\cdots \ll i_{\ell}\leq n }\fr{(-1)^{\ell}}{2^{\ell}} g_{P,\si;1}\ldots g_{P,\si; n} \times  \non\\
& &\times (R_{P,\si; i_1} \hat\pa_{i_1} \hat\pa_{i_1-1}  )\ldots 
(R_{P,\si; i_{\ell}}\hat\pa_{i_{\ell}} \hat\pa_{i_{\ell}-1}  )
\bigg\{(R_{P,\si; n}\La_{P,\si;n})\ldots (R_{P,\si; 1}\La_{P,\si;1})  \bigg\}\cpsi_{P,\si},
\label{hfn-formula-intro}
\eeqa
where $\La_{P,\si;i}:=\La_{P,\si}+\unk_{i-1}+\h k_i$, {\color{\red} the} notation $i\ll i'$ means $i+1<i'$, and
each operator $(R_{P,\si; i}\hat\pa_{i} \hat\pa_{i-1} )$ acts on the product in the
curly bracket by replacing $(R_{P,\si; i}\La_{P,\si;i})(R_{P,\si; i-1}\La_{P,\si;i-1})$
with $R_{P,\si; i}$.

Let us now compare  formulas~(\ref{hfn-formula-intro}) and (\ref{check-intro}).
First, by (\ref{f-P-functions}), (\ref{g-definition}), $g_{P,\si}$ and $v_{\alf}^{\si}$ have very similar behaviour at small $k$. Second, the summation over $\ell$ and
$i_1, \ldots, i_{\ell}$ merely leads to an additional factor $c^n$, which is allowed in (\ref{infrared-spectral-bound-zero}). Thus the main difference are the factors
$\La_{P,\si;i}$ present in formula (\ref{hfn-formula-intro}) even before computing the derivatives. We note that the term $\ell=0$ has the desirable alternating structure (i.e. resolvents separated by $\La_{P,\si}$) which is needed to use the spectral bounds (\ref{first-spectral-ingr}),
(\ref{eq:inequality-cont-1-intro-x}). However, {\color{\red}for $\ell>0$} this structure can be disturbed  by the action of the operators $(R_{P,\si; i}\hat\pa_{i} \hat\pa_{i-1} )$ 
which create terms with resolvents not separated by $\La_{P,\si}$.
\emph{The key point and the essence of the method is that the total number of resolvents in such terms is strictly less than $n$}.  This is due to the fact that $(R_{P,\si; i}\hat\pa_{i} \hat\pa_{i-1} )$  effectively removes one resolvent.  This missing resolvent can then be  replaced by 
the additional resolvent coming from $\pa_{P^j}R_{P,\si;i}=R_{P,\si;i}(\La_{P,\si}^i)^jR_{P,\si;i}$ so that simple bounds $\|R_{P,\si;i}\|\leq c|k|_{i}^{-1}$ lead to
the correct number of factors $|k_{i'}|^{-1}$ as required in 
(\ref{infrared-spectral-bound-zero}). We show in Subsection~\ref{derivatives} that this mechanism works in all the relevant cases.

For the sake of clarity, we illustrate this mechanism by a simple example. Let
us consider the following contribution to $\pa_{P^j}\pa_{P^{j'}}\hf^2$
\beqa
\pa_{P^j}\pa_{P^{j'}}\hf^2\y&\ni &\y(-1)P_{\sym}g_{P,\si; 1} g_{P,\si; 2} 
(\pa_{P^j}R_{P,\si; 2  }\hat\pa_{2 } \hat\pa_{1}  )
\bigg\{(R_{P,\si; 2}\La_{P,\si;2}) (R_{P,\si; 1}\La_{P,\si;1})  \bigg\}\pa_{P^{j'}}\cpsi_{P,\si}\non\\
\y&=&\y(-1)P_{\sym}g_{P,\si; 1} g_{P,\si; 2} 
\bigg\{  \pa_{P^j}R_{P,\si; 2  }  \bigg\}\pa_{P^{j'}}\cpsi_{P,\si}\non\\
\y&=&\y(-1)P_{\sym}g_{P,\si; 1} g_{P,\si; 2} 
\bigg\{ R_{P,\si; 2  }(\La^2_{P,\si})^j  R_{P,\si; 2  }\bigg\} 
R_{P,\si}(\La_{P,\si})^{j'}\cpsi_{P,\si}, \label{example}
\eeqa
where in the last step we applied (\ref{rules-of-the-game}). This expression has
a structure {\color{\red} similar } to (\ref{second-der-problem}), but the essential difference is that it is a contribution to the two-photon and not the one-photon wave function.  Thus
using the simple resolvent bounds $\|R_{P,\si; 2  }(\La^2_{P,\si})^j\|, \|R_{P,\si; 2 }\|\leq c|\unk|_2^{-1}$, the facts that $|\unk|_2^{-1}\leq |\unk|_1^{-1}$, 
$\|R_{P,\si}(\La_{P,\si})^{j'}\cpsi_{P,\si}\|\leq c\ir$  and $|g_{P,\si}(k)|\leq cv_{\alf}^{\si}(k)$, we obtain in agreement with our main result~(\ref{infrared-spectral-bound-zero})
\beqa
\|(\ref{example})\|\leq c\ir \fr{v_{\alf}^{\si}(k_1)}{|k_1|}\fr{v_{\alf}^{\si}(k_2)}{|k_2|}.
\eeqa

This paper is organized as follows: In Section~\ref{Preliminaries-and-results} we list some
preliminaries and state precisely our results. In Section~\ref{non-commutative section} we 
develop the theory of non-commutative recurrence relations in a form suitable for our investigation.
In Section~\ref{formulas-for-ground-state-wavefunctions} we use this theory to  derive 
formulas~(\ref{check-intro}), (\ref{hfn-formula-intro}) for the ground state wave functions.
{\color{\red}In Section~\ref{Derivatives-section} we identify convenient sufficient conditions for the 
bounds~(\ref{infrared-spectral-bound-zero}) on the derivatives of the wave functions to  hold.} These conditions are then verified in Section~\ref{verification-section}. More standard parts of the discussion are postponed to the appendices.

\vspace{0.5cm}

\noindent{\bf Acknowledgment:} A.P. thanks the Hausdorff Research Institute of Mathematics, Bonn, for hospitality. A.P. is supported by the NSF grant \#DMS-0905988.

W.D.   thanks the University of California Davis and the Hausdorff  Research Institute for  Mathematics, Bonn,  for hospitality. 
W.D. is  supported by the German Research Foundation (DFG) within the
grants SP181/25--2,  DY107/1--1, DY107/2--1. Moreover, he would like to acknowledge the support of 
the Danish Council for Independent Research, grant no. 09-065927 "Mathematical Physics", and of the Lundbeck Foundation.

\section{Preliminaries and results}\label{Preliminaries-and-results}
\setcounter{equation}{0}
In this section we list some preliminaries and state the main results of this paper which were announced already in Section~1.2 of \cite{DP12.1}.

Let $\mfh_{\fib}=L^2(\real^{3}, d^3k)$ be the single-photon subspace in the fiber picture.
Let $\Ga(\mfh_{\fib})$ be the symmetric Fock space over $\mfh_{\fib}$ and let us denote the
corresponding (improper) creation and annihilation operators  by $b^*(k)$ an $b(k)$. We will
denote by $\ka=1$ the ultraviolet cut-off and by $\si\in (0,\ka]$ the infrared cut-off.  The fiber
Hamiltonian of the massless Nelson model with these cut-offs is a self-adjoint operator on
a domain $D(H_{P,\si})\subset \Ga(\mfh_{\fib})$ {\color{\red}and it is} given by the formula
\beqa
H_{P,\si}:=\h(P-P_{\pho})^2+H_{\pho}+\int d^3k\, \vv^{\si}(k)\, ( b(k)+b^*(k) ), \label{infrared-cut-off-Hamiltonian}
\eeqa  
where $H_{\pho}=\int d^3k\, |k|\,b^*(k)b(k)$, $P_{\pho}:=\int d^3k\, k \, b^*(k)b(k)$ are the photon total energy and 
momentum operators. By the Kato-Rellich  theorem, we have  $D(H_{P,\si})=D(H_{P,\free})$, 
where $H_{P,\free}$ is given by (\ref{infrared-cut-off-Hamiltonian}) with the interaction term omitted.
To define  the form factor $\vv^{\si}$ we need to introduce some notation: Let $\mcB_r$ be the open ball of radius $r>0$ centered at zero.
Let $\chi_{r}\in C_0^{\infty}(\real^3)$ be a function which is rotationally invariant, non-increasing in the radial direction, supported in $\mcB_r$ 
and equal to one on $\mcB_{(1-\eps_0)r}$, for $0<\eps_0<1$.  Let $\g\neq 0$ be the coupling constant.  
 Then the form-factor is given by
\beqa
\vv^{\si}(k):=\g\fr{ \chi_{[\si,\ka)}(k)    |k|^{\alf} }{(2|k|)^\half}, \label{IR-cut-off-propagator}
\eeqa
where $\chi_{[\si,\ka)}(k):=\mathbf{1}_{  \mcB'_{\si}   }(k)\chi_{\ka}(k)$  and 
$\mcB_{\si}'=\real^3\backslash \mcB_{\si}$.  The parameter $0\leq \alf\leq 1/2$ was needed in \cite{DP12.1} and is kept
here only for consistency with this earlier work. We stress that all the results listed below remain true in the infrared singular
case $\alf=0$.

As our analysis concerns the  bottom of the spectrum of the fiber Hamiltonians, let us define
\beqa
   E_{P,\si}:=\inf \boldsymbol{\si}(H_{P,\si}),
\eeqa
where $\boldsymbol{\si}$ denotes the spectrum.
Under certain conditions, stated  in Theorem~\ref{preliminaries-on-spectrum} below,  {\color{\red}$E_{P,\si}$ is a non-degenerate isolated  eigenvalue and its normalized eigenvector is denoted by
 $\cpsi_{P,\si}$. The phase of the vector $\cpsi_{P,\si}$ is fixed by the definitions
\beqa \label{def-checkpsi}
\check{\psi}_{P,\sigma}:= W^*_{P,\sigma}\cphi_{P,\sigma}, \quad
 \check{\phi}_{P,\sigma}:=\frac{\oint_{\gamma}\frac{dw}{H_{P,\sigma}^{W}-w}\Omega}{\|\oint_{\gamma}\frac{dw}{H_{P,\sigma}^{W}-w}\Omega\|}, 
\eeqa
where $H_{P,\sigma}^{W}:=W_{P,\sigma}H_{P,\sigma}W_{P,\sigma}^{*}$ with $W_{P,\sigma}$ defined in (\ref{def-W}), $\Omega$ is the vacuum vector in $\Ga(\mfh_{\fib})$, and $\gamma$ is a circle of integration enclosing no other point of the spectrum of $H_{P,\sigma}^{W}$ apart form $E_{P,\sigma}$; for details see Section 5 in  \cite{DP17.1}.}
Since we are interested in small values of the total momentum $P$ at which the electron 
moves slower than the photons, we will consider $P$ from the set
\beqa
\PS:=\{\, P\in \real^3\,|\, |P|< 1/3\,\}. \label{S-definition}
\eeqa
As we  work in the weak coupling regime we restrict attention to $|\g|\in (0, \g_0]$, where $\g_0$ is  sufficiently small as specified in Theorem~\ref{preliminaries-on-spectrum}.

 The following theorem collects some results of the companion paper \cite{DP17.1}, which are relevant for the present investigation. 
We remark that regularity of $P\mapsto E_{P,\si}$ in infrared singular models was studied before in particular in \cite{AH12,
KM12, FP10}. Analyticity of ground state projections  was established in some infrared regular models  in \cite{FFS14}. 
However, to our knowledge, derivatives of ground state vectors and their dependence on the infrared cut-off  was not treated in infrared singular models before \cite{DP17.1}.
\begin{thm}\label{preliminaries-on-spectrum} Fix $0\leq \alf \leq 1/2$. Then there exists $\g_0>0$ s.t. for all $|\g|\in (0, \g_0]$ and $P\in S$,
  the following statements hold:
\begin{enumerate}[label = \textup{(\alph*)}, ref =\textup{(\alph*)},leftmargin=*]
\item\label{energy-part} 
$S\ni P\mapsto E_{P,\si}$ is analytic and strictly convex, for all $\si\in (0,\ka]$.  Moreover, 
\beqa
& &|\pa_{P}^{\be_1}E_{P,\si}|\leq c,\quad  |\pa_{P}^{\be_2}E_{P,\si}|\leq c, \quad  |\pa_{P}^{\be_3}E_{P,\si}|\leq c/\si^{\de_{\la_0}} 
\label{velocity-boundedness}
\eeqa
for multiindices $\be_\ell$ s.t. $|\be_\ell|=\ell$, $\ell\in\{1,2,3\}$.

\item\label{cut-off-part} For  $\si\in (0,\ka]$, $E_{P,\si}$ is a simple eigenvalue corresponding to a normalized eigenvector $\cpsi_{P,\si}$,
whose phase is fixed in (\ref{def-checkpsi}).
There holds
\beqa
\|\pa^{\be}_P\cpsi_{P,\si}\|\leq  c /\si^{\de_{\g_0}} \label{state-bound}
\eeqa
for  multiindices $\be$ s.t. $0<|\be|\leq 2$.

\end{enumerate}
The constant $c$ above is independent of $\si$, $P$, $\g$, $\alf$ within the assumed restrictions. Clearly, all statements above
remain true after replacing $\g_0$ by some $\ti\g_0\in (0,\g_0]$. The resulting function
  $\ti\g_0\mapsto \de_{\ti\g_0}$ can be chosen positive and s.t. $\lim_{\ti\g_0\to 0} \de_{\ti\g_0}=0$.
\end{thm}

As mentioned in the introduction, the regularity of the vector $\cpsi_{P,\si}$, established in Theorem~\ref{preliminaries-on-spectrum}, is not sufficient for scattering theory of two electrons. One also needs similar bounds for their wave functions $f^n_{P,\si}$.   
Clearly, each $f^{\mm}_{P,\si}$   is a square-integrable function symmetric in $\mm$ variables from $\real^3$. We introduce the following auxiliary functions:
\beqa
 g^{\mm}_{\si}(k_1,\ldots, k_{\mm}):=\prod_{i=1}^{\mm}\fr{ c\g  \chi_{[\si,\kas)}(k_i) |k_i|^{\alf} }{|k_i|^{3/2}},\quad \kas:=(1-\eps_0)^{-1}\ka, \label{def-g-ks}
\eeqa
where $\tic$ is some positive constant independent of $\mm,\si, P$ and $\g$ within the restrictions specified above. 
We also introduce the notation
\beqa
\mcA_{r_1,r_2}:=\{\, k\in \real^3 \,|\, r_1<|k|<r_2 \,\}, \label{A-set}
\eeqa
where $0\leq r_1< r_2$. 
Now we are ready to state the required properties of the functions $f^{\mm}_{P,\si}$.
\begin{thm}\label{main-theorem-spectral} 
Fix $0\leq \alf \leq 1/2$. Then there exists  $\lambda_0>0$ s.t. for all $|\lambda|\in (0, \lambda_0]$ and $P \in S$ 
  the following statements hold:
\begin{enumerate}[label = \textup{(\alph*)}, ref =\textup{(\alph*)},leftmargin=*]

\item \label{f-m-support} Let  $\{f^n_{P,\sigma}\}_{n\in\nat_0}$ be the $n$-particle components of $\cpsi_{P,\sigma}$
and let $\overline{\mathcal{A}}_{\sigma,\kappa}^{\times n}$ be defined  as the Cartesian product of $n$ copies of the closure of the set $\mathcal{A}_{\sigma,\kappa}$ introduced in (\ref{A-set}). Then,
  for any $P\in S$, the function $f^n_{P,\sigma}$ is supported in $\overline{\mathcal{A}}_{\sigma,\kappa}^{\times n}$.

\item \label{f-m-smoothness} The  function
\begin{equation}
S\times \mathcal{A}_{\sigma,\infty}^{\times n}\ni (P; k_1,\dots, k_n) \mapsto f^n_{P,\sigma}(k_1,\dots, k_n) \label{momentum-wave-functions}
\end{equation}
is twice continuously differentiable and extends by continuity, together with its derivatives, to the set 
$S\times \overline{\mathcal{A}}_{\sigma,\infty}^{\times n}$. 

\item\label{derivatives-bounds} For any multiindex $\beta$, $0\leq |\beta|\leq 2$, the function (\ref{momentum-wave-functions}) satisfies
\begin{eqnarray}
|\partial^{\beta}_{k_l}f^n_{P,\sigma}(k_1,\dots, k_n)|\y&\leq&\y \frac{1}{\sqrt{n!}} |k_l|^{-|\beta|}g^n_{\sigma}(k_1,\dots, k_n), 
\label{simple-spectral-bound}\\
|\partial^{\beta}_{P}f^n_{P,\sigma}(k_1,\dots, k_n)|\y&\leq& \y \frac{1}{\sqrt{n!}}  \frac{1}{\sigma^{\delta_{\lambda_0}}}   g^n_{\sigma}(k_1,\dots, k_n), 
\label{infrared-spectral-bound}\\
|\partial_{P^{j'}}\partial_{k_l^j}f^n_{P,\sigma}(k_1,\dots, k_n)|\y& \leq &\y  \frac{1}{\sqrt{n!}}\frac{1}{\sigma^{\delta_{\lambda_0}}}|k_l|^{-1}g^n_{\sigma}(k_1,\dots, k_n), \label{mixed-spectral-bound}
\end{eqnarray}
where the function $\tilde{\lambda}_0 \mapsto \delta_{\tilde{\lambda_0}}$ has the properties specified in Theorem~\ref{preliminaries-on-spectrum}, $g^n_{\sigma}$ is defined in~(\ref{def-g-ks}) and the factor  $\sigma^{-\delta_{\lambda_0}}$ in (\ref{infrared-spectral-bound}) can be omitted
for $|\be|=0$.
\end{enumerate}
\end{thm}

Parts \ref{f-m-support}, \ref{f-m-smoothness}  and   estimate~(\ref{simple-spectral-bound}) in 
\ref{derivatives-bounds}  can be extracted from \cite{Fr73,Fr74.1,Fr73.1} or proven using  methods from these papers. 
Estimate~(\ref{infrared-spectral-bound}) for $|\be|=1$ and (\ref{mixed-spectral-bound}) require only the standard formula
(\ref{check-intro}) for the wave functions and the spectral ingredient (\ref{first-spectral-ingr}). All these straightforward estimates
are proven in Appendix~\ref{standard-proofs}. The main part of the paper is devoted to the proof of (\ref{infrared-spectral-bound})
for $|\be|=2$ which requires both spectral ingredients (\ref{first-spectral-ingr}) and (\ref{eq:inequality-cont-1-intro-x}) and the novel
formula for the wave functions (\ref{hfn-formula-intro}).

\newpage

\noindent \bf Standing assumptions and conventions: \rm
\begin{enumerate}

\item The  parameter  $1/2\geq \alf\geq 0$, which appeared in (\ref{IR-cut-off-propagator}) is kept fixed in the remaining part of the paper.

\item The total momentum $P$ belongs to the set $S$ defined in (\ref{S-definition}). The photon momenta $k_1,\ldots, k_n$ take arbitrary
values from $\real^3$.

\item The maximal value of the coupling constant $\g_0>0$  is sufficiently small. In particular  s.t. Theorem~\ref{preliminaries-on-spectrum} holds.

\item $\ti\g_0\mapsto \de_{\ti\g_0}$  denotes a positive function of $\ti\g_0\in (0,\la_0]$, which may differ from line to line  and has the property $\lim_{\ti\g_0\to 0} \de_{\ti\g_0}=0$. (Cf. Theorem~\ref{preliminaries-on-spectrum}).

\item We will denote by $c,c',c''$ numerical constants which may change from line to line.

\item We reserve letters $j,j'=1,2,3$ to denote vector indices (as superscripts). Letters $i,i'$ usually count photon momenta.  E.g. $k^j_1, \ldots, k^j_i, \ldots k^j_n$.

\end{enumerate}

\noindent \bf Notation related to non-commutative recurrence relations: \rm
\begin{enumerate}

\item $\Op^{(n)}_{\hua, \hub}$ is the quantization map defined in formula~(\ref{Opmap}).

\item The operator $\hat\de_i=:\hat{b}_{i+1,i}\hat{\pa}_{i+1}\hat{\pa}_i$ defined on the range of the quantization map appears for the first time
below formula~(\ref{de-on-op})

\item The symmetrization operator $P_{\sym}$ is defined in (\ref{sym-def}).

\item For the operators $\hat a_{\mrm{ns}}$ and $\hat a:= P_{\sym} \hat a_{\mrm{ns}}$ see (\ref{hat-a-ns-zero})-(\ref{hat-a-ns}).

\item For $(R_{P,\si; i+1} \hat\pa_{i+1} \hat\pa_{i}  ), \ C_{\ell}, \ C^{2\ell}, \ \hat\de_{C_{\ell}}, \ \hat\de'_{C_{\ell}}, \ \hat\de''_{C_{\ell}}, \ I_{C_{\ell}}$ we refer
to (\ref{hat-I-relation})-(\ref{skipped-factors}). 

\item For $\al(n,\ldots, 1):=\Op^{(n)}_{\hua, \hub}(a_n,\ldots, a_1)=
(R_{P,\si;n}\La_{P,\si;n})\ldots (R_{P,\si;1}\La_{P,\si;1})$, see (\ref{alpha-definition}), (\ref{alpha-in-terms-of-RL}).

\item For $\pa_{R_{i}}I_{C_{\ell}}$, $\pa_{\La_{i}} I_{C_{\ell}}$ and $\pa_{\cpsi} I_{C_{\ell}}$ and 
second derivatives of this type see Section~\ref{Derivatives-section}.

\item $i\ll i'$ means $i+1<i'$.

\end{enumerate}

\noindent \bf Notation related to the Nelson model: \rm
\begin{enumerate}

\item $b_W(k):=b(k)+f_{P,\si}(k)$, where  $f_{P,\si}(k):=\fr{  \vv^{\si}(k)    }{|k|\al_{P,\si}(\hat k) }, \
\al_{P,\si}(\hat k):=(1-\hat k\cdot \nabla E_{P,\si})$ and $\hat k:=k/|k|$, appears in~(\ref{new-formula}).

\item  $g_{P,\si}(k):=f_{P,\si}(k)k$ appears in (\ref{g-defin}).

\item $\check{f}^m_{P,\si}(k_1,\ldots, k_m):=b(k_1)\ldots b(k_m)\cpsi_{P,\si}$ is introduced in  (\ref{vector-valued-distr}). 

\item  $\hat{f}^m_{P,\si}(k_1,\ldots, k_m):=b_{W}(k_1)\ldots b_{W}(k_m)\cpsi_{P,\si}$ is introduced in (\ref{main-vectors}).

\item Given $k_1,\ldots, k_n$  and a permutation $\pi\in S_n$ we write 
\beqa
& &\unk_i:= k_1+\cdots +k_i, \quad\ph{444} \unk_{\pi,i}:=k_{\pi(1)}+\cdots +k_{\pi(i)}, \\
& &|\unk|_i:=|k_1|+\cdots+|k_i|, \quad |\unk|_{\pi,i}:=|k_{\pi(1)}|+\cdots+|k_{\pi(i)}|.
\eeqa
\item We  use the following quantities:
\beqa
& &\La_{P,\si}:= \nabla E_{P,\si}-(P-P_{\pho}),\\
& &\La_{P,\si; i}:=\La_{P,\si}+\unk_{i-1}+\h k_{i}, \\
& &\La^i_{P,\si}:=\La_{P,\si}+\unk_{i}, \\
& &F_{P,\si}(\ti k,k):=-g_{P,\si}(\ti k){\cdot}\bigg(\La_{P,\si}+k+\h \ti k\bigg), 
\eeqa
which appear in (\ref{Lambda-def}), (\ref{lambda-first-appearance}), (\ref{first-resolvent-i}),  (\ref{F-defin}) respectively.

\item We  use the following quantities:
\beqa
& &H_{P,\si; k_1,\ldots, k_n }:=H_{P- (k_1+\cdots+k_n),\si}+| k_1|+\cdots+|k_n|, \\
& &R_{P,\si; k_1, \ldots, k_n}:=\fr{1}{H_{P,\si; k_1,\ldots, k_n }-E_{P,\si} }, 
\eeqa
which appear in (\ref{Hamiltonian-sum-one-x-x}) and (\ref{R-first-definition}).

\item We use the following resolvents
\beqa
& &R_{P,\si}:=\fr{1}{H_{P,\si}-E_{P,\si}},  \\
& &R_{P,\si;i}:=\fr{1}{H_{P-\unk_{i},\si}-E_{P,\si}+|\un k|_{i}}, \\ 
& &R_{P,\si}^i:=\fr{1}{H_{P,\si}-E_{P,\si}+ r(\unk)_{i} }, 
\eeqa
 where  $r(\unk)_{i}:=|\unk|_{i}+|\unk_{i}|^2/2-\nabla E_{P,\si}\cdot \unk_{i}$. 
They appear in (\ref{first-res}), (\ref{sec-res}), (\ref{where-res}), respectively.
We note that $R_{P,\si;i}=R_{P,\si; k_1,\ldots, k_i }$.

\item In Section~\ref{verification-section}  and Appendix~\ref{Appendix-trivial} we  drop  
subscripts $P,\si$ from most quantities considered above. For example
$\La^i:=\La_{P,\si}^i$.

\item In Appendix~\ref{Appendix-trivial} we consider the above quantities with
permuted photon momenta. Then we write e.g. $\La^i_{\pi}:=\La_{P,\si}+\unk_{\pi,i}$.

\end{enumerate}

\section{ Non-commutative recurrence relations}\label{non-commutative section}
\setcounter{equation}{0}
\subsection{General theory}\label{non-commutative-subsection}

Let $\una:=\{a_i\}_{i\in\nat}$, $\unb:=\{ b_{i+1,i} \}_{i\in\nat}$ be two sequences of real variables, $\Pol(\una,\unb)$ be the commutative algebra of all polynomials in $\una$ and $\unb$ and $\Pol(\una)$ the algebra of polynomials in $\una$.
We denote by $\pa_{i}:=\pa/\pa_{a_i}$ the operators on $\Pol(\una,\unb)$ acting by differentiation w.r.t. $a_i$. (Derivatives w.r.t. $b_{i+1,i}$ will not be used).  We also define the following operators on $\Pol(\una,\unb)$ for $i\in \nat$
\beqa
\de_i{\color{\red}:=}b_{i+1,i}\pa_{i+1}\pa_{i}, \label{new-derivative} 
\eeqa
where $b_{i+1,i}$ is understood as a multiplication operator. Furthermore, we define the following subspaces of $\Pol(\una,\unb)$ for $n\in\nat$
\beqa\label{gen}
\wt{\Pol}^{(n)}\!\!(\una,\unb):=\Span \{ \de_{n}^{\mu_n}\de_{n-1}^{\mu_{n-1}} \ldots \de_{1}^{\mu_1} (a_n a_{n-1}\ldots a_1)\,|\, \mu_i\in\{0,1\}\, \},
\eeqa
{\color{\red}where $\de_{j}^{\mu_j}=\de_j$ if $\mu_j=1$, and $\de_{j}^{\mu_j}=1$ if $\mu_j=0$.}
We note that $\wt{\Pol}^{(n)}\!\!(\una,\unb)$ is spanned by monomials of the form
\beqa
a_n\ldots b_{i+1,i}\ldots b_{i'+1,i'}\ldots a_1
\eeqa
which arise by removing pairs $a_{i+1}a_i$ from the monomial $a_n\ldots a_1$ and replacing them with $b_{i+1,i}$.

We can treat $\wt\Pol^{(n)}\!\!\!(\una,\unb)$ as a class of polynomial symbols and define
their quantization as follows: Let $\{X_i\}_{i\in\nat_0}$ be a family of vector spaces and $L(X_i, X_j)$ is the space of linear maps from $X_i$ to $X_j$.
Let $\hua:=\{\hat a_i\}_{i\in\nat}$, $\hub:=\{\hat b_{i+1,i}\}_{i\in\nat}$ be two sequences of linear operators
$\hat a_i: X_{i-1}\to X_{i}$, $\hat b_{i+1,i}: X_{i-1}\to X_{i+1}$.
We define the  linear map $\Op^{(n)}_{\hua, \hub}: \wt{\Pol}^{(n)}\!\!(\una,\unb) \to L(X_0, X_n)$ by the following specification on monomials
\beqa
\Op^{(n)}_{\hua, \hub}[ a_n\ldots b_{i+1,i}\ldots b_{i'+1,i'}\ldots a_1]=\ha_n\ldots \hb_{i+1,i}\ldots \hb_{i'+1,i'}\ldots \ha_1.  \label{Opmap}
\eeqa
{\color{\red}The case with $X_i=X$, $\ha_i=\ha$ and $\hb_{i+1,i}=\hb$ is of particular importance. In this case} we simply denote the quantization map
by $\Op^{(n)}_{\ha, \hb}$.

Clearly, various operations on $\wt{\Pol}^{(n)}\!\!\!(\una,\unb)$  naturally carry over to operators from the ranges of the quantization maps.
For example, we can define
\beqa
\hat\de_i\Op_{\hua, \hub}[p(\una,\unb)]:=\Op_{\hua, \hub}[\de_i p(\una,\unb)], \quad p(\una,\unb)\in \wt{\Pol}^{(n)}\!\!(\una,\unb).
\label{de-on-op}
\eeqa
If there is a need for more explicit notation, we will write $\hat\de_i=:\hat{b}_{i+1,i}\hat{\pa}_{i+1}\hat{\pa}_i$. We note, however,
that $\hat{\pa}_i$ is not well defined, since $\pa_i$ does not leave $ \wt{\Pol}^{(n)}\!\!\!(\una,\unb)$ invariant.

Now we can prove the main result of this section:
\begin{thm} Let $X$ be a vector  space, $\{x_n\}_{n\in\nat_0}$ a sequence of vectors from $X$ and $\hat a, \hat b\in L(X)$.
Suppose  the following recurrence relation holds
\beqa
x_n=\hat a x_{n-1}+\hat b x_{n-2} \label{abstract-recurrence}
\eeqa
with the initial conditions $x_0$ and $x_1=\hat a x_0$.  Then, for $n\geq 1$ 
\beqa
x_n=\Op^{(n)}_{\hat a, \hat b}[\exp(\sum_{i=1}^{n-1}b_{i+1,i}\pa_{i+1}\pa_{i}   ) a_n \ldots a_1]x_0,
\label{recurrence-solution-x}
\eeqa
where $\exp$ is defined by its power series and for $n=1$ the sum in the exponent above is equal to zero by convention. {\color{\red}(We recall that the symbol $\Op^{(n)}_{\hat a, \hat b}$ implies that all the $\hat{a}_n$ are equal to $\hat{a}$ and all $\hat{b}_{i+1,i}$ are equal to $\hat{b}$.)}
\end{thm}
\proof We asume $n\geq 2$, write as before $\de_{i}:=b_{i+1,i} \pa_{i+1} \pa_{i} $ and set $d_{n-1}:=\sum_{i=1}^{n-1}\de_i$. 
Now we  compute 
\beqa
\exp(d_{n-1}) a_n \ldots a_1\y&=&\y\{\exp(d_{n-2}) \exp(\de_{n-1})a_na_{n-1}\ldots a_1\}\non\\
\y&=&\y \{\exp(d_{n-2}) a_n a_{n-1}\ldots a_1\}+\{\exp(d_{n-2})(\de_{n-1}) a_{n}a_{n-1}\ldots a_1\}\non\\
\y&=&\y a_n\{\exp(d_{n-2}) a_{n-1}\ldots a_1\}+ b_{n,n-1}\{\exp(d_{n-3}) a_{n-2}\ldots a_1\}. \label{poly-computation}
\eeqa
 Here in the second step we used that $(\de_{n-1})^2a_{n}a_{n-1}\ldots a_1=0$ and in the last step we could replace
$d_{n-2}$ with $d_{n-3}$ since $\de_{n-2} a_{n-2}\ldots a_n=0$. (For $n=2$  we have $a_{n-2}\ldots a_n\equiv 1$ by convention).  Relation~(\ref{poly-computation}) gives
\beqa
\Op^{(n)}_{\hat a, \hat b}[\exp(d_{n-1}) a_n \ldots a_1]=\hat a\, \Op^{(n-1)}_{\hat a, \hat b}[\exp(d_{n-2}) a_{n-1}\ldots a_1]
+\hat b\,  \Op^{(n-2)}_{\hat a, \hat b}[\exp(d_{n-3}) a_{n-2}\ldots a_1]. \ \   \label{operator-relation}
\eeqa
Acting with both sides of (\ref{operator-relation}) on $x^0$
we obtain the recurrence relation (\ref{abstract-recurrence}). With the specified initial conditions it determines
$x_n$, $n\geq 2$, uniquely. \qed

{\color{\red}In the next lines,} we derive a variant of formula~(\ref{recurrence-solution-x}) which is  convenient for our applications. Still using the notation   
$\de_i:=b_{i+1,i} \pa_{i+1}\pa_{i} $ {\color{\red}and the property $(\de_{i})^2=0$,} {\color{\red}  we observe that} the multinomial formula implies
\beqa
(\sum_{i=1}^{n-1}\de_i    )^{\ell}  a_n \ldots a_1=\sum_{\mu: |\mu|=\ell} \fr{\ell!}{\mu!} \de^{\mu}
  a_n \ldots a_1, \label{multinomial}
\eeqa
where $\mu{\color{\red}:=}(\mu(1),\dots,\mu(n-1))$ is an $n-1$-index, $\mu!=\mu(1)!\ldots \mu(n-1)!$, $|\mu|=\sum_{i=1}^{n-1}\mu(i)$ and $\de^{\mu}{\color{\red}:=}\de_1^{\mu(1)}\ldots \de_{n-1}^{\mu(n-1)}$. Now we
note that $\de_i^2(a_n \ldots a_1)=0$, thus only for $\mu(i)\in \{0,1\}$ the corresponding terms can be non-zero. 

\noindent
{\color{\red} Next, since $\de_i^2(a_n \ldots a_1)=0$ and $\de_i\de_{i+1}(a_n \ldots a_1)=0$}  we can write
\beqa
x_n= \sum_{\ell=0}^{[n/2]}\fr{1}{\ell!}
\Op_{\hat a, \hat b}^{(n)} [ (\sum_{i=1}^{n-1}\de_i    )^{\ell}  a_n \ldots a_1]x_0. 
\label{first-rec-rearrangement} 
\eeqa
Hence (\ref{first-rec-rearrangement}), (\ref{multinomial}) give
\beqa
x_n=\sum_{\ell=0}^{[n/2]}\sum_{\mu: |\mu|=\ell} \Op^{(n)}_{\hat a, \hat b} [ (\de)^{\mu}  a_n \ldots a_1]x_0= 
\sum_{\ell=0}^{[n/2]}\sum_{1\leq i_1 \ll\cdots \ll i_{\ell}\leq n-1 } \Op^{(n)}_{\hat a, \hat b} [ \de_{i_1}\ldots \de_{i_{\ell}}  a_n \ldots a_1]x_0, \label{x-n-formula}
\eeqa
where $i\ll i'$ means $i+1<i'$. 

Now {\color{\red} we consider a class of non-commutative recurrence relations for which the solution can be expressed in terms of the generalized framework in (\ref{Opmap}). Suppose} that $X=\bigoplus_{i\in\nat_0} X_i$, set $\ha_i:=(\ha\restriction X_{i-1})$, $\ha_i:=(\hb\restriction X_{i-1})$ and assume
that $\ha_i: X_{i-1}\to X_i$, $\,\hb_{i+1,i}: X_{i-1}\to X_{i+1}$ and $x_0\in X_0$. Thus setting $\hua:=\{\ha_i\}_{i\in\nat}$, $\hub:=\{\hb_{i+1,i}\}_{i\in\nat}$ we have
\beqa
x_n=\sum_{\ell=0}^{[n/2]}\sum_{1\leq i_1 \ll\cdots \ll i_{\ell}\leq n-1 } \hat\de_{i_1}\ldots \hat\de_{i_{\ell}}\Op^{(n)}_{\hua, \hub} [   a_n \ldots a_1]x_0. \label{solution-of-the-recurrence}
\eeqa

\subsection{Fock space combinatorics}\label{FS-comb}

{\color{\red}The formula  in (\ref{solution-of-the-recurrence}) will be applied to the recurrence relation derived in (\ref{recurrence-relation-W}) for the vectors $\hf^n(k_{1},\ldots, k_{n})$ by which we express the ground state wave functions $f^n_{P,\si}(k_1,\ldots,k_n)$ in (\ref{f-n-hat-f}). For the given sequence, it is not immediate to identify the families of symmetrized (in the variables $k_1,\dots, k_n$) operators $\ha_i$ and $\hb_{i+1,i}$ that enter the solution according to the formula in (\ref{solution-of-the-recurrence}). To this end, it is  convenient to start from non-symmetrized operators acting on non-symmetric tensor products of Hilbert spaces which are related to  $\ha_i$ and $\hb_{i+1,i}$. The definition of the non-symmetrized operators is dictated by the specific recurrence relation  in  (\ref{recurrence-relation-W}) though in our discussion below  this is not made explicit.  }

We start with an elementary observation, which will be important in the later
part of this section.

\begin{lem}\label{symmetrisation-lemma} Let $f$ be a function of $m$ variables $k_1,\ldots, k_m\in \real^3$ with values in some set $Y$. 
Denote its
symmetrization in $m'$ consecutive variables $k_{i_1},\ldots, k_{i_{m'}}$ by
\beqa
[f(k_1,\ldots, k_m)]^{(i_1,i_{m'}) }_{\sym}:=\fr{1}{m'!}\sum_{\pi\in S_{m'}} 
f(k_{1},\ldots, k_{\pi{(i_1)}},\ldots, k_{\pi(i_{m'}) },   \ldots, k_{m}),
\eeqa
where  $S_{m'}$ is the set of permutations of $m'$ variables.  Then
\beqa
 [[f(k_1,\ldots, k_m)]^{(i_1,i_{m'}) }_{\sym}]_{\sym}=[f(k_1,\ldots, k_m) ]_{\sym},
\eeqa
where  $[\ldots ]_{\sym}:=[\ldots ]^{(1,m)}_{\sym}$.
\end{lem}
\proof We write
\beqa
[[f(k_1,\ldots, k_m)]^{(i_1,i_{m'}) }_{\sym}]_{\sym}\y &=& \y
\fr{1}{m'!}\sum_{\pi\in S_{m'}} \fr{1}{m!}\sum_{\rho\in S_{m}}   
f(k_{\rho(1)},\ldots, k_{\rho(\pi{(i_1)}) },\ldots, k_{\rho(\pi(i_{m'}))  },   \ldots, k_{\rho(m)})\non\\
\y &=&\y \fr{1}{m'!}\sum_{\pi\in S_{m'}} \fr{1}{m!}\sum_{\rho'_{{\color{\red}\pi}}\in S_{m}}   
f(k_{\rho'_{{\color{\red}\pi}}(1)},\ldots, k_{\rho'_{{\color{\red}\pi}}(i_1) },\ldots, k_{  \rho'_{{\color{\red}\pi}}(i_{m'})  },   \ldots, k_{\rho'_{{\color{\red}\pi}}(m)})\non\\
\y &=&\y {\color{\red}[[f(k_1,\ldots, k_m) ]_{\sym}]^{(i_1,i_{m'}) }_{\sym}}\non\\
\y &=&\y [f(k_1,\ldots, k_m) ]_{\sym},
\eeqa
where we set $\rho'_{{\color{\red}\pi}}=\rho\circ \ti\pi$ and $\ti\pi$ is the extension of $\pi$ from $(i_1,\ldots,i_{m'})$ to $(1,\ldots,m)$ by {\color{\red}the}  identity. \qed\\
Let $\hil$ be a Hilbert space, 
$\Ga_{\ns}(L^2(\real^3)):=\bigoplus_{m\in\nat_0} L^2(\real^3)^{\otimes m}$ be the
non-symmetrized Fock space and $X_{\ns}:=\Ga_{\ns}(L^2(\real^3))\otimes\hil$.
The elements of $X_{\ns}$ have the form $x_{\ns}=\{x_{\ns}^m\}_{m\in\nat_0}$, where  $x_{\ns}^m$ are 
$\hil$-valued square integrable functions of $m$ variables from $\real^3$. 
Let $P_{\sym}$ be the symmetrization operator on $X_{{\color{\red}\ns}}$ given by
\beqa
(P_{\sym}x_{\ns})^{m}(k_1,\ldots, k_m)=[ x_{\ns}^m(k_{1},\ldots, k_{m})]_{\sym}, \label{sym-def}
\quad x_{{\color{\red}\ns}}\in X_{\ns}.
\eeqa
The symmetric Fock space is given by $\Ga(L^2(\real^3)):=\bigoplus_{m\in\nat_0} P_{\sym}L^2(\real^3)^{\otimes m}$ and  
$X:=\Ga(L^2(\real^3))\otimes \hil$.
The elements of $X$ will be denoted by $x=\{x^m\}_{m\in\nat_0}$ and we treat $X$ as a subspace of $X_{\ns}$. We denote by
$X_{\ns,\fin}$, $X_{\fin}$ dense subspaces of $X_{\ns}, X$, respectively, consisting of
vectors with only a finite number of non-zero Fock space components.

Let $\ha_{\ns}=\{\ha_{\ns,m}\}_{m\geq r}$ be a {\color{\red} sequence 
of bounded functions {\color{\red}of $3m$ real variables}
with values in $B(\hil)$, supported in some compact sets $C_m\subset \real^{3m}$
and norm continuous in their interiors.
It defines an operator on $X_{\ns,\fin}$ as follows
\beqa
& &(\ha_{\ns}x_{{\color{\red}\ns}})^m(k_1, \ldots, k_m){\color{\red}:=}\ha_{\ns,m}(k_1,\ldots, k_{m-r},k_{m-r+1},\ldots,k_m)x_{{\color{\red}\ns}}^{m-r}(k_1, \ldots, k_{m-r}) \ \textrm{for} \ m\geq r, \quad
\label{hat-a-ns-zero}\\
& &(\ha_{\ns}x_{{\color{\red}\ns}})^m(k_1,\ldots,k_m){\color{\red}:=0} \quad \textrm{for} \quad m<r. \label{hat-a-ns}
\eeqa
Similarly we define $\ha:=P_{\sym} \ha_{\ns}$ as an operator on $X$ and call $r\in\nat_0$ the order of the operators $\ha_{\ns}, \ha$. 
The following fact will be helpful for identifying  $\ha$, $\ha_{\ns}$ in examples.
\begin{lem}\label{sum-rep} Let $\ha_{\ns}$ be an operator on $X_{\ns}$ of order $r$  defined as in (\ref{hat-a-ns-zero})-(\ref{hat-a-ns}). 
Suppose that the functions $\ha_{\ns,m}$ are symmetric in the collections of variables $(k_1,\ldots,k_{m-r})$
and $(k_{m-r+1},\ldots, k_m)$ separately. Then
\beqa
& &(\ha x)^{m}(k_1,\ldots, k_m)= \fr{(m-r)!r!}{m!} \sum_{1\leq i_1<\cdots <i_r\leq m} 
\ha_{\ns, m}( k_1, \ldots\check{i_1}\ldots\ldots\check{i_r}\ldots, k_m; k_{i_1},\ldots, k_{i_r})\times\non\\
& &\ph{444444444444444444444}\times x^{m-r}(k_1, \ldots\check{i_1}\ldots\ldots\check{i_r}\ldots, k_m) \label{sum-formula-x-x}
\eeqa
where we separated the two groups of variables by a semicolon for clarity.
\end{lem}
\proof We write
\beqa
(\ha x)^{m}(k_1,\ldots, k_m)\y&=&\y \fr{1}{m!} \sum_{\pi\in S_m }\ha_{\ns,m}(k_{\pi(1)}, \ldots, k_{\pi(m-r)}; k_{\pi(m-r+1)}, \ldots, k_{\pi(m)} )x^{m-r}(k_{\pi(1)},\ldots, k_{\pi(m-r)})\non\\
\y&=&\y \fr{1}{m!}  \sum_{\substack{i_1,\ldots, i_r \\ i_{j_1}\neq i_{j_2}    }} \sum_{\pi\in S_m^{(i_1, \ldots, i_r)} }
\ha_{\ns,m}( k_{\pi(1)}, \ldots, k_{\pi(m-r)}; k_{i_1}, \ldots, k_{i_r}  )x^{m-r}(k_{\pi(1)},\ldots, k_{\pi(m-r)})\non\\
\y&=&\y\fr{(m-r)!}{m!}   \sum_{\substack{i_1,\ldots, i_r \\ i_{j_1}\neq i_{j_2}    }}     
\ha_{\ns,m}(k_1, \ldots\check{i_1}\ldots\ldots\check{i_r}\ldots, k_m; 
k_{i_1}, \ldots, k_{i_r} )\times\non\\
& &\ph{444444444444444444444444}\times x^{m-r}(k_1, \ldots\check{i_1}\ldots\ldots\check{i_r}\ldots, k_m   )\non\\
\y&=&\y\fr{(m-r)!r!}{m!}   \sum_{1\leq i_1<\cdots < i_r \leq m  }     \ha_{\ns,m}( k_{i_1}, \ldots, k_{i_r};  k_1, \ldots\check{i_1}\ldots\ldots\check{i_r}\ldots, k_m)\non\\
& &\ph{444444444444444444444444}\times x^{m-r}(k_1, \ldots\check{i_1}\ldots\ldots\check{i_r}\ldots, k_m  ),
\eeqa
where $S_m^{(i_1,\ldots i_r)}$ is the set of permutations from $S_m$ {\color{\red}such that} $\pi(m-r+1)=i_1, \ldots, \pi(m)=i_r$. \qed
\subsection{Non-commutative recurrence relations on Fock space}\label{Fock-space-recurrence}

Given a family $\ha_{\ns}^{(f)}, \ldots, \ha_{\ns}^{(1)}$ of  operators defined in (\ref{hat-a-ns-zero})-(\ref{hat-a-ns}), 
with orders $r_f, \ldots, r_1$, {\color{\red}respectively, using the definitions in (\ref{hat-a-ns-zero}) and (\ref{hat-a-ns})}   we have by iteration
\beqa
& &(\ha^{(f)}_{\ns}\ldots  \ha^{(1)}_{\ns}x_{{\color{\red}\ns}})^m(k_1,\ldots, k_m)=\ha^{(f)}_{\ns,m}(k_1,\ldots,k_m)( \ha^{(f-1)}_{\ns}\ldots  \ha_{\ns}^{(1)}  x_{{\color{\red}\ns}})^{m-r_f}(k_1,\ldots, k_{m-r_f})\non\\
& &=\ha^{(f)}_{\ns,m}(k_1,\ldots,k_m) \ha^{(f-1)}_{\ns,m-r_f}(k_{1},\ldots, k_{m-r_f})  \ldots  
\ha_{\ns,m-(r_2+\cdots+r_{f})}^{(1)} (k_1,\ldots, k_{m-(r_2+\cdots+r_{f})  }) \times \non\\
& &\ph{444444444444444444444444444444444444444}\times x_{{\color{\red}\ns}}^{m-(r_1+\cdots+r_f)}(k_1, \ldots,  k_{m-(r_1+\cdots+r_f)}),\quad\quad\quad \label{ns-multiplication}
\eeqa 
for $m\geq (r_1+\cdots+r_f)$ and zero otherwise. Next, for $\ha^{(j)}:=P_{\sym} \ha^{(j)}_{\ns}$  we have by Lemma~\ref{symmetrisation-lemma} and formula~(\ref{ns-multiplication}) 
\beqa
\ha^{(f)}\ldots \ha^{(1)}x=P_{\sym}\ha^{(f)}_{\ns}\ldots  \ha^{(1)}_{\ns}x, \quad x\in X.    \label{multiplication-sym}
\eeqa
Let us now write formula~(\ref{ns-multiplication}) in the case $x_{\ns}=(x^0_{\ns},0,\ldots)=:x_{0}$. (We skip the index $\ns$, since symmetrization acts trivially on this vector). Then there is 
only one non-trivial component, namely {\color{\red}the one corresponding to}  $m=(r_1+\cdots+r_f)$:
\beqa
& &(\ha^{(f)}_{\ns}\ldots  \ha^{(1)}_{\ns}x_{0})^m(k_1,\ldots, k_m)\non\\
& &=\ha^{(f)}_{\ns,r_1+\cdots+r_f}(k_1,\ldots,k_{r_1+\cdots+r_{f}})\ldots \ha^{(j)}_{\ns,r_1+\cdots+r_{j}}
(k_1, \ldots,  k_{r_1+\cdots+r_{j}} )  \ldots  \ha_{\ns,r_1 }^{(1)} (k_{1},\ldots,k_{r_1})x_{0}. \label{x-zero-formula}
\eeqa
To make contact with the discussion below equation~(\ref{x-n-formula}), suppose that $r_{\ell}\in \{1,2\}$ and all operators $\ha^{(\ell)}_{\ns}$ of a given order are equal. For $r_{\ell}=1$ we set $\ha_{\ns}:=\ha_{\ns}^{(\ell)}$ and for $r_\ell=2$ we write $\hb_{\ns}:=\ha_{\ns}^{(\ell)}$. 
Let us now define $X_{\ns,i}:=L^2(\real^3)^{\otimes i}\otimes \hil$ so that $X_{\ns}=\bigoplus_{i\in\nat_0} X_{\ns,i} $ and write 
$\ha_{\ns, i} :=\ha \restriction X_{\ns,i-1}$ and $\hb_{\ns, i+1,i}:= \hb_{\ns} \restriction X_{\ns,i-1}$. Clearly $\ha_{\ns,i}: X_{\ns,i-1}\to X_{\ns,i}$ and $\hb_{\ns,i+1,i}: X_{\ns,i-1}\to X_{\ns,i+1}$. More explicitly, we have
\beqa
& &(\ha_{\ns,i}x_{\ns,i-1})(k_1,\ldots, k_{i-1}, k_i)=\ha_{\ns,i}(k_1,\ldots,k_{i-1}, k_i)x_{\ns,i-1}(k_1,\ldots, k_{i-1}), \\ 
& & (\hb_{\ns,i+1,i}x_{\ns,i-1})(k_1,\ldots,  k_{i-1}, k_i, k_{i+1})=  \hb_{\ns,i+1,i}(k_1,\ldots, k_{i-1}, k_i, k_{i+1})x_{\ns,i-1}(k_1,\ldots,  k_{i-1})
\label{formulas-in-variables}
\eeqa
for $x_{\ns,i-1}\in X_{\ns,i-1}$.
With the notation of Subsection~\ref{non-commutative-subsection} we can rewrite (\ref{x-zero-formula}) as
\beqa
\ha^{(f)}_{\ns}\ldots  \ha^{(1)}_{\ns}x_{0}\y&=&\y(\ha_{\ns,m}\ldots \hb_{\ns,i+1,i}\ldots \hb_{\ns,i'+1,i'}\ldots \ha_{\ns,1})x_{0}\non\\
\y&=&\y\Op^{(m)}_{\huans,\hubns}[a_m\ldots b_{i+1,i}\ldots b_{i'+1,i'}\ldots a_1]x_{0}\non\\
\y&=&\y\hat\de_{i}\ldots \hat\de_{i'}\Op^{(m)}_{\huans,\hubns}[a_m \ldots a_1]x_{0}, \label{f-operator-formula}
\eeqa
where $\huans:=\{\ha_i\}_{i\in\nat}$ and $\hubns:=\{\hb_{i+1,i}\}_{i\in\nat}$. 

Now let $\ha=P_{\sym}\ha_{\ns}$, $\hb=P_{\sym}\hb_{\ns}$ and consider the non-commutative recurrence relation
\beqa
x_{n}=\ha x_{n-1}+\hb x_{n-2}. \label{intermediate-recurrence}
\eeqa
We have by formulas~(\ref{x-n-formula}), (\ref{multiplication-sym}), 
(\ref{f-operator-formula})
\beqa
x_{n}\y&=&\y \sum_{\ell=0}^{[n/2]}\sum_{1\leq i_1 \ll\cdots \ll i_{\ell}\leq n-1 } \Op^{(n)}_{\hat a, \hat b} [ \de_{i_1}\ldots \de_{i_{\ell}}  a_n \ldots a_1]x_{0}\non\\
\y&=&\y P_{\sym} \sum_{\ell=0}^{[n/2]}\sum_{1\leq i_1 \ll\cdots \ll i_{\ell}\leq n-1 } \Op^{(n)}_{\hat a_{\ns}, \hat b_{\ns}} [ \de_{i_1}\ldots \de_{i_{\ell}}  a_n \ldots a_1]x_{0}\non\\
\y&=&\y P_{\sym} \sum_{\ell=0}^{[n/2]}\sum_{1\leq i_1 \ll\cdots \ll i_{\ell}\leq n-1 }  \hat\de_{i_1}\ldots \hat\de_{i_{\ell}}\Op^{(n)}_{\huans, \hubns} [  a_n \ldots a_1]x_{0}\non\\
\y&=&\y  P_{\sym} \sum_{\ell=0}^{[n/2]}\sum_{1\leq i_1 \ll\cdots \ll i_{\ell}\leq n-1 }   (\hat b_{\ns,i_1+1,i_1}\hat\pa_{i_1+1} \hat\pa_{i_1}  )\ldots 
(\hat b_{\ns, i_{\ell}+1,i_{\ell}}\hat\pa_{i_{\ell}+1} \hat\pa_{i_{\ell}}  )
\Op^{(n)}_{\huans, \hubns} [  a_n \ldots a_1]x_{0},\ \  \label{x-n-formula-x}
\eeqa
where in the last step we use the notation introduced below (\ref{de-on-op}). 

{\color{\red}As anticipated at the beginning of Section \ref{FS-comb}, the solution $x_n$ --  that we have constructed in a slightly general setting concerning the sequences of operator valued functions  $\ha_{\ns}=\{\ha_{\ns,m}\}_{m\geq r}$ -- will be applied to the recursive relation (\ref{recurrence-relation-W}) for the vectors $\hf^n(k_{1},\ldots, k_{n})$. For this purpose,} now let us consider the case of $\ha_{\ns,i}$, $\hb_{\ns,i+1,i}$ of the following form:
\beqa
& &\ha_{\ns,i}(k_1,\ldots,k_{i-1}, k_i)=\sum_{j=1}^3g^{(j)}(k_i)\ca^{(j)}_{\ns,i}(k_1,\ldots,k_{i-1}, k_i), \\
& &\hb_{\ns,i+1,i}(k_1,\ldots,k_{i-1}, k_i, k_{i+1})=\bigg(\sum_{j=1}^3 g^{(j)}(k_i) g^{(j)}(k_{i+1}) \bigg) \cb_{\ns,i+1,i}(k_1,\ldots,k_{i-1}, k_i, k_{i+1}), \label{scalar-product}
\eeqa
where $\ca^{(j)}_{\ns,i}, \cb_{\ns,i+1,i}$ are $B(\hil)$-valued functions as specified above (\ref{hat-a-ns-zero})  and $g^{(j)}$ are complex valued, bounded measurable functions on $\real^3$.
By substituting this to formula~(\ref{x-n-formula-x}), we get
\beqa
x_{n}\y&=&\y P_{\sym} \sum_{\ell=0}^{[n/2]}\sum_{1\leq i_1 \ll\cdots \ll i_{\ell}\leq n-1 }\sum_{j_1,\ldots,j_n} g_1^{(j_1)}\ldots g_n^{(j_n)}\times  \non\\
& &\times (\cb_{\ns,i_1+1,i_1}\hat\pa_{i_1+1} \hat\pa_{i_1}  )\ldots 
(\cb_{\ns, i_{\ell}+1,i_{\ell}}\hat\pa_{i_{\ell}+1} \hat\pa_{i_{\ell}}  )
\Op^{(n)}_{\cuans, \cubns} [  a_n^{(j_n)} \ldots a^{(j_1)}_1]x_{0}, \label{x-n-g}
\eeqa
where $(g_1^{(j_1)}\ldots g_n^{(j_n)})(k_1,\ldots,k_n):= g^{(j_1)}(k_1)\ldots g^{(j_n)}(k_n)$, $\cuans:=\{\ca_{\ns,i}^{(j_i)}\}_{i\in\nat, j_i=1,2,3}$, $\cubns:=\{\cb_{\ns, i+1,i}^{(j_i)}\}_{i\in\nat, j_i=1,2,3}$.
As a slight generalization of our  notations,  the action of  $\cb_{\ns, i+1,i}  \hat\pa_{i+1} \hat\pa_{i}$ is meant to
replace $\ca^{(j_{i+1})}_{\ns,i+1} \ca^{(j_{i})}_{\ns,i}$ with $\cb_{\ns, i+1,i} \de_{j_{i+1}, j_{i}  }$, (where $\de_{j_{i+1}, j_{i}}$ is the Kronecker delta, not to be confused with (\ref{new-derivative})), and thus recovers the scalar product $\sum_{j=1}^3 g^{(j)}(k_i) g^{(j)}(k_{i+1})$ from (\ref{scalar-product}). In applications we will not write the summation over $j_i$ explicitly in our notation, since it has  no influence on our estimates.

\section{Ground state  wave functions  of the Nelson model}\label{formulas-for-ground-state-wavefunctions}
\setcounter{equation}{0}

\subsection{Pull through identities}\label{pull-through-subsection}

In Lemmas~\ref{standard-pull-through}, \ref{standard-pull-through-two} below we recall
standard pull-through identities from \cite{Fr73}. 
{\color{\red} They are} used here to establish new pull-through identities in Lemmas~\ref{pull-through-proposition}-\ref{n-particle-pull-through-x} below which are
then exploited in Subsection~\ref{novel-subsection} to derive a novel formula for wave functions, more 
suitable for the analysis of infrared properties.

We define  $C^{\infty}(H_{P,\free}):=\bigcap_{\ell\geq 0} D(H^{\ell}_{P,\free})$  and state the following lemma which is obtained by a standard computation. 
\begin{lem}\label{standard-pull-through}\cite{Fr73} For $\psi\in C^{\infty}(H_{P,\free})$  we have in the sense of vector valued distributions in $\ti k$
\beqa
b(\ti k)H_{P-k,\si}\psi=(H_{P-k-\ti k,\si}+|\ti k|) b(\ti k)\psi+\vv^{\si}(\ti k)\psi. \label{Froehlich-pull-through-formula}
\eeqa
\end{lem}
\nin Now we introduce the notation 
\begin{equation}\label{def-PK...K}
{\color{\red}H_{P,\si; k_1,\ldots, k_n}:=H_{P- (k_1+\cdots+k_n),\si}+| k_1|+\cdots+|k_n|}
\end{equation} and note the following
simple consequence of Lemma~\ref{standard-pull-through}. 
\begin{lem}\label{standard-pull-through-two}\cite{Fr73} For $\psi\in C^{\infty}(H_{P,\free})$ we have 
in the sense of vector valued distributions in $k_1,\ldots, k_n$
\beqa
b(k_n)\ldots b(k_1)H_{P,\si}\psi=H_{P,\si; k_1,\ldots, k_n}b(k_n)\ldots b(k_1)\psi+
\sum_{i=1}^n\vv^{\si}(k_i)b(k_n)\ldots {\color{\red}\check{i}} \ldots b(k_1)\psi. \label{n-part-pull-x}
\eeqa
\end{lem}
\proof  For the reader's convenience we provide a proof. By Lemma~\ref{standard-pull-through}, equality~(\ref{n-part-pull-x}) holds for $n=1$. Suppose the equality holds for $n$. 
Then we can compute
\beqa
b(k_{n+1})b(k_{n})\ldots b(k_1)H_{P,\si}\psi \y &=&\y b(k_{n+1})H_{P,\si; k_1,\ldots, k_n}b(k_{n})\ldots b(k_1)\psi\non\\
& &+b(k_n)\sum_{i=1}^{n}\vv^{\si}(k_i)b(k_{n-1})\ldots \check{i} \ldots b(k_1)\psi\non\\
\y &=&\y H_{P,\si; k_1,\ldots, k_{n+1}}b(k_{n+1})b(k_{n-1})\ldots b(k_1)\psi \non\\
& &+\sum_{i=1}^{n+1}\vv^{\si}(k_i)b(k_{n+1})\ldots \check{i} \ldots b(k_1)\psi,
\eeqa
where we used Lemma~\ref{standard-pull-through} with $k=k_1+\cdots+k_{n}$ and smearing in $k_1,\ldots, k_{n+1}$ is understood. \qed

Let us now proceed to new pull through estimates which have better infrared properties. For this purpose we define
\beqa
& &f_{P,\si}(k):=\fr{  \vv^{\si}(k)    }{|k|\al_{P,\si}(\hat k) }, \textrm{ where }
\quad \al_{P,\si}(\hat k):=(1-\hat k\cdot \nabla E_{P,\si}), \quad \hat k:=k/|k|
\label{new-formula}\\ 
& & g_{P,\si}(k):=f_{P,\si}(k) k,\\
& &b_W(k):=b(k)+f_{P,\si}(k), \label{b_W}\\
& &\La_{P,\si}:=\nabla E_{P,\si}-(P-P_{\pho}),  \label{Lambda-def}
\eeqa
 and note in passing that $b_W(k)=W_{P,\si}^* b(k) W_{P,\si}$ for 
 \begin{equation}
 W_{P,\si}:=e^{b^*(f_{P,\si})-b(f_{P,\si})} \label{def-W}
 \end{equation}We have the following facts:
\begin{lem}\label{pull-through-proposition}  For $\psi\in C^{\infty}(H_{P,\free})$ 
we have in the sense of vector-valued distributions in $k,\ti k$
\beqa
b_{W}(\ti k) H_{P-k,\si}\psi\y &=&\y (H_{P-(\ti k+k),\si}+|\ti k|)b_{W}(\ti k)\psi-g_{{\color{\red}P,\si}}(\ti k)\cdot\bigg(\La_{P,\si}+k+\h \ti k\bigg)\psi,\label{pull-through-formula}\\
b_{W}(\ti k)\La_{P,\si}\psi\y&=&\y (\La_{P,\si}+\ti k)b_{W}(\ti k)\psi-g_{P,\si}(\ti k)\psi.  
\label{second-pull-through-formula}
\eeqa
\end{lem}
\proof First, {\color{\red}using the definition in (\ref{Lambda-def}),} we note the following identity on $C^{\infty}(H_{P,\free})$: 
\beqa
H_{P-(k+\ti k),\si}=H_{P-k,\si}+(\La_{P,\si}+k+\h \ti k)\cdot \ti k-\nabla E_{P,\si}\cdot \ti k.
\label{Hamiltonina-shift}
\eeqa
Omitting $\psi$ {\color{\red}and using (\ref{b_W})}, the l.h.s. of (\ref{pull-through-formula}) has the following form
\beqa
\textrm{l.h.s.} \, \, (\ref{pull-through-formula})=b(\ti k)H_{P-k,\si}+f_{P,\si}(\ti k)H_{P-k,\si}.  \label{l.h.s.}
\eeqa
Now we rearrange the r.h.s. of (\ref{pull-through-formula}) making use of (\ref{Hamiltonina-shift}) and of the fact that $f_{P,\si}(\ti k)\ti k=:g_{P,\si}(\ti k)$.
\beqa
\textrm{r.h.s.}\,\, (\ref{pull-through-formula})\y&=&\y (H_{P-(\ti k+k),\si}+|\ti k|)(b(\ti k)+f_{P,\si}(\ti k))-g_{P,\si}(\ti k)\cdot \bigg(\La_{P,\si}+k+\h \ti k\bigg)\non\\
\y &=&\y (H_{P-(\ti k+k),\si}+|\ti k|)b(\ti k)+ (H_{P-(\ti k+k),\si}+|\ti k|) f_{P,\si}(\ti k)
-g_{P,\si}(\ti k)\cdot \bigg(\La_{P,\si}+k+\h \ti k\bigg)\non\\
\y&=&\y (H_{P-(\ti k+k),\si}+|\ti k|)b(\ti k)
+ \bigg(H_{P-k,\si}+(\La_{P,\si}+k+\h \ti k)\cdot \ti k-\nabla E_{P,\si}\cdot \ti k   +|\ti k|\bigg) f_{P,\si}(\ti k)\non\\
& &\ph{444444444444444444444444444444444444444}-g_{P,\si}(\ti k)\cdot \bigg(\La_{P,\si}+k+\h \ti k\bigg)\non\\
\y&=&\y(H_{P-(\ti k+k),\si}+|\ti k|)b(\ti k)
+\big(H_{P-k,\si}-\nabla E_{P,\si}\cdot \ti k   +|\ti k|\big) f_{P,\si}(\ti k).
\label{r.h.s.}
\eeqa
Now it is easy to see, making use of 
$(|\ti k|-\nabla E_{P,\si}\cdot \ti k)f_{P,\si}(\ti k)=\vv^{\sigma}(\ti k)$,
that the equality between (\ref{l.h.s.}) and (\ref{r.h.s.}) is a consequence of Lemma~\ref{standard-pull-through}.

Now we show (\ref{second-pull-through-formula}). Its l.h.s. has the form (omitting the vector $\psi$)
\beqa
\textrm{l.h.s.} \,\, (\ref{second-pull-through-formula})=(b(k)+f_{P,\si}(k))\La_{P,\si}=(b(k)+f_{P,\si}(k))(\nabla E_{P,\si}-P+P_{\pho}).
\label{l.h.s.-second}
\eeqa
The r.h.s. is given by 
\beqa
\textrm{r.h.s.}\,\, (\ref{second-pull-through-formula})&=&(\nabla E_{P,\si}-P+P_{\pho}+\ti k)(b(\ti k)+f_{P,\si}(\ti k))-g_{P,\si}(\ti k)\non\\
&=&(\nabla E_{P,\si}-P+P_{\pho})(b(\ti k)+f_{P,\si}(\ti k)) +\ti k b(\ti k),
\label{r.h.s.-second}
\eeqa
where we made use again of the fact that $f_{P,\si}(\ti k)\ti k=:g_{P,\si}(\ti k)$. Now the equality
of (\ref{l.h.s.-second}) and (\ref{r.h.s.-second}) follows from $b(\ti k)P_{\pho}=
(P_{\pho}+\ti k)b(\ti k)$. \qed\\
\nin In the next lemma we will use the following definitions:
\beqa
& &G_{P,\si}(k',\ti k):=g_{P,\si}(k')\cdot g_{P,\si}(\ti k), \\
& &F_{P,\si}(\ti k,k):=-g_{P,\si}(\ti k)\cdot \bigg(\La_{P,\si}+k+\h \ti k\bigg).
\eeqa
In these terms, we immediately get from Lemma~\ref{pull-through-proposition}:
\begin{lem}\label{second-pull-through-lemma} For $\psi\in C^{\infty}(H_{P,\free})$
we have in the sense of vector-valued distributions in $\ti k, k_1,\ldots, k_n$
\beqa
& &b_{W}(\ti k)H_{P,\si; k_1,\ldots, k_n }\psi=H_{P,\si; \ti k,k_1,\ldots, k_n} b_W(\ti k)\psi+F_{P,\si}(\ti k, k_1+\cdots+k_n)\psi,
\label{Hamiltonian-pull-through-one}\\
& &b_W(\ti k)F_{P,\si}(k', k)\psi= 
F_{P,\si}(k',\ti k+k)b_W(\ti k)\psi+G_{P,\si}(k',\ti k)\psi,
\label{Gamma-pull-through-one}
\eeqa
{\color{\red}where $H_{P,\si; k_1,\ldots, k_n }$ is defined in (\ref{def-PK...K}).}
\end{lem}
\proof Equality~(\ref{Hamiltonian-pull-through-one}) is obtained from (\ref{pull-through-formula})
by setting $k=k_1+\cdots+k_n$. As for (\ref{Gamma-pull-through-one}) we have
\beqa
& &b_W(\ti k)F_{P,\si}(k', k)=b_W(\ti k)(-)g_{P,\si}(k'){\color{\red}\cdot }\bigg(\La_{P,\si}+k+\h  k'\bigg)\non\\
& &\ph{4444444444444}=(-)g_{P,\si}(k')\cdot \bigg((\La_{P,\si}+\ti k+k+\h  k')b_W(\ti k)-g_{P,\si}(\ti k)\bigg)\non\\
& &\ph{4444444444444}=F_{P,\si}(k',\ti k+k)b_W(\ti k)+G_{P,\si}(k',\ti k),
\eeqa
where we made use of (\ref{second-pull-through-formula}). \qed\\
\nin Now we can prove the n-particle pull-through identity:
\begin{lem} \label{n-particle-pull-through-x} For $\psi\in C^{\infty}(H_{P,\free})$
we have in the sense of vector-valued distributions in $k_1,\ldots, k_n$
\beqa
b_W(k_n)\ldots b_W(k_1)H_{P,\si}\psi\y&=&\y H_{P,\si; k_n,\ldots,k_1}b_{W}(k_n)\ldots b_{W}(k_1)\psi\non\\
& &+\sum_{i=1}^nF_{P,\si}(k_i,k_1+\cdots \check{i} \cdots +k_n)b_{W}(k_n)\cdots \check{i} \cdots b_{W}(k_1)\psi\non\\
& &+\sum_{1\leq i<i'\leq n} G_{P,\si}(k_i,k_{i'})b_{W}(k_n)\cdots \check{i}'\cdots \check{i} \cdots b_{W}(k_1)\psi. \label{n-part-pull-through}
\eeqa
(For $n=1$ the $\sum_{i<i'}$ part should be dropped and $k_1+\cdots \check{i} \cdots +k_n=0$).
\end{lem}
\proof We proceed by induction: For $n=1$ the statement follows from (\ref{Hamiltonian-pull-through-one}). Suppose the equality holds for $n$. Then
\beqa
& &b_W(k_{n+1})b_W(k_n)\ldots b_W(k_1)H_{P,\si}
=b_W(k_{n+1})H_{P,\si; k_n,\ldots, k_1}b_{W}(k_n)\ldots b_{W}(k_1)\label{first-term-p-t}\\
& &\ph{44444444444}+\sum_{i=1}^nb_W(k_{n+1})F_{P,\si}(k_i,k_1+\cdots \check{i} \cdots +k_n)b_{W}(k_n)\cdots \check{i} \cdots b_{W}(k_1) \label{second-term-pt} \\
& &\ph{44444444444}+\sum_{1\leq i<i'\leq n} G_{P,\si}(k_i,k_{i'})b_W(k_{n+1})b_{W}(k_n)\cdots \check{i}'\cdots \check{i} \cdots b_{W}(k_1). \label{third-term-pt}
\eeqa
Now we obtain by (\ref{Hamiltonian-pull-through-one}), (\ref{Gamma-pull-through-one})
\beqa
(\ref{first-term-p-t})\y&=&\y H_{P,\si; k_{n+1},k_n,\ldots,k_1 } b_W(k_{n+1})b_{W}(k_n)\ldots b_{W}(k_1) \\   
& &\ph{44444444444444444}+F_{P,\si}(k_{n+1},k_1+\cdots+k_n) b_{W}(k_n)\ldots b_{W}(k_1),\label{last-step-induction}\\
(\ref{second-term-pt})\y&=&\y \sum_{i=1}^n F_{P,\si}(k_i,k_1+\cdots \check{i} \cdots +k_n+k_{n+1})   
b_{W}(k_{n+1}) b_{W}(k_n)\cdots \check{i} \cdots b_{W}(k_{1}) \label{last-step-induction-one}\\
& &\ph{44444444444444444444}+\sum_{i=1}^n G_{P,\si}(k_i, k_{n+1})  b_{W}(k_n)\cdots \check{i} \cdots b_{W}(k_1),
\label{last-step-induction-three} \\
(\ref{third-term-pt})\y&=&\y \sum_{1\leq i<i'\leq n} G_{P,\si}(k_i,k_{i'})b_{W}(k_{n+1})   b_{W}(k_{n}) \cdots \check{i}'\cdots \check{i} \cdots b_{W}(k_1).
\label{last-step-induction-four}
\eeqa
It is clear that
\beqa
(\ref{last-step-induction})+(\ref{last-step-induction-one})\y&=&\y\sum_{i=1}^{n+1} F_{P,\si}(k_i,k_1+\cdots \check{i} \cdots+k_{n+1})   
b_{W}(k_{n+1})\cdots \check{i} \cdots b_{W}(k_{1}),\\
(\ref{last-step-induction-three})+(\ref{last-step-induction-four})\y&=&\y\sum_{1\leq i<i'\leq n+1} G_{P,\si}(k_i,k_{i'})b_{W}(k_{n+1})\cdots \check{i}'\cdots \check{i} \cdots b_W(k_{1}),
\eeqa
which concludes the proof. \qed

\subsection{Standard formula for the ground state wave functions}
\label{standard-subsection} 

As a simple application of the formalism of non-commutative recurrence relations, we first rederive the standard formula  for the ground state wave functions of the Nelson model from \cite{Fr73}. 

First, we define the following vector valued distributions
\beqa
\check{f}^n_{P,\si}(k_1,\ldots, k_n):=b(k_1)\ldots b(k_n)\cpsi_{P,\si} \label{vector-valued-distr}
\eeqa
To analyse (\ref{vector-valued-distr}), we define for $P\in S$ and $k_1,\ldots, k_n\in \real^3\backslash \{0\}$
\beqa
& &H_{P,\si; k_1,\ldots, k_n }:=H_{P- (k_1+\cdots+k_n),\si}+| k_1|+\cdots+|k_n|, \label{Hamiltonian-sum-one-x-x}\\
& &R_{P,\si; k_1, \ldots, k_n}:=\fr{1}{H_{P,\si; k_1,\ldots, k_n }-E_{P,\si} }, \label{R-first-definition}
\eeqa
where the resolvent is well defined 
by Lemma~\ref{elem-lemma} below.  The following proposition is a consequence of  of (\ref{n-part-pull-x}). We refer to Lemma D.3 of \cite{DP17.1} for a detailed argument. 
\begin{prop}\label{froehlich-relation}   $\check{f}_{P,\si}^n$ are elements of $L^2(\real^{3n})\otimes \Ga(\mfh_{\fib})$ and satisfy
the following recurrence relation for $n\geq 1$ 
\beqa\label{recur}
\check{f}_{P,\si}^n(k_1,\ldots, k_n)=(-)R_{P,\si; k_1,\ldots, k_n}\sum_{i=1}^n \vv^{\si}(k_i)\check{f}_{P,\si}^{n-1}(k_1,\ldots\check{i}\ldots, k_n) \label{simple-recurrence}
\eeqa
with the initial condition $\check{f}_{P,\si}^{0}=\cpsi_{P,\si}$.
\end{prop}
{\color{\red}We observe that formula (\ref{recur})}   is a  non-commutative recurrence relation on $X=\Ga(L^2(\real^3))\otimes \Ga(\mfh_{\mrm{fi}})$
of the form (\ref{intermediate-recurrence}) with $\hb=0$.
With the help of Lemma~\ref{sum-rep}, we identify 
\beqa
& &\ha_{\ns,i}(k_1, \ldots, k_{i-1},k_i)=i  (-)R_{P,\si;k_1, \ldots, k_i}\vv^{\si}(k_i), \\
& &\hb_{\ns,i+1,i}(k_1,\ldots, k_{i-1}, k_i,k_{i+1})=0.
\eeqa
Setting $R_{P,\si; i}(k_1,\ldots, k_i):=R_{P,\si;k_1, \ldots, k_i}$, we get from formula~(\ref{x-n-formula-x})
\beqa
\check{f}^n_{P,\si}=n! P_{\sym}
\Op^{(n)}_{\huans, \hubns} [  a_n \ldots a_1]\cpsi_{P,\si}=n!(-1)^n P_{\sym}(v^{\si}_{\alf;1}\ldots v^{\si}_{\alf;n})
(R_{P,\si; n}\ldots R_{P,\si; 1})\cpsi_{P,\si}, \label{hfn-formula-x-x}
\eeqa
where $(v^{\si}_{\alf;1}\ldots v^{\si}_{\alf;n})(k_1,\ldots, k_n):=\vv^{\si}(k_1)\ldots \vv^{\si}(k_n)$.
Now consider the wave functions $\{ f^n_{P,\si}\}_{n\in\nat_0}$ of the ground state vector $\cpsi_{P,\si}$.
Clearly, 
\beqa
f^n_{P,\si}(k_1,\ldots,k_n)\y&=&\y\fr{1}{\sqrt{n!}} \lan \Om,b(k_1)\ldots b(k_n)\cpsi_{P,\si}\ran=\fr{1}{\sqrt{n!}} \lan \Om,  \check{f}_{P,\si}^n(k_1,\ldots, k_n)\ran \label{functions-equality}\\
\y&=&\y\fr{1}{\sqrt{n!}} \sum_{\pi\in S_n}(-1)^n \lan \Om, \prod_{i=n}^1\fr{1}{H_{P,\si; k_{\pi(1)},\ldots, k_{\pi(i)} }-E_{P,\si}}
\vv^{\si}(k_{\pi(i)}) \cpsi_{P,\si}\ran, \label{Froehlich-form}
\eeqa
where in the last step we used (\ref{hfn-formula-x-x}),(\ref{Hamiltonian-sum-one-x-x}). This is the standard formula for the wave functions of $\cpsi_{P,\si}$ from \cite{Fr73}. \qed

\subsection{Novel formula for the ground state wave functions}
\label{novel-subsection}

In this subsection we derive a novel formula for the ground state wave functions $f^n_{P,\si}$ which facilitates the study of infrared properties. 
For this purpose we recall  that
$b_W(k):=b(k)+f_{P,\si}(k)$,   where $f_{P,\si}$ is given by (\ref{new-formula}), 
and set
\beqa
\hf^n(k_{1},\ldots, k_{n}):=b_W(k_1)\ldots b_W(k_n)\cpsi_{P,\si}, 
\label{main-vectors}
\eeqa
The $\Ga(\mfh_{\fib})$-valued functions $\hf^n$ will be the main objects of our analysis. We know already {\color{\red}from 
Proposition~\ref{froehlich-relation} } that they are elements of $L^2(\real^{3n})\otimes \Ga(\mfh_{\fib})$. They are related to the wave 
functions $f^n_{P,\si}$ as follows: 
\beqa
f^n_{P,\si}(k_1,\ldots, k_n)
\y&=&\y\fr{1}{\sqrt{n!}}
\lan \Om, (b_{W}(k_1)-f_{P,\si}(k_1))\ldots  (b_{W}(k_n) -f_{P,\si}(k_n)) \cpsi_{P,\si}\ran\non\\
\y&=&\y\fr{1}{\sqrt{n!}} \sum_{\ell=0}^{n} (-1)^{n-\ell} \sum_{\substack{1\leq i_1<\ldots<i_{\ell}\leq n \\ 1\leq i_{\ell+1}<\ldots<i_{n}\leq n }} 
  \lan \Om,\hf^n(k_{i_1},\ldots, k_{i_{\ell}}) \ran   f_{P,\si}(k_{i_{\ell+1}}) \ldots f_{P,\si}(k_{i_{n}}). \label{f-n-hat-f} \quad\quad
\eeqa
We have the following sufficient conditions for  estimates~(\ref{infrared-spectral-bound}).  
\begin{prop}\label{first-proposition-functions} Suppose that 
\beqa
& &\|\hf^n(k_1,\ldots, k_n)\|\leq   g^n_{\si}(k_1,\ldots, k_n), \label{assumption-one-estimate}\\
& &\|\pa_{P^j}\hf^n(k_1,\ldots, k_n)\|  \leq  \ir g^n_{\si}(k_1,\ldots, k_n), \label{assumption-two-estimate}\\
& &\|\pa_{P^j}\pa_{P^{j'}}\hf^n(k_1,\ldots, k_n)\|\leq \ir  g^n_{\si}(k_1,\ldots, k_n). \label{assumption-three-estimate} 
\eeqa 
Then  
$\sqrt{n!}| f^n_{P,\si}(k_1,\ldots, k_n)|$,  $\sqrt{n!}|\pa_{P^j} f^n_{P,\si}(k_1,\ldots, k_n)|$,   
$\sqrt{n!}|\pa_{P^j}\pa_{P^{j'}} f^n_{P,\si}(k_1,\ldots, k_n)|$
also satisfy the bounds (\ref{assumption-one-estimate}), (\ref{assumption-two-estimate}), (\ref{assumption-three-estimate}), respectively.
\end{prop}
\proof This lemma follows immediately from formula~(\ref{f-n-hat-f}),  the bounds
\beqa
& &|f_{P,\si}(k)|, \  |\pa_{P^j}f_{P,\si}(k)| \leq c\frac{ \vv^{\sigma}( k) }{|k|}, \quad
  |\pa_{P^j}\pa_{P^{j'}}f_{P,\si}(k)|\leq  c\ir\frac{ \vv^{\sigma}( k) }{|k|},
\eeqa
which are a consequence of (\ref{velocity-boundedness}) and a simple combinatorial
identity
\beqa
\sum_{\ell=0}^{n} \sum_{\substack{1\leq i_1<\ldots<i_{\ell}\leq n \\ 1\leq i_{\ell+1}<\ldots<i_{n}\leq n }}  1= \sum_{\ell=0}^{n} {n  \choose \ell}=2^n. \ \ \textrm{\qed}
\eeqa

Let us now recall that $\La_{P,\si}:=\nabla E_{P,\si}-(P-P_{\pho})$, introduce the following definitions
\beqa
& &g_{P,\si}(k):=f_{P,\si}(k)k, \label{g-defin}\\
& &F_{P,\si}(\ti k,k):=-g_{P,\si}(\ti k){\cdot}\bigg(\La_{P,\si}+k+\h \ti k\bigg), 
\label{F-defin}
\eeqa
and keep in mind definitions (\ref{Hamiltonian-sum-one-x-x}), (\ref{R-first-definition}). By standard energy bounds $R_{P,\si; k_1,\ldots, k_n}\La_{P,\si}$
are bounded operators on $\Ga(\mfh_{\fib})$. Thus we can formulate
the following proposition, which is a consequence of (\ref{n-part-pull-through}). We refer to Appendix~\ref{pull-through-appendix} for a proof. 
\begin{prop} \label{pull-through-new}  $\hat{f}_{P,\si}^m$ are elements of $L^2(\real^{3m})\otimes \Ga(\mfh_{\fib})$ and satisfy
the following recurrence relation for $n\geq 2$ 
\beqa
\hf^n (k_1,\ldots, k_n)\y&=&\y
(-)R_{P,\si; k_1,\ldots k_n}  \bigg(\sum_{i=1}^nF_{P,\si}(k_i,k_1+\cdots \check{i} \cdots +k_n)
\hf^{n-1}(k_1,\ldots \check{i} \ldots, k_n)\non\\
& &\ph{44444444444}+\sum_{1\leq i<i'\leq n} g_{P,\si}(k_i)\cdot g_{P,\si}(k_{i'})\hf^{n-2}(k_1,\ldots \check{i}\ldots \check{i'} \ldots, k_n)\bigg)
\label{recurrence-relation-W}
\eeqa
with the initial conditions $\hf^0=\cpsi_{P,\si}$, $\hf^1(k)=(-)R_{P,\si;k}F_{P,\si}(k,0)\cpsi_{P,\si}$.
\end{prop}

Equation~(\ref{recurrence-relation-W})  describes a non-commutative recurrence relation on $X_{\fin}\subset X=\Ga(L^2(\real^3))\otimes \Ga(\mfh_{\mrm{fi}})$
as discussed in Subsection~\ref{Fock-space-recurrence}. With the help of Lemma~\ref{sum-rep}, we identify 
\beqa
& &\ha_{\ns,i}(k_1, \ldots, k_{i-1},k_i):=i g_{P,\si}(k_i)\cdot \bigg\{R_{P,\si; k_1, \ldots, k_i} \big(\La_{P,\si}+(k_1+\cdots+k_{i-1})+\h  k_i\big)\bigg\}, \non\\
& &\ph{4444444444444444}=i g_{P,\si}(k_i)\cdot \bigg\{R_{P,\si; i} \big(\La_{P,\si}+\unk_{i-1}+\h  k_i\big)\bigg\}=i g_{P,\si}(k_i)\cdot 
\big\{R_{P,\si;i} \La_{P,\si; i}\big\}, \label{lambda-first-appearance}\\
& &\hb_{\ns,i+1,i}(k_1,\ldots, k_{i-1}, k_i,k_{i+1}):=\fr{(i+1)i}{2}(-)g_{P,\si}(k_{i})\cdot g_{P,\si}(k_{i+1}) \big\{R_{P,\si;k_1, \ldots, k_{i+1}} \big\} \non\\
& &\ph{4444444444444444444444}=\fr{(i+1)i}{2}(-)g_{P,\si}(k_{i})\cdot g_{P,\si}(k_{i+1}) \big\{R_{P,\si; i+1} \big\}, 
\eeqa
where we set $\unk_i:=k_1+\cdots+k_i$,\  \ $\La_{P,\si; i}:=\La_{P,\si}+\unk_{i-1}+\h  k_i$, \  \ $R_{P,\si; i}:=R_{P,\si; k_1, \ldots, k_i}$ {\color{\red}(defined in (\ref{R-first-definition}))}
and call the expressions in curly brackets $\ca_{\ns,i}:=R_{P,\si; i} \La_{P,\si;i},\ \  \cb_{\ns, i+1,i}:=R_{P,\si;i+1}$. Now formula~(\ref{x-n-g}) gives
\beqa
\hf^n \y&=&\y n! P_{\sym}\sum_{\ell=0}^{[n/2]}\sum_{2\leq i_1 \ll\cdots \ll i_{\ell}\leq n }\fr{(-1)^{\ell}}{2^{\ell}} g_{P,\si;1}\ldots g_{P,\si; n} \times  \non\\
& &\times (R_{P,\si; i_1} \hat\pa_{i_1} \hat\pa_{i_1-1}  )\ldots 
(R_{P,\si; i_{\ell}}\hat\pa_{i_{\ell}} \hat\pa_{i_{\ell}-1}  )
\Op^{(n)}_{\cuans, \cubns} [  a_n \ldots a_1]\cpsi_{P,\si}, \label{hfn-formula}
\eeqa
where we skipped the summation over vector indices $j=1,2,3$ in the notation, shifted the indices $i_{i'}$  and set 
$(g_{P,\si; 1} \ldots g_{P,\si; n})(k_1,\ldots, k_n):=g_{P,\si}(k_1)\ldots g_{P,\si}(k_n)$. 
 Relations~(\ref{hfn-formula}), (\ref{f-n-hat-f}) constitute our novel formula for the ground state wave functions $f^n_{P,\si}$.

Since $\cuans, \cubns$ will not change in the future, we can simplify the notation  further by writing:
\beqa
\al(n,\ldots,1):=\Op^{(n)}_{\cuans, \cubns} [  a_n \ldots a_1]. \label{alpha-definition}
\eeqa
Clearly, we have $\al(i)=R_{P,\si; i}\La_{P,\si; i}$ and the following equality of operators holds:
\beqa
\al(n,\ldots,1)=R_{P,\si; n}\La_{P,\si;n}\ldots R_{P,\si;1}\La_{P,\si; 1}. 
\label{alpha-in-terms-of-RL}
\eeqa
However, we need to remember definition~(\ref{alpha-definition}) in terms of the quantisation map to act on $\al(n,\ldots, 1)$ with  $\cb_{\ns, i_{\ell}+1,i_{\ell}}\hat\pa_{i_{\ell}+1} \hat\pa_{i_{\ell}}$. 
Thus the  relevant expressions to study are
\beqa
I_{i_1,\ldots, i_{\ell} }:=(R_{P,\si; i_1+1} \hat\pa_{i_1+1} \hat\pa_{i_1}  )\ldots (R_{P,\si;i_{\ell}+1}\hat\pa_{i_{\ell}+1} \hat\pa_{i_{\ell}}  ) \al(n,\ldots,1)\cpsi_{P,\si}. \label{hat-I-relation}
\eeqa
Let us now set 
\beqa
C_{\ell}\y&:=&\y (i_{1},\ldots, i_{\ell}), \textrm{ where }2\leq i_{1}\ll\ldots \ll i_{\ell}\leq n,\\ 
C^{2\ell}\y&:=&\y (i_{1}-1,i_{1},\ldots, i_{\ell}-1, i_{\ell}), \label{c2l}\\
\hde_{C_{\ell}}\y&:=&\y (R_{P,\si; i_1} \hpa_{i_1} \hpa_{i_1-1})   \ldots  (  R_{P,\si; i_\ell}  \hpa_{i_\ell}\hpa_{i_\ell-1}), \\
I_{C_{\ell}}\y&:=&\y I_{i_1,\ldots,i_{\ell}}=\hde_{C_{\ell}}\al(n,\ldots, 1)\cpsi_{P,\si}, 
\label{I-C-Def}
\eeqa
so that formula (\ref{hfn-formula}) reads
\beqa
  \hf^n= n! P_{\sym}\sum_{\ell=0}^{[n/2]} \fr{(-1)^{\ell}}{2^{\ell}} \sum_{C_{\ell}}g_{C_{\ell}} I_{C_{\ell}}=
 n! P_{\sym}\sum_{\ell=0}^{[n/2]} \fr{(-1)^{\ell}}{2^{\ell}} \sum_{C_{\ell}}g_{C_{\ell}} \hde_{C_{\ell}}\al(n,\ldots, 1)\cpsi_{P,\si}, \label{I-f-formulas}
\eeqa
where $g_{C_{\ell}}(k_1,\ldots, k_n):=g_{P,\si}(k_1)\ldots g_{P,\si}(k_n)$ with the contraction pattern of indices $j_i$ dictated by $C_{\ell}$.  The reader should keep in mind that  $g_{C_{\ell}}$, $I_{C_{\ell}}$, $\al(\ldots)$ depend on $P,\si$, although this is suppressed in the notation.
For future reference we also introduce the decompositions
\beqa
\hde_{C_{\ell}}=:\hde'_{C_{\ell}}(R_{P,\si;i}\hde_{i}\hde_{i-1}),\quad  \hde_{C_{\ell}}=:\hde''_{C_{\ell}}(R_{P,\si;i}\hde_{i}\hde_{i-1}) (R_{P,\si;\hat i}\hde_{\hat i}\hde_{\hat i-1}), \label{skipped-factors}
\eeqa
where $\hde'_{C_{\ell}}$ (resp.  $\hde''_{C_{\ell}})$ denotes $\hde_{C_{\ell}}$ with one (resp. two) factors omitted as indicated
in (\ref{skipped-factors}). Which factors are omitted will be clear from the context.

\section{Derivatives of $\hf^n$ w.r.t. $P$}\label{Derivatives-section}
\setcounter{equation}{0}
{\color{\red} In Proposition \ref{first-proposition-functions} we proved the desired bounds on the ground state wave functions $f^n_{P,\si}(k_1,\ldots, k_n)$ assuming suitable estimates on  the derivatives in $P$ up to degree $2$ of $\hf^n(k_1,\ldots, k_n)$. In the present section, namely in Subsections~\ref{f-der}-\ref{s-der},  we compute and organize the different contributions to the derivatives in $P$ of $\hf^n(k_1,\ldots, k_n)$. In Proposition \ref{intermediate-proposition} we show that the assumptions upon which Proposition \ref{first-proposition-functions} is based follow from some estimates of the contributions  computed in Subsections~\ref{f-der}-\ref{s-der}. The lengthy verification of the latter estimates (assumed in Proposition \ref{intermediate-proposition}) is deferred to Section~\ref{verification-section}.}

\subsection{First derivatives}\label{f-der}

In order to analyse $\pa_P I_{C_{\ell}}$ it is convenient to introduce operators $\pa_{R_{i}}$, $\pa_{\La_{i}}$ and $\pa_{\cpsi}$, which act as $\pa_{P}$ only via {\color{\red}the indicated expressions ({\color{\red}$R_{i}$, $\La_{i}$, and $\cpsi $, respectively}).}\footnote{$\pa_P$ in the following
denotes $\pa_{P^j}$  and $\pa_P^2$ denotes $\pa_{P^{j'}}\pa_{P^j}$ for some $j, j'=1,2,3$ which we do not write explicitly.} To define them, let us introduce for a moment a more explicit notation 
\beqa
I_{C_{\ell}}=:I_{C_{\ell}}(R_{P,\si; n},\ldots, R_{P,\si; 1}; \La_{P,\si;n},\ldots, \La_{P,\si; 1};   \cpsi_{P,\si}).
\eeqa
Now  we set
\beqa
& &\pa_{R_{i}}I_{C_{\ell}}:=\pa_{\ti P}I_{C_{\ell}}( R_{P,\si; n},\ldots,R_{\ti P,\si; i }, \ldots R_{P,\si;1}; \La_{P,\si; n},\ldots, \La_{P,\si;1};   \cpsi_{P,\si}   )|_{\ti P=P}, \label{indi-1}\\
& &\pa_{\La_{i}}I_{C_{\ell}}:=\pa_{\ti P}I_{C_{\ell}}(R_{P,\si; n},\ldots, R_{P,\si; 1}; \La_{P,\si; n},\ldots,  \La_{\ti P,\si; i}, 
\ldots,\La_{P,\si; 1};   \cpsi_{P,\si}  )|_{\ti P=P}, \label{indi-2}\\
& &\pa_{\cpsi}I_{C_{\ell}}:=\pa_{\ti P}I_{C_{\ell}}( R_{P,\si; n},\ldots, R_{P,\si, 1}; \La_{P,\si; n},\ldots, \La_{P,\si; 1};   \cpsi_{\ti P,\si}   )|_{\ti P=P},\label{indi-3}
\eeqa
so that obviously we have 
\beqa
\pa_P I_{C_{\ell}}=\sum_{i=1}^{n}(\pa_{R_{i}}I_{C_{\ell}}+\pa_{\La_{i}}I_{C_{\ell}})+\pa_{\cpsi}I_{C_{\ell}}. \label{first-der}
\eeqa
Let us now see in more detail how these derivatives act:  For $i\notin C^{2\ell}$ {\color{\red}(defined in (\ref{c2l}))} we have
\beqa
\pa_{R_{i}} I_{C_{\ell}}\y&=&\y\hde_{C_{\ell}}\al(n,\ldots, i+1) (\pa_{P}R_{P,\si;i})\La_{P,\si; i} \hal(i-1,\ldots, 1)\cpsi_{P,\si}\non\\
\y&=&\y\hde_{C_{\ell}}\hal(n,\ldots, i+1)R_{P,\si;i}\La_{P,\si}^iR_{P,\si; i}\La_{P,\si; i} \hal(i-1,\ldots, 1)\cpsi_{P,\si} \non\\
\y&=&\y\hde_{C_{\ell}}\hal(n,\ldots, i+1)R_{P,\si;i}\La_{P,\si}^i \hal(i,\ldots, 1)\cpsi_{P,\si}, \label{first-resolvent-i}
\eeqa
where we made use of (\ref{R-derivative}) and set $\La_{P,\si}^i:=\La_{P,\si}+\unk_{i}$. 
Now suppose that $i\in C^{2\ell}$. Then, we can
assume that $i\in C_{\ell}$, since otherwise the resolvent $R_{P,\si; i}$ does not appear in the expression and the result is zero. For $i\in C_{\ell}$ we have
\beqa
\pa_{R_{i}} I_{C_{\ell}}\y&=&\y\hde'_{C_{\ell}}((\pa_PR_{P,\si;i})\hde_{i}\hde_{i-1})   \hal(n,\ldots, 1)\cpsi_{P,\si}\non\\
\y&=&\y\hde'_{C_{\ell}}(R_{P,\si;i}\La^i_{P,\si} R_{P,\si;i}\hde_{i}\hde_{i-1})   \hal(n,\ldots, 1)\cpsi_{P,\si}\non\\
\y&=&\y\hde'_{C_{\ell}} \hal(n,\ldots, i+1) R_{P,\si;i}\La_{P,\si}^i R_{P,\si;i}   \hal(i-2,\ldots, 1)\cpsi_{P,\si}. \label{first-resolvent-i-i}
\eeqa
Now we consider $\pa_{\La_i}$. When computing the first derivative we can set $i\notin C^{2\ell}$, since 
otherwise the expression does not contain $\La_{P,\si; i}$ and the result is zero. We have
\beqa
\pa_{\La_{i}} I_{C_{\ell}}\y&=&\y\hde_{C_{\ell}}\hal(n,\ldots, i+1) R_{P,\si; i} (\pa_P\La_{P,\si;i}) \al(i-1,\ldots, 1)\cpsi_{P,\si}\non\\
\y&=&\y(\pa_P\La_{P,\si;i})\hde_{C_{\ell}}\hal(n,\ldots, i+1) R_{P,\si; i}  \al(i-1,\ldots, 1)\cpsi_{P,\si}, 
\label{first-lambda-i}
\eeqa
where we made use of the fact that $\pa_P\La_{P,\si;i}$ is a number. Finally $\pa_{\cpsi}$ acts in an obvious manner
\beqa
\pa_{\cpsi} I_{C_{\ell}}\y&=&\y \hde_{C_{\ell}}\hal(n,\ldots, 1)\pa_P\cpsi_{P,\si}\non\\
\y&=&\y \hde_{C_{\ell}}\hal(n,\ldots, 1)R_{P,\si}\La_{P,\si}\cpsi_{P,\si}, \label{first-cpsi-i}
\eeqa
where in the second step we used that $\pa_P\cpsi_{P,\si}=R_{P,\si}\La_{P,\si}\cpsi$, $R_{P,\si}=(H_{P,\si}-E_{P,\si})^{-1}$ (see formula~(\ref{derivative-formula})).

\subsection{Second derivatives} \label{s-der}

Let us now proceed to $\pa^2_P I_{C_{\ell}}$.  To analyse it, we will define the second derivatives 
$\pa_{R_{i'}} \pa_{R_i} I_{C_{\ell}}$, $\pa_{\La_{i'}}\pa_{R_i}I_{C_{\ell}}$, $\pa_{\cpsi}\pa_{R_i}I_{C_{\ell}}$, 
$\pa_{R_{i'}} \pa_{\La_i} I_{C_{\ell}} $, $\pa_{\La_{i'}} \pa_{\La_i} I_{C_{\ell}}$, $\pa_{\cpsi}\pa_{\La_i} I_{C_{\ell}}$,
$\pa_{\cpsi}^2 I_{C_{\ell}}$, {\color{\red} by which we mean that the two derivatives $\pa_{P}$ act via the indicated expressions $R_i$, $\La_{i}$, and $\cpsi$  as we explain below formula (\ref{supplement}.}{\color{\red} We supplement these definitions with }
\beqa \label{supplement}
{\color{\red}\pa_{R_i}\pa_{\cpsi} I_{C_{\ell}}:=\pa_{\cpsi}\pa_{R_i} I_{C_{\ell}}, \quad \pa_{\La_i}\pa_{\cpsi} I_{C_{\ell}}:=\pa_{\cpsi}\pa_{\La_i} I_{C_{\ell}}.}
\eeqa
Informally speaking, the previous derivatives are defined by iterating the computational rules (\ref{first-resolvent-i})-(\ref{first-cpsi-i}) and taking the following specifications into account:
\begin{enumerate}
\item  $\pa_{\La_{i}}$ acts also on $\La_{P,\si}^i$ appearing in (\ref{first-resolvent-i}) and (\ref{first-resolvent-i-i}).
\item  $\pa_{\La_{i}}$ acts also on the numerical function $(\pa_P\La_{P,\si;i})$ appearing in (\ref{first-lambda-i}), changing it into $(\pa_P^2\La_{P,\si;i})$.
\item  $\pa_{\La_{i}}$ ignores $\La_{P,\si}$ appearing in (\ref{first-cpsi-i}), as this contribution is taken care of by
the second application {\color{\red}of} $\pa_{\cpsi}$.
\end{enumerate}
Now we give formal definitions starting with the derivatives of  $\pa_{R_{i}}I_{C_{\ell}}$:  Given formulas~(\ref{first-resolvent-i}), (\ref{first-resolvent-i-i}) we can introduce for a moment an explicit notation
\beqa
\pa_{R_{i}}I_{C_{\ell}}=:\pa_{R_{i}}I_{C_{\ell}}(R_{P,\si; n}, \ldots, R_{P,\si; i+1}, R_{P,\si; i}\La^i_{P,\si}, R_{P,\si; i}, \ldots R_{P,\si; 1}; \La_{P,\si; n},\ldots, \La_{P,\si; 1};   \cpsi_{P,\si}   ),
\eeqa
which is meaningful both for $i\notin C^{2\ell}$ and $i\in C_{\ell}$. (In the latter case  there is no dependence on $R_{P,\si; i-1}$ and $\La_{P,\si; i-1}$, however). Now we define:
\beqa
\pa_{ R_{i'} }  \pa_{R_{i}} I_{C_{\ell}}:=\left\{ \begin{array}{ll}
 \pa_{\ti P}\pa_{R_{i}}I_{C_{\ell}}(R_{P,\si; n},\ldots, R_{\ti P,\si;  i'},\ldots, R_{P,\si; i}\La^i_{P,\si}, R_{P,\si; i}, \ldots R_{P,\si; 1}; \La_{P,\si; n},\ldots, \La_{P,\si; 1};   \cpsi_{P,\si}   )|_{\ti P=P}, &  \non\\
\pa_{\ti P}\pa_{R_{i}}I_{C_{\ell}}(R_{P,\si; n}, \ldots , R_{\ti P,\si; i}\La^i_{P,\si}, R_{\ti P,\si; i}, \ldots R_{P,\si; 1}; \La_{P,\si; n},\ldots, \La_{P,\si; 1};   \cpsi_{P,\si}   )|_{\ti P=P},  &  \non
\end{array}\right. 
\eeqa
\beqa
\pa_{ \La_{i'} }  \pa_{R_{i}} I_{C_{\ell}}:=\left\{ \begin{array}{ll}
 \pa_{\ti P}\pa_{R_{i}}I_{C_{\ell}}(R_{P,\si;n},\ldots, R_{P,\si; i}\La^i_{P,\si}, R_{P,\si; i}, \ldots, R_{P,\si;1}; \La_{P,\si; n},
\ldots,\La_{\ti P,\si; i'},\ldots, \La_{P,\si; 1};   \cpsi_{P,\si}   )|_{\ti P=P}, &  \non\\
\pa_{\ti P}\pa_{R_{i}}I_{C_{\ell}}(R_{P,\si; n}, \ldots , R_{P,\si; i}\La^i_{\ti P,\si}, R_{P,\si; i}, \ldots, R_{P,\si;1}; \La_{P,\si;n},
\ldots,\La_{\ti P,\si,i},\ldots, \La_{P,\si;1};   \cpsi_{P,\si}   )|_{\ti P=P},  & 
\end{array}\right.
\eeqa
where the upper line in each of the two definitions holds for $i'\neq i$ and the lower line for $i'=i$. Furthermore, we set
\beqa
\pa_{ \cpsi }  \pa_{R_{i}} I_{C_{\ell}}:=
 \pa_{\ti P}\pa_{R_{i}}I_{C_{\ell}}(R_{P,\si; n},\ldots, R_{P,\si; i}\La^i_{P,\si}, R_{P,\si; i}, \ldots, R_{P,\si; 1}; \La_{P,\si; n},
\ldots,\La_{\ti P,\si; j},\ldots, \La_{P,\si; 1};   \cpsi_{\ti P,\si}   )|_{\ti P=P}.\ph{444444444} \non
\eeqa

Next, we proceed to the derivatives of $\pa_{\La_{i}}I_{C_{\ell}}$ for $i \notin C_{\ell}$. (We recall that 
$\pa_{\La_{i}}I_{C_{\ell}}=0$ otherwise). Given formula~(\ref{first-resolvent-i-i}) we can introduce for a moment an explicit notation:
\beqa
\pa_{\La_i}I_{C_{\ell}}=:
\pa_{\La_i}I_{C_{\ell}}(R_{P,\si; n},\ldots, R_{P,\si; 1}; \La_{P,\si; n},\ldots,\pa_P\La_{P,\si; i},\ldots, \La_{P,\si; 1};   \cpsi_{P,\si}).
\eeqa
We define
\beqa
\pa_{R_{i'}}\pa_{\La_i}I_{C_{\ell}}:=\pa_{\ti P}\pa_{\La_i}I_{C_{\ell}}(R_{P,\si; n},\ldots, R_{\ti P,\si; i'} ,\ldots, R_{P,\si; 1}; \La_{P,\si; n},\ldots,\pa_P\La_{P,\si; i},\ldots, \La_{P,\si; 1};   \cpsi_{P,\si})|_{\ti P=P}, \,\, 
\eeqa
\beqa
\pa_{\La_{i'}}\pa_{\La_i}I_{C_{\ell}}:=\left\{ \begin{array}{ll}
\pa_{\ti P}\pa_{\La_i}I_{C_{\ell}}(    
R_{P,\si;n},\ldots, R_{P,\si;1}; \La_{P,\si; n},\ldots, \La_{\ti P,\si; i'} ,\ldots ,\pa_P\La_{P,\si;i},\ldots, \La_{P,\si;1};   \cpsi_{P,\si})|_{\ti P=P}, &  \\
\pa_{\ti P}\pa_{\La_i}I_{C_{\ell}}(    
R_{P,\si; n},\ldots, R_{P,\si;1}; \La_{P,\si; n},\ldots,\pa_{\ti P}\La_{\ti P,\si; i},\ldots, \La_{P,\si; 1};   \cpsi_{P,\si})|_{\ti P=P}, & 
\end{array}\right.\non
\eeqa
where the upper line  holds for $i'\neq i$ and the lower line for $i'=i$. Also, we write
\beqa
\pa_{\cpsi}\pa_{\La_i}I_{C_{\ell}}:= \pa_{\ti P}\pa_{\La_i}I_{C_{\ell}}(R_{P,\si; n},\ldots, R_{P,\si; 1}; \La_{P,\si; n},\ldots,
\pa_P\La_{P,\si; i},\ldots, \La_{P,\si;1}; \cpsi_{\ti P,\si})|_{\ti P=P}.
\eeqa
Finally, given formula~(\ref{first-lambda-i}), we write
\beqa
\pa_{\cpsi}I_{C_{\ell}}=:I_{C_{\ell}}(R_{P,\si; n},\ldots, R_{P,\si; 1}; \La_{P,\si; n},\ldots, \La_{P,\si; 1};   \pa_{P}\cpsi_{P,\si}),
\eeqa
and define in an obvious manner
\beqa
\pa_{\cpsi}^2I_{C_{\ell}}:=\pa_{\ti P}\pa_{\cpsi}I_{C_{\ell}}(R_{P,\si; n},\ldots, R_{P,\si; 1}; \La_{P,\si; n},\ldots, \La_{P,\si;1};  
 \pa_{\ti P}\cpsi_{\ti P,\si})|_{\ti P=P}.
\eeqa
To conclude, 
we note that $\pa_{R_i}\pa_{\La_{i'}} I_{C_{\ell}}=\pa_{\La_{i'}}\pa_{R_i} I_{C_{\ell}}$ for $i\neq i'$ but for $i=i'$
this equality may fail due to specification~1. above.  We also observe that
$\pa_{R_i}\pa_{R_{i'}} I_{C_{\ell}}=\pa_{R_{i'}}\pa_{R_{i}} I_{C_{\ell}}$ and
$\pa_{\La_i}\pa_{\La_{i'}} I_{C_{\ell}}=\pa_{\La_{i'}}\pa_{\La_{i}} I_{C_{\ell}}$
for all $i,i'$ and recall~(\ref{supplement}).

By inspection of the definitions in this and {\color{\red}the} previous subsection we see that the appearances of $\ti P$ cover all possible
dependencies of $I_{C_{\ell}}$ on $P$. Therefore
\beqa
\pa^2_P I_{C_{\ell}}\y&=&\y\bigg(\sum_{i'=1}^n( \pa_{R_{i'}}+\pa_{\La_{i'}})+\pa_{\cpsi} \bigg)\bigg(\sum_{i=1}^{n}(\pa_{R_{i}}+\pa_{\La_{i}} )+\pa_{\cpsi}\bigg)I_{C_{\ell}}\non\\
\y&=&\y\sum_{i,i'=1}^n( \pa_{R_{i'}}+\pa_{\La_{i'}})( \pa_{R_{i}}+\pa_{\La_i})I_{C_{\ell}}+2\sum_{i=1}^n( \pa_{R_{i}}+\pa_{\La_i})\pa_{\cpsi}I_{C_{\ell}}+\pa_{\cpsi}^2I_{C_{\ell}}. \label{second-der-decomp}
\eeqa
\subsection{Sufficient condition for estimates (\ref{assumption-one-estimate}), (\ref{assumption-two-estimate}), (\ref{assumption-three-estimate})}
\begin{prop}\label{intermediate-proposition}  Suppose that
\beqa
& &\|  I_{C_{\ell}}\|\leq c^n \prod_{m'=1}^n |\unk|^{-1}_{m'}, \label{zero-derivative}\\
& &\|\pa_{R_{i}}  I_{C_{\ell}}\|, \ \|\pa_{\La_{i}}  I_{C_{\ell}}\|,\ \|\pa_{\cpsi}  I_{C_{\ell}}\| \leq c^n\ir \prod_{m'=1}^n |\unk|^{-1}_{m'}, 
\label{first-derivatives}\\
& &\|\pa_{R_{i'}}\pa_{R_{i}}  I_{C_{\ell}}\|, \ \|\pa_{R_{i'}}\pa_{\La_{i}}  I_{C_{\ell}}\|,\ \|\pa_{R_{i'}}\pa_{\cpsi}  I_{C_{\ell}}\| \leq c^n\ir \prod_{m'=1}^n |\unk|^{-1}_{m'}, 
\label{second-derivatives}\\
& &\|\pa_{\La_{i'}}\pa_{\La_{i}} I_{C_{\ell}}\|, \ \|\pa_{\La_{i'}}\pa_{R_{i}}  I_{C_{\ell}}\|,
\ \|\pa_{\La_{i'}}\pa_{\cpsi}  I_{C_{\ell}}\|, \ \|\pa^2_{\cpsi} I_{C_{\ell}}\| \leq c^n\ir 
\prod_{m'=1}^n |\unk|^{-1}_{m'}, 
\label{second-derivatives-continued}
\eeqa
where $|\unk|_{m'}:=|k_1|+\cdots+|k_{m'}|$.
Then the assumptions of Proposition~\ref{first-proposition-functions} are satisfied, namely estimates 
(\ref{assumption-one-estimate}), (\ref{assumption-two-estimate}), (\ref{assumption-three-estimate}) hold.
\end{prop}
\proof We collect several preparatory facts: First, 
we recall formula~(\ref{I-f-formulas})
\beqa \label{fn}
\hf^n= n! P_{\sym}\sum_{\ell=0}^{[n/2]} \fr{(-1)^{\ell}}{2^{\ell}} \sum_{C_{\ell}}g_{C_{\ell}} I_{C_{\ell}}.
\eeqa
Next, since $g_{C_{\ell}}(k_1,\ldots,k_n):=g(k_1)\ldots g(k_n)$ and 
$g(k):= (\vv^{\si}(k)  k) / (|k|\al_{P,\si}(\hat k)) $ by definition~(\ref{g-defin}) we have
\beqa
& &|g_{C_{\ell}}(k_1,\ldots, k_n)|,\  |\pa_{P}g_{C_{\ell}}(k_1,\ldots, k_n)|\leq c^n \prod_{j=1}^{n}  \vv^{\sigma}(k_{j}),  \label{derivatives-G} \\
& &|\pa^2_{P}g_{C_{\ell}}(k_1,\ldots, k_n)|\leq c^n \ir \prod_{j=1}^{n}  \vv^{\sigma}(k_{j}), 
\eeqa
where we made use of (\ref{velocity-boundedness}). Moreover, we note that
\beqa
& &\sum_{\ell=0}^{[n/2]} \fr{1}{2^\ell} \sum_{C_{\ell}}1\leq \sum_{\ell=0}^n{n \choose \ell}=2^n, \quad
 \sum_{\pi\in S_n}\prod_{m'=1}^n |\unk|^{-1}_{\pi,m'}=\prod_{m'=1}^n |k_{m'}|^{-1}, \label{froehlich-combinatorics}
\eeqa
where $|\unk|_{\pi,m'}=|k_{\pi(1)}|+\cdots+|k_{\pi(m')}|$ and the latter equality in (\ref{froehlich-combinatorics}) is a combinatorial identity from \cite{Fr73}, which can be proven by induction. 

From the information above and (\ref{zero-derivative}) we immediately obtain (\ref{assumption-one-estimate}) {\color{\red}because the $n!$ in (\ref{fn}) is canceled by $\frac{1}{n!}$ from the symmetrization symbol $P_{\sym}$ while the sum over the permutations is controlled by the identity on the left of (\ref{froehlich-combinatorics})}.
Let us now prove (\ref{assumption-two-estimate}). We have
\beqa
\pa_{P}\hf^n\y&=&\y n!P_{\sym} \sum_{\ell=0}^{[n/2]} \fr{(-1)^{\ell}}{2^\ell}   \sum_{C_{\ell}}  (\pa_{P}g_{C_{\ell}})
     I_{C_{\ell}} \label{first-derivative-first}\\
& &+n!P_{\sym}\sum_{\ell=0}^{[n/2]} \fr{(-1)^{\ell}}{2^\ell}\sum_{C_{\ell}}   g_{C_{\ell}}
    \pa_{P} I_{C_{\ell}}. \label{first-derivative-second}
\eeqa
Term~(\ref{first-derivative-first}) is immediately estimated as before. We consider the second term:
\beqa
(\ref{first-derivative-second})=n!P_{\sym}  \sum_{\ell=0}^{[n/2]} \fr{(-1)^{\ell}}{2^\ell}   \sum_{C_{\ell}}   g_{C_{\ell}}
    \big\{ \sum_{i=1}^n\big(\pa_{R_{i}} I_{ C_{\ell}}+\pa_{\La_{i}} I_{ C_{\ell}}\big)+\pa_{\cpsi} I_{C_{\ell}} \big\},
\eeqa
which is also easily estimated using (\ref{derivatives-G}), assumption (\ref{first-derivatives}) and the combinatorial 
relations~(\ref{froehlich-combinatorics}). Finally, we analyse the second derivative. We have 
(schematically\footnote{Note that   $2(\pa_{P}g_{C_\ell})    \pa_{P} I_{C_{\ell}}$ in 
(\ref{second-derivative-second}) actually means  $(\pa_{P^{j}}g_{C_\ell})    \pa_{P^{j'}} I_{C_{\ell}}+\{j\leftrightarrow j'\}$.  })
\beqa
\pa^2_{P}\hf^n\y&=&\y n!P_{\sym}\sum_{\ell=0}^{[n/2]} \fr{{\color{\red}(-1)^{\ell}}}{2^\ell}  (\pa^2_{P}g_{C_\ell}) \sum_{C_{\ell}}  
     I_{C_{\ell}} \label{second-derivative-first}\\
& &+n!P_{\sym}\sum_{\ell=0}^{[n/2]} \fr{{\color{\red}(-1)^{\ell}}}{2^\ell}   \sum_{C_{\ell}}  
2(\pa_{P}g_{C_\ell})    \pa_{P} I_{C_{\ell}} \label{second-derivative-second} \\
& &+n!P_{\sym}\sum_{\ell=0}^{[n/2]} \fr{{\color{\red}(-1)^{\ell}}}{2^\ell}   \sum_{C_{\ell}}  g_{C_\ell}
    \pa^2_{P} I_{C_{\ell}}. \label{second-derivative-third}
\eeqa
Terms (\ref{second-derivative-first}) and (\ref{second-derivative-second}) are estimated as before. 
To bound~(\ref{second-derivative-third})  we use formula~(\ref{second-der-decomp}) and assumptions~(\ref{second-derivatives}), (\ref{second-derivatives-continued}). \qed

\section{Verification of assumptions of Proposition~\ref{intermediate-proposition}} \label{verification-section}
\setcounter{equation}{0}

\subsection{Spectral ingredients}\label{spectral-ingredients-sub}
\setcounter{equation}{0}

In this subsection we collect spectral ingredients which constitute the key input needed to verify
the assumptions of Proposition~\ref{intermediate-proposition}.  We start with a list of relevant defnitions,
some of which appeared before. For the sake of clarity in later parts of this section, we also  limit
the appearance of the subscript $P,\si$ in our notation.
\beqa
R\y&:=&\y R_{P,\si}=\fr{1}{H_{P,\si}-E_{P,\si}}, \label{first-res} \\
R_{i}\y&:=&\y R_{P,\si;i}:=\fr{1}{H_{P-\unk_{i},\si}-E_{P,\si}+|\un k|_{i}}, \label{sec-res}\\ 
R^{i}\y&:=&\y R_{P,\si}^{i}:=\fr{1}{H_{P,\si}-E_{P,\si}+ r(\unk)_{i} }, \textrm{ where } r(\unk)_{i}:=|\unk|_{i}+|\unk_{i}|^2/2-\nabla E_{P,\si}\cdot \unk_{i}, 
\label{where-res} \\
\La\y&:=&\y \La_{P,\si}= \nabla E_{P,\si}-(P-P_{\pho}),\label{def-La}\\
\La_{i}\y&:=&\y\La+\unk_{i-1}+\h k_{i},\label{lai}  \\
\La^i\y&:=&\y\La+\unk_{i},\\
\cpsi\y&:=&\y\cpsi_{P,\si}, \\ 
Q\y&:=&\y|\cpsi\ran \lan \cpsi|,
\eeqa
where $\unk_i:=k_1+\cdots+k_i$ and $|\unk|_i:=|k_1|+\cdots+|k_i|$. 
Of course $R$ is only defined on
the range of $Q^{\bot}:=1-Q$. The following lemma treats the  existence of the other resolvents. 
\begin{lem}\label{elem-lemma} We have the following bounds
\beqa
& & \|R_{i}\|, \ \  \|R_{i}\La\|,  \  \  \|R_{i}Q^{\bot}\La\| \leq \fr{c}{|\unk|_{i}}, \label{elem-resolv} \\
& &\|Q \La\|\leq c.
\eeqa
The same estimates hold if $R_i$ is replaced with $R^i$. 
\end{lem}
\proof  The estimate on $\|R_i\|$
uses formula (3.6) of \cite{DP17.1}. $\|R_i\La\|$ is easily reduced to estimating $\|R_i(P-P_{\pho}-\unk_i)^2R_i\|$
which can be handled by Lemma A.3 of \cite{DP17.1}. To bound $\|Q\La\|$ we write 
$Q\La=(E_{P,\si}+i)Q(H_{P,\si}+i)^{-1}\La$ and apply again  Lemma A.3 of \cite{DP17.1}. The estimate
on  $\|R_{i}Q^{\bot}\La\|$  follows from the bounds on $\|Q \La\|$ and $\|R_{i}\La\|$.  The bound on $\|R^i\|$ follows from the fact that  $r(\unk)_{i}\geq \h |\unk|_{i}$ which holds true due to $|\nabla E_{P,\si}|\leq 1/3+c|\la|$ (cf. formula (3.7) of \cite{DP17.1}). Other bounds involving $R^i$ follow by similar considerations as above. \qed\\
Furthermore, we note the following relations {\color{\red}and estimates}:
\beqa
R_{i}\y&=&\y R^{i}-R_{i}(\unk_{i}\cdot \La)R^{i}, \label{R-shift-bound}\\
\pa_{P}R_{i}\y&=&\y R_{i}\La^{i}R_{i}, \label{R-derivative} \\
\pa_{P}\La\y&=&\y O(1), \label{deriv-la} \\
\pa_{P}^2\La\y &=&\y O(\ir), \label{second-deriv-la} \\
\pa_{P}\cpsi\y&=&\y R\La\cpsi. \label{derivative-formula}
\eeqa
Here (\ref{R-shift-bound}), (\ref{R-derivative}) are obtained by  straightforward computations,
(\ref{deriv-la}) and (\ref{second-deriv-la}) follow from (\ref{velocity-boundedness}) and (\ref{derivative-formula})
was derived in Section~5 of \cite{DP17.1}.

Next we collect more sophisticated  spectral estimates. The following theorem
follows from the main technical results of \cite{DP17.1} by maximal modulus principle arguments (explained in the same reference).
\begin{thm}\label{spectr-theorem}\cite{DP17.1} Let $R(\De):=(H_{P,\si}-E_{P,\si}+\De)^{-1}$. Then 
\beqa
& &\  \ \ \sup_{\De\geq 0}\|R(\De)\La\cpsi\|\leq c\ir, \label{spectral-theorem-first} \\
& &\sup_{\De_1,\De_2\geq 0 }\|R(\De_1)Q^{\bot}\La R(\De_2)\La \cpsi\|\leq c\ir.\label{spectral-theorem-second}
\eeqa
\end{thm}
\nin Since $\La\cpsi$ is orthogonal to $\cpsi$, so the l.h.s. of (\ref{spectral-theorem-first}), (\ref{spectral-theorem-second}) are well defined also for $\De=\De_2=0$.  
  Lemma~\ref{elem-lemma} and Theorem~\ref{spectr-theorem} have the following corollary. 
\bec\label{key-corollary} With the above definitions, we have for $i\geq i'\geq  i''$
\beqa
& &\|R\La \cpsi\|\leq c\ir, \label{first-and-half-consequence} \\
& &\|R_{i} \La_{i}\cpsi\|\leq c\ir, \label{first-consequence} \\
& &\|R_{i}Q^{\bot}\La_{i}R\La\cpsi\|\leq c\ir, \label{fourth-consequence} \\
& &\|R_{i}Q^{\bot}\La_{i}R_{i'}\La_{i'}\cpsi\|\leq c\ir, \label{second-consequence} \\
& &\|R_{i}\La_{i}R_{i'}\cpsi\|\leq \ir \fr{c}{|\unk|_{i'}  }, \label{third-consequence} \\
& &\|R_{i}\La_{i} R_{i'}\La_{i'} R_{i''} \La_{i''}\cpsi\|\leq \ir\fr{c}{|\unk|_{i'} }, \label{last-consequence}\\
& &\|R_{i}\La_{i} R_{i'}\La_{i'} R \La \cpsi\|\leq \ir \fr{c}{|\unk|_{i'} }. \label{last-last-consequence}
\eeqa
The results remain true if some of $\La_{i}$ are replaced with $\La^i$ or $\La$.
\eec
\proof Property (\ref{first-and-half-consequence}) follows directly from (\ref{spectral-theorem-first}). To show~(\ref{first-consequence}) we  shift the resolvent {\color{\red}(see (\ref{R-shift-bound})), take (\ref{lai}) into account,} and make use of (\ref{spectral-theorem-first}) and (\ref{elem-resolv}):
\beqa
R_{i}\La_{i}\cpsi=R^{i}\La_{i}\cpsi-R_{i}(\unk_{i}\cdot \La)R^{i}\La_{i}\cpsi=O(\ir).
\eeqa
In order to prove (\ref{fourth-consequence})  we again make use of the shift of the resolvent{\color{\red}, of the definition in (\ref{lai}), }
and of (\ref{spectral-theorem-second}):
\beqa
R_{i}Q^{\bot}\La_{i}R\La\cpsi=R^{i}Q^{\bot}\La_{i}R\La\cpsi
-R_{i}(\unk_{i}\cdot \La)R^{i}Q^{\bot}\La_{i}R\La\cpsi=O(\ir).
\eeqa
{\color{\red}Concerning} (\ref{second-consequence}), we will repetitively use (\ref{spectral-theorem-second}) and (\ref{first-consequence}):  We {\color{\red}implement a resolvent shift}
\beqa
R_{i}Q^{\bot}\La_{i}R_{i'}\La_{i'}\cpsi
&=&R_{i}Q^{\bot}\La_{i}R^{i'}\La_{i'}\cpsi \label{first-term-corr}\\
& &-R_{i}Q^{\bot}\La_{i}R_{i'}(\unk_{i'}\cdot \La)R^{i'}\La_{i'}\cpsi \label{second-term-corr}.
\eeqa
We consider the last two terms {\color{\red}(\ref{first-term-corr}) and (\ref{second-term-corr}) separately. In the first one we implement a shift on the first resolvent from the left, and make use of the fact that $\La_{i'}=\La+O(|\unk|_{i'})$}:
\beqa
& &(\ref{first-term-corr})=R^{i}Q^{\bot}\La_{i}R^{i'}\La
\cpsi-R_{i}(\unk_{i}\cdot \La) R^{i}Q^{\bot}\La_{i}R^{i'}\La\cpsi\non\\
& &\ph{4444444}+R^{i}Q^{\bot}\La_{i}R^{i'}O(|\unk|_{i'})
\cpsi-R_{i}(\unk_{i}\cdot \La) R^{i}Q^{\bot}\La_{i}R^{i'}O(|\unk|_{i'})\cpsi=O(\ir) \ \  \ \ 
\eeqa
{\color{\red}With regard to (\ref{second-term-corr}) we split the identity operator in front of $R_{i'}$  into $Q+Q^{\bot}$:}
\beqa
& &(\ref{second-term-corr})=-R_{i}Q^{\bot}\La_{i}R_{i'}Q^{\bot}(\unk_{i'}\cdot \La)R^{i'}\La_{i'}\cpsi
\label{second-term-corr-one}\\
& &\ph{4444444}-R_{i}Q^{\bot}\La_{i}R_{i'}Q(\unk_{i'}\cdot \La)R^{i'}\La_{i'}\cpsi.\label{second-term-corr-two}
\eeqa
The last two terms give
\beqa
(\ref{second-term-corr-one})\y&=&\y-(R_{i}\unk_{i'})Q^{\bot}\La_{i}R^{i'}Q^{\bot} \La R^{i'}\La_{i'}\cpsi\non\\
& &+
(R_{i} \unk_{i'})  Q^{\bot}\La_{i}R_{i'}(\unk_{i'} \cdot \La )R^{i'}Q^{\bot} \La R^{i'}\La_{i'}\cpsi=O(\ir),\label{1}\\
(\ref{second-term-corr-two})\y&=&\y-R_{i}Q^{\bot}\La_{i} (\unk_{i'}R^{i'})Q \La R^{i'}\La_{i'}\cpsi\non\\
& &+
(R_{i} \unk_{i'}) Q^{\bot}\La_{i}  R_{i'}(\unk_{i'}\cdot \La  )R^{i'}Q \La R^{i'}\La_{i'}\cpsi=O(\ir),\label{2}
\eeqa
where we made use of (\ref{spectral-theorem-second}) and (\ref{first-consequence}) {\color{\red}for (\ref{1}) and (\ref{2}), respectively.} 

Now we show (\ref{third-consequence}): We shift the resolvent next to $\cpsi$ and make use of (\ref{first-consequence}):
\beqa
R_{i}\La_{i}R_{i'}\cpsi\y&=&\y
R_{i}\La_{i}R^{i'}\cpsi-R_{i}\La_{i} R_{i'} (\unk_{i'}\cdot\La) R^{i'}\cpsi\non\\
\y&=&\y O(|\unk|_{i'}^{-1})R_{i}\La_{i}\cpsi-O(|\unk|_{i'}^{-1})(\unk_{i'} R_{i} )\La_{i} R_{i'}\La\cpsi
=O(|\unk|_{i'}^{-1} \ir), \ \ \ \  \ \ \ \  \ \ 
\eeqa
where  in the last step we made use of (\ref{first-consequence})
and $(\unk_{i'}R_{i}\La_{i})=O(1)$.

To prove (\ref{last-consequence}), we write 
\beqa
R_{i}\La_{i} R_{i'}\La_{i'} R_{i''} \La_{i''}\cpsi=R_{i}\La_{i} R_{i'} Q^{\bot}\La_{i'} R_{i''} \La_{i''}\cpsi+R_{i}\La_{ i} R_{i'} Q \La_{i'} R_{i''} \La_{i''}\cpsi
\eeqa
apply (\ref{first-consequence}), (\ref{second-consequence}), (\ref{third-consequence}) and 
the fact that $|\unk|_{i}^{-1}\leq |\unk|^{-1}_{i'}$. 

To verify (\ref{last-last-consequence}) we proceed analogously. We write
\beqa
R_{i}\La_{i} R_{i'}\La_{i'} R \La\cpsi=R_{i}\La_{i} R_{i'} Q^{\bot}\La_{i'} R \La\cpsi
+R_{i}\La_{i} R_{i'} Q \La_{i'} R \La\cpsi
\eeqa 
apply (\ref{second-consequence}), (\ref{third-consequence}), (\ref{first-and-half-consequence}) and the fact that 
$|\unk|_{i}^{-1}\leq |\unk|^{-1}_{i'}$. \qed

\subsection{Estimates on derivatives} \label{derivatives}

\begin{prop} There hold the bounds for $i\geq i'$ 
\beqa
\|  I_{ C_{\ell}}\|\y&\leq&\y c^n \prod_{m'=1}^n |\unk|^{-1}_{m'}, \label{R-zero-derivative}\\
\|\pa_{R_{i}}  I_{C_{\ell}}\|\y&\leq&\y c^n\ir \prod_{m'=1}^n |\unk|^{-1}_{m'}, \label{R-first-derivative}\\
\|\pa_{R_{i'}}\pa_{R_{i}}  I_{C_{\ell}}\|\y&\leq&\y c^n\ir \prod_{m'=1}^n |\unk|^{-1}_{\m'}.
\eeqa
\end{prop}
\proof  Estimate (\ref{R-zero-derivative}) is obvious from formula~(\ref{I-C-Def}), standard resolvent bounds (Lemma~\ref{elem-lemma}) and the fact that $\hde_{C_{\ell}}$ can only decrease the number of resolvents. So  we can proceed to  the first derivative and divide the proof into several cases: \\

\nin\textbf{Case 1:} $i\notin  C^{2\ell}$. 
\beqa
\pa_{R_{i}} I_{C_{\ell}}= \hde_{C_{\ell}}  
\pa_{R_{i}} \hal(n,\ldots, 1)\cpsi
= \hde_{C_{\ell}}     
 \hal(n,\ldots, i+1)R_{i}\La^i  \hal(i,\ldots, 1)\cpsi. \label{first-derivative-start}
\eeqa
We consider several sub-cases. \\
\nin$\bullet$  $1\notin  C^{2\ell}$.
\beqa
\pa_{R_{i}} I_{ C_{\ell}}=\hde_{C_{\ell}}     
 \hal(n,\ldots, i+1)R_{i}\La^i  \hal(i,\ldots, 2)R_{1}\La_{1}\cpsi
=O(c^n\ir \prod_{m'=1}^n |\unk|^{-1}_{m'}),
\eeqa
where we made use of standard resolvent bounds (Lemma~\ref{elem-lemma}), $R_{1}\La_{1}\cpsi=O(\ir)$, $|\unk|_{i}^{-1}\leq |\unk|_{1}^{-1}$ for $i\geq 1$  and  the fact that the action of $ \hde_{C_{\ell}} $ can only decrease the number of resolvents.\\
\nin$\bullet$  $1\in  C^{2\ell}$.
\beqa
\pa_{R_{i}} I_{ C_{\ell}}\y&=&\y \hde'_{C_{\ell}}   (R_{2}  \hpa_{2}\hpa_{1})     
 \hal(n,\ldots, i+1)R_{i}\La^i  \hal(i,\ldots, 3)R_{2}\La_{2}R_{1}\La_{1}\cpsi\non\\
\y&=&\y  \hde'_{C_{\ell}}        
 \hal(n,\ldots, i+1)R_{i}\La^i  \hal(i,\ldots, 3)R_{2}\cpsi
=O(c^n\ir \prod_{m'=1}^n |\unk|^{-1}_{m'}),
\eeqa
where we made use of Lemma~\ref{elem-lemma},  $|\unk|_{i}^{-1}\leq |\unk|_{1}^{-1}$ for $i\geq 1$  and  the fact that the action of 
$\hde'_{C_{\ell}}$ can only decrease the number of resolvents.\\
\nin\textbf{Case 2:} $i\in  C^{2\ell}$. Clearly, for $i\in C^{2\ell}\backslash C_{\ell}$ the $i$-th resolvent is eliminated from the
expression by the action of $(R_{i+1}\hpa_{i+1}\hpa_{i})$ so $\pa_{R_{i}} I_{C_{\ell}}=0$. Thus we can assume
 $i\in  C_{\ell}$.
\beqa
\pa_{R_{i}} I_{C_{\ell}}=\hde'_{C_{\ell}}  (R_{i}\La^i R_{i}  \hpa_{i}\hpa_{i-1})     
 \hal(n,\ldots, 1)\cpsi
=O(c^n\ir \prod_{m'=1}^n |\unk|^{-1}_{m'}),
\eeqa
where we made use of Lemma~\ref{elem-lemma} and  $|\unk|_{i}^{-1}\leq 
|\unk|_{i-1}^{-1}$. This concludes the analysis
of the first derivative and we proceed to the second derivative.\\

\nin\textbf{Case (1.1):} $i, i'\notin  C^{2\ell}$. Starting from formula~(\ref{first-derivative-start}), we have 
\beqa
& &\pa_{R_{i'}}\pa_{R_{i}} I_{C_{\ell}} 
= \hde_{C_{\ell}}     
 \hal(n,\ldots, i+1)R_{i}\La^i  \hal(i,\ldots, i'+1) R_{i'}\La^{i'}   \hal(i',\ldots, 1)\cpsi, \label{first-formula-s-d}
\eeqa
where $\hal(i,\ldots, i'+1)=1$ is understood for $i=i'$.
We consider several sub-cases here:\\ 
$\bullet$ $i'=1$, $i\in \{1,2,3\}$. Consider the case\footnote{Expression (\ref{subcase-one}) is somewhat schematic.
$2R_{1} \La^1 R_{1} \La^1 R_{1}$ stands for $R_{1} (\La^1)^j R_{1} (\La^1)^{j'} R_{1}+\{j\leftrightarrow j'\}$, but the indices $j,j'=1,2,3$,
coming from $\pa_{P^j}\pa_{P^{j'}}$ are suppressed in our notation.} $i=1$:
\beqa
\pa_{R_{i'}}\pa_{R_{i}} I_{C_{\ell}}
=\hde_{C_{\ell}}  \hal(n,\ldots,  2) 2R_{1} \La^1 R_{1} \La^1 R_{1}\La_{1}\cpsi=O(c^n\ir \prod_{m'=1}^n |\unk|^{-1}_{m'}), \label{subcase-one}
\eeqa
where we made use of (\ref{last-consequence}). The cases $i=2$ and $i=3$ can be treated analogously, making use
of the fact that $|\unk|^{-1}_{2}, |\unk|^{-1}_{3}\leq |\unk|^{-1}_{1}$ and noting that for $i'=1$ and  $i\in \{1,2,3\}$
and $i,i'\notin C^{2\ell}$ we necessarily have $i_{1}-1>i$, so $\hde_{C_{\ell}}$ does not interfere with the above argument. \\
$\bullet$ $i'=1$, $i>3$.
\beqa
\pa_{R_{i'}}\pa_{R_{i}} I_{C_{\ell}}
=\hde_{C_{\ell}}  \hal(n,\ldots, i+1)R_{i}\La^i  \hal(i,\ldots, 2) R_{1}\La^1R_{1}\La_{1}\cpsi.
\label{second-case}
\eeqa
First, suppose $2\notin C^{2\ell}$. Then we can write  
$\hal(i,\ldots, 2)=\hal(i,\ldots, 3)R_{2}\La_{2}$
apply (\ref{last-consequence}) as in (\ref{subcase-one}) and use 
$|\unk|_{i}^{-1}\leq |\unk|_{2}^{-1}$ to obtain the required
bound. Next, suppose $2\in C^{2\ell}$. Then $\hde_{C_{\ell}}$ eliminates $R_{2}$ from expression (\ref{second-case})
so we can use  $|\unk|_{i}^{-1}\leq |\unk|_{2}^{-1}$ and (\ref{first-consequence}) to obtain a bound as in (\ref{subcase-one}). \\
$\bullet$ $i'=2{\color{\red}, i\geq 2}$. 
\beqa
\pa_{R_{i'}}\pa_{R_{i}} I_{C_{\ell}}
=\hde_{C_{\ell}}  \hal(n,\ldots, i+1)R_{i}\La^i  \hal(i,\ldots, 3) (R_{2}\La^{2}R_{2})\La_{2}R_{1}\La_{1}\cpsi
\label{second-case-one}
\eeqa
Here we make use of (\ref{last-consequence}) and $|\unk|^{-1}_{i}\leq 
|\unk|_{1}^{-1}$. \\ 
$\bullet$ $i'=3{\color{\red}, i\geq 3}$. 
\beqa
& &\pa_{R_{i'}}\pa_{R_{i}} I_{C_{\ell}} 
=\hde_{C_{\ell}}     
 \hal(n,\ldots, i+1)R_{i}\La^i  \hal(i,\ldots, 4) R_{3}\La^{3}   \hal(3,2,1)\cpsi. \quad\quad
\eeqa
Suppose first that $1\notin C^{2\ell}$. Then we write  $\hal(3,2,1)\cpsi=R_{3}\La_{3} R_{2}\La_{2} R_{1}\La_{1}\cpsi$ and apply (\ref{last-consequence}) and 
$|\unk|_{i}^{-1}\leq 
|\unk|_{1}^{-1}$ to obtain the
required bound. Now  suppose that $1\in C^{2\ell}$. Writing $\hde_{C_{\ell}}=\hde'_{C_{\ell}} (R_{2}\hpa_{2}\hpa_1)$,
we have
\beqa
& &\pa_{R_{i'}}\pa_{R_{i}} I_{C_{\ell}} 
= \hde'_{C_{\ell}}     
 \hal(n,\ldots, i+1)R_{i}\La^i  \hal(i,\ldots, 4) R_{3}\La^{3}R_{3}\La_{3}R_{2}\cpsi. \quad\quad
\eeqa
By (\ref{third-consequence}) we have $R_{3}\La_{3}R_{2}\cpsi=O(\ir |\unk|_{2}^{-1})$. Thus by
$|\unk|_{i}^{-1}\leq |\unk|_{1}^{-1}$ we obtain the required bound. \\
\nin $\bullet$ $i'>3{\color{\red}, i\geq  i'}$. We come back to formula (\ref{first-formula-s-d}): 
\beqa
& &\pa_{R_{i'}}\pa_{R_{i}} I_{C_{\ell}} 
= \hde_{C_{\ell}}     
 \hal(n,\ldots, i+1)R_{i}\La^i  \hal(i,\ldots, i'+1) R_{i'}\La^{i'}   \hal(i',\ldots, 1)\cpsi.
\eeqa

If $2, 3, 4 \notin C_{\ell}$, we write $\hal(i',\ldots, 1)\cpsi=\hal(i',\ldots, 4)R_{3}\La_{3}R_{2}\La_{2}R_{1}\La_{1}\cpsi$ and apply (\ref{last-consequence}) and 
$|\unk|_{i}^{-1}\leq |\unk|_{1}^{-1}$, $|\unk|_{i'}^{-1}\leq |\unk|_{3}^{-1}$to obtain the required bound.

If $2\in C_{\ell}$  (hence, automatically, $3\notin C_{\ell}$) and $4\notin C_{\ell}$ we have
\beqa
\pa_{R_{i'}}\pa_{R_{i}} I_{C_{\ell}} 
= \hde'_{C_{\ell}}     
 \hal(1,\ldots, i+1)R_{i}\La^i  \hal(i,\ldots, i'+1) R_{i'}\La^{i'}   
\hal(i',\ldots, 4)R_{3}\La_{3}R_{2}\cpsi.
\eeqa
In this case we apply (\ref{third-consequence}), which gives $R_{3}\La_{3}R_{2}\cpsi=O(\ir |\unk|_{2}^{-1})$
and we bound $|\unk|_{i}^{-1}\leq |\unk|_{1}^{-1}$, $|\unk|_{i'}^{-1}\leq 
|\unk|_{3}^{-1}$ to obtain the required estimate.

If $2, 4 \in C_{\ell}$, then, automatically,  $i'>4$ and  we obtain
\beqa
\pa_{R_{i'}}\pa_{R_{i}} I_{C_{\ell}} 
= \hde''_{C_{\ell}}     
 \hal(n,\ldots, i+1)R_{i}\La^i  \hal(i,\ldots, i'+1) R_{i'}\La^{i'}   \hal(i',\ldots, 5)R_{4}R_{2}\cpsi,
\eeqa
where $\hde_{C_{\ell}}=:\hde''_{C_{\ell}} (R_{4}\hpa_{4}\hpa_{3}) (R_{2}\hpa_{2}\hpa_{1})$.
This expression can be handled by standard resolvent bounds and 
$|\unk|_{i}^{-1}\leq |\unk|_{1}^{-1}$, $|\unk|_{i'}^{-1}\leq |\unk|_{3}^{-1}$.

If $3\in C_{\ell}$ (and therefore $2,4\notin C_{\ell}$)  we have
\beqa
\pa_{R_{i'}}\pa_{R_{i}} I_{C_{\ell}} 
= \hde'_{C_{\ell}}     
 \hal(n,\ldots, i+1)R_{i}\La^i  \hal(i,\ldots, i'+1) R_{i'}\La^{i'}   \hal(i',\ldots, 4) R_{3} R_{1}\La_{1}\cpsi, 
\eeqa
where $\hde_{C_{\ell}}=:\hde'_{C_{\ell}}(R_{3}\hpa_{3}\hpa_{2})$. This expression is handled making use of
$R_{1}\La_{1}\cpsi=O(\ir)$, standard resolvent bounds and 
$|\unk|_{i}^{-1}\leq |\unk|_{1}^{-1}$, $|\unk|_{i'}^{-1}\leq |\unk|_{2}^{-1}$.

\nin\textbf{Case (1.2):} $i \notin  C^{2\ell}$, $i'\in C^{2\ell}$. We can assume that $i'\in C_{\ell}$ since otherwise
$R_{i'}$ does not appear in the expression and the result is zero. We set $\hde_{C_{\ell}}=: \hde'_{C_{\ell}} (R_{i'}\hpa_{i'}\hpa_{i'-1})$
and write, making use of (\ref{first-derivative-start})
\beqa
\pa_{R_{i'}}\pa_{R_{i}} I_{C_{\ell}}
= \hde'_{C_{\ell}}  (R_{i'}\La^{i'}R_{i'}\hpa_{i'}\hpa_{i'-1})   
 \hal(n,\ldots, i+1)R_{i}\La^i  \hal(i,\ldots, 1)\cpsi. 
\eeqa
By our assumptions, $i>i'$. We study several sub-cases:\\
$\bullet$ ${\color{\red}i>}i'=2$. 
\beqa
\pa_{R_{i'}}\pa_{R_{i}} I_{C_{\ell}}
= \hde'_{C_{\ell}}   
 \hal(n,\ldots, i+1)R_{i}\La^i  \hal(i,\ldots, 3)R_{2}\La^{2}R_{2}\cpsi. 
\eeqa
In this case we apply (\ref{third-consequence}) which gives $R_{2}\La^{2}R_{2}\cpsi=O(\ir |\unk|_{2}^{-1})$
and estimate $|\unk|_{i}^{-1}\leq |\unk|_{1}^{-1}$ to obtain the required bound.\\
$\bullet$ $i>i'>2$ and $2\notin C_{\ell}$.
\beqa
\pa_{R_{i'}}\pa_{R_{i}} I_{C_{\ell}}
= \hde'_{C_{\ell}}   
 \hal(n,\ldots, i+1)R_{i}\La^i  \hal(i,\ldots, i'+1) R_{i'}\La^{i'}R_{i'}  
\hal(i'-2,\ldots, 2)R_{1}\La_{1}\cpsi,
\eeqa
where $\hal(i'-2,\ldots, 2)=1$ is understood for $i'=3$.
In this case we apply (\ref{first-consequence}) which gives $R_{1}\La_{1}\cpsi=O(\ir)$ and 
$|\unk|_{i}^{-1}\leq |\unk|_{1}^{-1}$, $|\unk|_{i'}^{-1}\leq |\unk|_{i'-1}^{-1}$ to obtain the required bound.\\
$\bullet$ $i>i'>2$ and $2\in C_{\ell}$ (hence $i'>3$).
\beqa
\pa_{R_{i'}}\pa_{R_{i}} I_{C_{\ell}}
= \hde''_{C_{\ell}}   
 \hal(n,\ldots, i+1)R_{i}\La^i  \hal(i,\ldots, i'+1) R_{i'}\La^{i'}R_{i'}  
\hal(i'-2,\ldots, 3)R_{2}\cpsi,
\eeqa
where $\hal(i'-2,\ldots, 3)=1$ is understood for $i'=4$.
This case is handled by standard resolvent bounds  and 
$|\unk|_{i}^{-1}\leq |\unk|_{1}^{-1}$, $|\unk|_{i'}^{-1}\leq |\unk|_{i'-1}^{-1}$.

\nin\textbf{Case (2.1):} $i \in  C^{2\ell}$, $i'\notin C^{2\ell}$ {\color{\red} and $i>i'$}. As in the previous case, we can assume that $i\in C_{\ell}$.
We set $\hde_{C_{\ell}}= \hde'_{C_{\ell}} (R_{i}\hpa_{i}\hpa_{i-1})$
and write, making use of (\ref{first-derivative-start})
\beqa
\pa_{R_{i'}}\pa_{R_{i}} I_{C_{\ell}}
\y&=&\y \hde'_{C_{\ell}}  (R_{i}\La^{i}R_{i}\hpa_{i}\hpa_{i-1})   
 \hal(n,\ldots, i'+1)R_{i'}\La^{i'}  \hal(i',\ldots, 1)\cpsi\non\\ 
\y&=&\y \hde'_{C_{\ell}}    
 \hal(n,\ldots, i+1)R_{i}\La^{i}R_{i} \hal(i-2,\ldots, i'+1)R_{i'}\La^{i'}  \hal(i',\ldots, 1)\cpsi, \ \ \ 
\eeqa
where $\hal(i-2,\ldots, i'+1)=1$ for $i-2<i'+1$.
We study two sub-cases:\\
$\bullet$ $2\notin C_{\ell}$.
\beqa
\pa_{R_{i'}}\pa_{R_{i}} I_{C_{\ell}}= \hde'_{C_{\ell}}    
 \hal(n,\ldots, i+1)R_{i}\La^{i}R_{i} \hal(i-2,\ldots, i'+1)R_{i'}\La^{i'}  \hal(i',\ldots, 2)R_{1}\La_{1}\cpsi.
\eeqa
This is handled using $R_{1}\La_{1}\cpsi=O(\ir)$  and 
$|\unk|_{i}^{-1}\leq |\unk|_{i-1}^{-1}$, $|\unk|_{i'}^{-1}\leq |\unk|_{1}^{-1}$. \\
$\bullet$ $2\in C_{\ell}$.
\beqa
\pa_{R_{i'}}\pa_{R_{i}} I_{C_{\ell}}= \hde''_{C_{\ell}}    
 \hal(n,\ldots, i+1)R_{i}\La^{i}R_{i} \hal(i-2,\ldots, i'+1)R_{i'}\La^{i'}  \hal(i',\ldots, 3)R_{2}\cpsi. \quad
\eeqa
Here we use again $|\unk|_{i}^{-1}\leq |\unk|_{i-1}^{-1}$, $|\unk|_{i'}^{-1}\leq |\unk|_{1}^{-1}$. \\
\nin\textbf{Case (2.2):} $i,i' \in  C^{2\ell}$. As before, we can actually assume that $i,i' \in  C_{\ell}$. We study
several sub-cases:\\
$\bullet$ $i=i'$, $i=2$\footnote{Expression (\ref{schematic-next}) is somewhat schematic.
$2R_{i} \La^i R_{i} \La^i R_{i}$ stands for $R_{i} (\La^i)^j R_{i} (\La^i)^{j'} R_{i}+\{j\leftrightarrow j'\}$, but the indices $j,j'=1,2,3$,
coming from $\pa_{P^j}\pa_{P^{j'}}$ are suppressed in our notation.}.
\beqa
\pa_{R_{i'}}\pa_{R_{i}} I_{C_{\ell}}
\y&=&\y \hde'_{C_{\ell}}(2R_{i}\La^iR_{i}\La^{i}R_{i}\hpa_{i}\hpa_{i-1})  \hal(n,\ldots,  1)\cpsi\non\\
\y&=&\y \hde'_{C_{\ell}}  \hal(n,\ldots,  3) 2R_{2}\La^{2}R_{2}\La^{2}R_{2}\cpsi. \label{schematic-next}
\eeqa 
This is estimated using (\ref{third-consequence}) and $|\unk|_{2}^{-1}\leq 
|\unk|_{1}^{-1}$. \\
$\bullet$ $i=i'$, $i>2$, $2\notin C_{\ell}$.
\beqa
\pa_{R_{i'}}\pa_{R_{i}} I_{C_{\ell}}
= \hde'_{C_{\ell}}(2R_{i}\La^iR_{i}\La^{i}R_{i}\hpa_{i}\hpa_{i-1})  \hal(n,\ldots,  2)R_{1}\La_{1}\cpsi.
\eeqa
This is estimated using $R_{1}\La_{1}\cpsi=O(\ir)$, $|\unk|_{i}^{-1}\leq |\unk|_{i-1}^{-1}$, 
$|\unk|_{i}^{-1}\leq |\unk|_{1}^{-1}$.\\
$\bullet$ $i=i'$, $i>2$, $2\in C_{\ell}$.
\beqa
\pa_{R_{i'}}\pa_{R_{i}} I_{C_{\ell}}
= \hde''_{C_{\ell}}(2R_{i}\La^iR_{i}\La^{i}R_{i}\hpa_{i}\hpa_{i-1})  \hal(n,\ldots,  3)R_{2}\cpsi.
\eeqa
Here we use again $|\unk|_{i}^{-1}\leq |\unk|_{i-1}^{-1}$, 
$|\unk|_{i}^{-1}\leq |\unk|_{1}^{-1}$.\\ 
$\bullet$ $i{\color{red}>}i'$.  
\beqa
\pa_{R_{i'}}\pa_{R_{i}} I_{C_{\ell}}
= \hde''_{C_{\ell}}(R_{i}\La^{i}R_{i}\hpa_{i}\hpa_{i-1})   (R_{i'}\La^{i'}R_{i'}\hpa_{i'}\hpa_{i'-1})   
 \hal(n,\ldots,  1)\cpsi. 
\eeqa 
Here we use standard resolvent bounds and $|\unk|_{i}^{-1}\leq 
|\unk|_{i-1}^{-1}$, $|\unk|_{i'}^{-1}\leq |\unk|_{i'-1}^{-1}$. \qed
\begin{prop} There hold the bounds  
\beqa
\|\pa_{\La_{i}}  I_{ C_{\ell}}\|\y&\leq&\y c^n \prod_{m'=1}^n |\unk|^{-1}_{m'}, \label{La-derivative}\\
\|\pa_{\La_{i'}}  \pa_{\La_{i}}  I_{ C_{\ell}}\|\y&\leq&\y c^n \prod_{m'=1}^n 
|\unk|^{-1}_{m'}, \label{La-second-derivative}\\
\|\pa_{R_{i'}}  \pa_{\La_{i}}  I_{ C_{\ell}}\|\y&\leq&\y c^n \ir \prod_{m'=1}^n 
|\unk|^{-1}_{m'},  \label{La-R} \\
\|  \pa_{\La_{i}} \pa_{R_{i'}} I_{ C_{\ell}}\|\y&\leq&\y c^n \ir \prod_{m'=1}^n 
|\unk|^{-1}_{m'}.  \label{R-La}
\eeqa
\end{prop}
\proof We first show (\ref{La-derivative}).  Since for $i\in C^{2\ell}$ the factor  $R_{i}\La_{i}$ does not appear in $ I_{ C_{\ell}}$, we can assume $i\notin C^{2\ell}$. Then
\beqa
\pa_{\La_{i}} I_{C_{\ell}}\y&=&\y\hde_{ C_{\ell} }     
\pa_{\La_{i}} \hal(n,\ldots, 1)\cpsi\non\\
\y&=&\y (\pa_P\La_{i})\hde_{ C_{\ell} }    
 \hal(n,\ldots, i+1)  R_{i} \hal(i-1,\ldots, 1)\cpsi  \label{La-first-derivative-formula}\\
\y&=&\y O(c^n \prod_{m'=1}^n |\unk|^{-1}_{m'}), \label{La-first-derivative-bound}
\eeqa
where we used (\ref{deriv-la}) and the fact that the action of $\pa_{\La_{i}}$ 
does not change the number of resolvents.\\

\nin Now we consider (\ref{La-second-derivative}). As before we can assume that $i, i'\notin C^{2\ell}$. Let us 
first set $i\geq i'$
\beqa
\pa_{\La_{i'}}\pa_{\La_{i}} I_{C_{\ell}}\y&=&\y \hde_{ C_{\ell} } \pa_{\La_{i'}}\pa_{\La_{i}} \hal(n,\ldots, 1)\cpsi\non\\
\y&=&\y (\pa_P\La_{i}) (\pa_P\La_{i'})\hde_{ C_{\ell} } \hal(n,\ldots, i+1)R_{i} 
\hal(i-1,\ldots,  i'+1)R_{i'} \hal(i'-1,\ldots,  1)   \cpsi \non\\
\y&=&\y O(c^n \prod_{m'=1}^n |\unk|^{-1}_{m'}),
\eeqa
where we argued as in (\ref{La-first-derivative-bound}). For $i<i'$ the argument is analogous. For $i=i'$ we have
\beqa
\pa_{\La_{i'}}\pa_{\La_{i}} I_{C_{\ell}}=(\pa^2_P\La_{i})\hde_{ C_{\ell} }    
 \hal(n,\ldots, i+1)  R_{i} \hal(i-1,\ldots, 1)\cpsi= O(c^n\ir \prod_{m'=1}^n |\unk|^{-1}_{m'}), \
\eeqa
where we argued as before and used that $\pa^2_P\La_{i}=O(\ir)$.\\

\nin To estimate~(\ref{La-R}) we can assume as before $i\notin C^{2\ell}$, come
back to formula~(\ref{La-first-derivative-formula}), denote $\pa_{P}\La_{i}$ by $O(1)$ and consider several cases: \\
\nin\textbf{Case 1:} $i,i'\notin C^{2\ell}, i'>i$. 
\beqa
\pa_{R_{i'}}  \pa_{\La_{i}}  I_{ C_{\ell}}\y&=&\y O(1)\hde_{ C_{\ell} } \pa_{R_{i'}}    
 \hal(n,\ldots, i+1)  R_{i} \hal(i-1,\ldots, 1)\cpsi \non\\
\y &=&\y O(1)\hde_{ C_{\ell} }    
 \hal(n,\ldots, i'+1) R_{i'}\La^{i'}  \hal(i',\ldots, i+1)  R_{i} \hal(i-1,\ldots, 1)\cpsi. 
\eeqa
We consider several sub-cases.\\
$\bullet$ $1\notin C^{2\ell}, i\neq 1$.
\beqa
\pa_{R_{i'}}  \pa_{\La_{i}}  I_{ C_{\ell}}\y&=&\y O(1)\hde_{ C_{\ell} }    
 \hal(n,\ldots, i'+1) R_{i'}\La^{i'}  \hal(i',\ldots, i+1)  R_{i} \hal(i-1,\ldots, 2)R_{1}\La_{1}\cpsi\non\\
\y &=&\y O(c^n \ir \prod_{m'=1}^n |\unk|^{-1}_{m'}).
\eeqa
Here we used standard resolvent bounds, $R_{1}\La_{1}\cpsi=O(\ir)$ and $|\unk|_{i'}^{-1}\leq |\unk|_{1}^{-1}$.\\ 
$\bullet$ $1\notin C^{2\ell}, i=1$.
\beqa
\pa_{R_{i'}}  \pa_{\La_{i}}  I_{ C_{\ell}}=O(1)\hde_{ C_{\ell} }    
 \hal(n,\ldots, i'+1) R_{i'}\La^{i'}  \hal(i',\ldots, 2)  R_{1} \cpsi. \label{some-subcase-expression}
\eeqa
First, suppose that $i'>2$. Then (\ref{some-subcase-expression}) gives
\beqa
\pa_{R_{i'}}  \pa_{\La_{i}}  I_{ C_{\ell}}=O(1)\hde_{ C_{\ell} }    
 \hal(n,\ldots, i'+1) R_{i'}\La^{i'}  \hal(i',\ldots, 3)R_{2}\La_{2} R_{1} \cpsi
\eeqa
and we obtain from (\ref{third-consequence}) and $|\unk|_{i'}^{-1}\leq 
|\unk|_{2}^{-1}$ the required bound. 

Next, suppose $i'=2$. Then (\ref{some-subcase-expression}) gives
\beqa
\pa_{R_{i'}}  \pa_{\La_{i}}  I_{ C_{\ell}}=O(1)\hde_{ C_{\ell} }    
 \hal(n,\ldots, 3)R_{2}\La^{2}R_{2}\La_{2} R_{1} \cpsi
\eeqa
and we obtain from (\ref{third-consequence}) the required bound. \\
$\bullet$ $1\in C^{2\ell}$ (and therefore $i>2$).
\beqa
\pa_{R_{i'}}  \pa_{\La_{i}}  I_{ C_{\ell}}=O(1)\hde'_{ C_{\ell} }    
 \hal(n,\ldots, i'+1) R_{i'}\La^{i'}  \hal(i',\ldots,i+1)R_{i}  \hal(i-1,\ldots, 3) R_{2}\cpsi. \ \ \ 
\eeqa
This is estimated using standard resolvent bounds and $ |\unk|_{i'}^{-1}\leq 
|\unk|_{1}^{-1}$. \\
\nin\textbf{Case 2:} $i,i'\notin C^{2\ell}, i'=i$. 
\beqa
\pa_{R_{i}}  \pa_{\La_{i}}  I_{C_{\ell}}=O(1)\hde_{ C_{\ell} }    
 \hal(n,\ldots, i+1)  R_{i} \La^i R_{i} \hal(i-1,\ldots, 1)\cpsi.
\eeqa
We consider several sub-cases:\\
$\bullet$ $1\notin C^{2\ell}, i\neq 1$.
\beqa
\pa_{R_{i}}  \pa_{\La_{i}}  I_{ C_{\ell}}=O(1)\hde_{ C_{\ell} }    
 \hal(n,\ldots, i+1)  R_{i} \La^i R_{i} \hal(i-1,\ldots, 2)R_{1}\La_{1}\cpsi.
\eeqa
Here we use standard resolvent bounds, $R_{1}\La_{1}\cpsi=O(\ir)$ and $|\unk|_{i}^{-1}\leq |\unk|_{1}^{-1}$.\\ 
$\bullet$ $1\notin C^{2\ell}, i=1$.
\beqa
\pa_{R_{i}}  \pa_{\La_{i}}  I_{ C_{\ell}}=O(1)\hde_{ C_{\ell} }    
    \hal(n,\ldots, 2)R_{1} \La^1 R_{1}\cpsi.
\eeqa
This is estimated using (\ref{third-consequence}).\\
$\bullet$ $1\in C^{2\ell}$ (and therefore $i>2$).
\beqa
\pa_{R_{i}}  \pa_{\La_{i}}  I_{ C_{\ell}}=O(1)\hde'_{ C_{\ell} }    
 \hal(n,\ldots, i+1)  R_{i} \La^i R_{i} \hal(i-1,\ldots, 3)R_{2}\cpsi.
\eeqa
{\color{\red}where $\hal(i-1,\ldots, 3)$ is understood to be $1$ for $i=3$.} This is estimated using standard resolvent bounds and $ |\unk|_{i}^{-1}\leq 
|\unk|_{1}^{-1}$.\\ 
\nin\textbf{Case 3:} $i,i'\notin C^{2\ell}, i>i'$. 
\beqa
\pa_{R_{i'}}  \pa_{\La_{i}}  I_{ C_{\ell}}
=
O(1)\hde_{ C_{\ell} }    
\hal(n,\ldots, i+1) R_{i}\hal(i-1,\ldots, i'+1) R_{i'}\La^{i'}  \hal(i',\ldots, 1)\cpsi. 
\eeqa
We consider several sub-cases.\\
$\bullet$ $1\notin C^{2\ell},  i'\neq 1$.
\beqa
\pa_{R_{i'}}  \pa_{\La_{i}}  I_{ C_{\ell}}=O(1)\hde_{ C_{\ell} }    
\hal(n,\ldots, i+1) R_{i}\hal(i-1,\ldots, i'+1) R_{i'}\La^{i'}  \hal(i',\ldots, 2)
R_{1}\La_{1}\cpsi.
\eeqa
Here we use standard resolvent bounds, $R_{1}\La_{1}\cpsi=O(\ir)$ and 
$|\unk|_{i'}^{-1}\leq |\unk|_{1}^{-1}$.\\ 
$\bullet$ $1\notin C^{2\ell}, i'=1$.
\beqa
\pa_{R_{i'}}  \pa_{\La_{i}}  I_{ C_{\ell}}
=O(1)\hde_{ C_{\ell} }    
\hal(n,\ldots, i+1) R_{i}\hal(i-1,\ldots, 2) R_{1}\La^{1}R_{1}\La_{1}\cpsi. 
\eeqa
Here we use standard resolvent bounds and $R_{1}\La_{1}\cpsi=O(\ir)$. \\
$\bullet$ $1\in C^{2\ell}$ (and therefore $i'>2$).
\beqa
\pa_{R_{i'}}  \pa_{\La_{i}}  I_{ C_{\ell}}
=O(1)\hde'_{ C_{\ell} }    
\hal(n,\ldots, i+1) R_{i}\hal(i-1,\ldots, i'+1) R_{i'}\La^{i'}  \hal(i',\ldots, 3)R_{2} \cpsi. \ \  \
\eeqa
This is estimated using standard resolvent bounds and $ |\unk|_{i'}^{-1}\leq 
|\unk|_{1}^{-1}$.\\
\nin\textbf{Case 4:} $i \notin C^{2\ell}, i'\in C^{2\ell}$.  (Clearly we can assume that $i'\in C_{\ell}$).
\beqa
\pa_{R_{i'}}  \pa_{\La_{i}}  I_{ C_{\ell}}=O(1)\hde'_{ C_{\ell} }    (R_{i'}\La^{i'}R_{i'}\hpa_{i'}\hpa_{i'-1})
 \hal(n,\ldots, i+1)  R_{i} \hal(i-1,\ldots, 1)\cpsi. 
\eeqa
This is estimated using standard resolvent bounds and $ |\unk|_{i'}^{-1}\leq 
|\unk|_{i'-1}^{-1}$.\\

\nin Finally, we prove (\ref{R-La}). In contrast to the proof of (\ref{La-R}) we have to consider the case $i\in C^{2\ell}$.
(In fact, for $i=i'$  the action of  $\pa_{R_{i}}$ generates $\La^{i}$ also for $i\in  C_{\ell}$).  \\
\nin\textbf{Case 1':} $i, i'\notin  C^{2\ell}$ $i\neq i'$. In this case $ \pa_{R_{i}}\pa_{\La_{i'}}=\pa_{\La_{i'}}\pa_{R_{i}}$
so the respective expressions are the same as in Case 1 and Case 3 above.\\
\nin\textbf{Case 2':} $i, i'\notin  C^{2\ell}$, $i=i'$. This situation differs from Case 2 above. In fact, since $\pa_{\La_{i}}$ acts
both on $\La_{i}$ and $\La^i$, we have
\beqa
\pa_{\La_{i}}\pa_{R_{i}}    I_{ C_{\ell}}\y&=&\y \hde_{ C_{\ell} }    
\pa_{\La_{i}} \hal(n,\ldots, i+1)  R_{i} \La^i R_{i} \La_{i}\hal(i-1,\ldots, 1)\cpsi\non\\
\y&=&\y O(1)\hde_{ C_{\ell} }     \hal(n,\ldots, i+1)  R_{i}  R_{i} \La_{i}\hal(i-1,\ldots, 1)\cpsi \\
& &+O(1)\hde_{ C_{\ell} }     \hal(n,\ldots, i+1)  R_{i} \La^i R_{i} \hal(i-1,\ldots, 1)\cpsi.
\eeqa
We consider several sub-cases: \\
$\bullet$ $1\notin C^{2\ell}$, $i\neq 1$.
\beqa
\pa_{\La_{i}}\pa_{R_{i}}    I_{ C_{\ell}}\y &=&\y O(1)\hde_{ C_{\ell} }     \hal(n,\ldots, i+1)  
R_{i}  R_{i} \La_{i}\hal(i-1,\ldots, 2)R_{1}\La_{1}\cpsi \\
& &+O(1)\hde_{ C_{\ell} }     \hal(n,\ldots, i+1)  R_{i} \La^i R_{i} \hal(i-1,\ldots, 2)R_{1}\La_{1}\cpsi.
\eeqa
This is estimated using standard resolvent bounds, $R_{1}\La_{1}\cpsi=O(\ir)$ and 
$|\unk|_{i}^{-1}\leq |\unk|_{1}^{-1}$. \\
$\bullet$ $1\notin C^{2\ell}$, $i=1$. 
\beqa
\pa_{\La_{i}}\pa_{R_{i}}   I_{ C_{\ell}}\y &=&\y
O(1)\hde_{ C_{\ell} }     \hal(n,\ldots, 2)  R_{1}  R_{1} \La_{1} \cpsi \label{n=i-one}\\
& &+O(1)\hde_{ C_{\ell} }     \hal(n,\ldots, 2)  R_{1} \La^1 R_{1} \cpsi. \label{n=i-two}
\eeqa
{\color{\red} In addition to standard resolvent bounds,} here (\ref{n=i-one}) is handled using (\ref{first-consequence}), whereas (\ref{n=i-two}) using (\ref{third-consequence}).\\
$\bullet$ $1\in C^{2\ell}$, (and therefore $i>2$).
\beqa
\pa_{\La_{i}}\pa_{R_{i}}   I_{ C_{\ell}}
\y &=&\y O(1)\hde'_{ C_{\ell} }     \hal(n,\ldots, i+1)  R_{i}  R_{i} \La_{i}\hal(i-1,\ldots, 3)R_{2}\cpsi \\
& &+O(1)\hde'_{ C_{\ell} }     \hal(n,\ldots, i+1)  R_{i} \La^i R_{i} \hal(i-1,\ldots, 3)R_{2}\cpsi
\eeqa
{\color{\red}where $\hal(i-1,\ldots, 3)$ is understood to be $1$  for $i=3$.} We estimate both expressions using standard resolvent bounds and $|\unk|_{i}^{-1}\leq |\unk|_{1}^{-1}$.\\
\nin\textbf{Case 3':} $i \notin  C^{2\ell}$, $i'\in C^{2\ell}$. This gives the same expression as Case 4 above.\\
\nin\textbf{Case 4':} $i,i' \in  C^{2\ell}$. Clearly is this case we can assume that $i,i'\in C_{\ell}$ and $i=i'$.
\beqa
 \pa_{\La_{i}} \pa_{R_{i}}  I_{ C_{\ell}}=O(1)\hde'_{ C_{\ell} }    (R_{i}R_{i}\hpa_{i}\hpa_{i-1})
 \hal(n,\ldots, 1)\cpsi. 
\eeqa
This is estimated using $|\unk|_{i}^{-1}\leq |\unk|_{i-1}^{-1}$. \qed

\begin{prop} There hold the bounds
\beqa
& &\|\pa_{\cpsi}  I_{ C_{\ell}}\|\leq c^n\ir \prod_{m'=1}^n |\unk|^{-1}_{m'}, \label{first-derivative-vector}\\
& &\|\pa^2_{\cpsi}  I_{ C_{\ell}}\|\leq c^n\ir \prod_{m'=1}^n |\unk|^{-1}_{m'}, \label{second-derivative-vector}\\
& &\|\pa_{\cpsi}\pa_{R_{i}}  I_{ C_{\ell}}\|\leq c^n\ir \prod_{m'=1}^n |\unk|^{-1}_{m'}, \label{R-vector-derivative} \\
& &\|\pa_{\cpsi} \pa_{\La_{i}} I_{C_{\ell}}\| \leq c^n\ir \prod_{m'=1}^n |\unk|^{-1}_{m'}. \label{La-vector-derivative}
\eeqa
\end{prop}
\proof To prove~(\ref{first-derivative-vector}), (\ref{second-derivative-vector}), we write for $q=1,2$
\beqa
\pa_{\cpsi}^q  I_{ C_{\ell}}=\hde_{ C_{\ell} } \hal(n,\ldots, 1)\pa^q_{P}\cpsi
=O(c^n \ir \prod_{m'=1}^n |\unk|^{-1}_{m'}),
\eeqa
where we applied standard resolvent bounds and estimates (\ref{state-bound}). \\

\nin Now we proceed 
to~(\ref{R-vector-derivative}) following the lines of the proof of (\ref{R-first-derivative}).  

\nin\textbf{Case 1:} $i\notin  C^{2\ell}$. 
\beqa
\pa_{\cpsi}\pa_{R_{i}} I_{C_{\ell}}\y&=&\y \hde_{C_{\ell}}  
\pa_{R_{i}} \hal(n,\ldots, 1)\pa_{P}\cpsi\non\\
\y&=&\y \hde_{C_{\ell}}     
 \hal(n,\ldots, i+1)R_{i}\La^i  \hal(i,\ldots, 1)\pa_{P}\cpsi. \label{first-derivative-start-one}
\eeqa
We consider several sub-cases. \\
\nin$\bullet$  $1, 2\notin  C^{2\ell}$, $i\geq 2$.
\beqa
\pa_{\cpsi}\pa_{R_{i}} I_{C_{\ell}}\y&=&\y \hde_{C_{\ell}}     
 \hal(n,\ldots, i+1)R_{i}\La^i  \hal(i,\ldots, 2)R_{1}\La_{1}\pa_{P}\cpsi\non\\
\y&=&\y \hde_{C_{\ell}}     
 \hal(n,\ldots, i+1)R_{i}\La^i  \hal(i,\ldots, 3)R_{2}\La_{2}R_{1}\La_{1}R\La\cpsi\non\\
\y&=&\y O(c^n\ir \prod_{m'=1}^n |\unk|^{-1}_{m'}),
\eeqa
where we made use of $\pa_{P}\cpsi=R\La\cpsi$,  (\ref{last-last-consequence}) and $|\unk|_{i}^{-1}\leq |\unk|_{2}^{-1}$ 
and $\hal(i,\ldots, 3)=1$ is understood for $i=2$.
\\
\nin$\bullet$  $1, 2\notin  C^{2\ell}$, $i=1$.
\beqa
\pa_{\cpsi}\pa_{R_{i}} I_{ C_{\ell}}\y&=&\y \hde_{C_{\ell}}     
 \hal(n,\ldots, 2)R_{1}\La^1R_{1}\La_{1}\pa_{P}\cpsi\\
\y&=&\y \hde_{C_{\ell}}     
 \hal(n,\ldots, 2)R_{1}\La^1R_{1}\La_{1} R\La\cpsi.
\eeqa
This is estimated using (\ref{last-last-consequence}).\\
\nin$\bullet$  $1, 2\in  C^{2\ell}$ (and therefore $i>2$).
\beqa
\pa_{\cpsi}\pa_{R_{i}} I_{ C_{\ell}}=\hde'_{C_{\ell}}     
 \hal(n,\ldots, i+1)R_{i}\La^i  \hal(i,\ldots, 3)R_{2} \pa_{P}\cpsi.
\eeqa
This is estimated using standard resolvent bounds, $\pa_{P}\cpsi=O(\ir)$ and $|\unk|_{i}^{-1}\leq |\unk|_{1}^{-1}$.\\
\nin$\bullet$  $1\notin  C^{2\ell}, 2\in C^{2\ell}$, $i>3$.  
\beqa
\pa_{\cpsi}\pa_{R_{i}} I_{ C_{\ell}}=\hde'_{C_{\ell}}     
 \hal(n,\ldots, i+1)R_{i}\La^i  \hal(i,\ldots, 4)R_{3}R_{1}\La_{1}\pa_{P}\cpsi.
\eeqa
This is estimated using standard resolvent bounds, $\pa_{P}\cpsi=O(\ir)$ and 
$|\unk|_{i}^{-1}\leq |\unk|_{2}^{-1}$.\\
\nin$\bullet$  $1 \notin  C^{2\ell}, 2 \in C^{2\ell}$, $i=1$. 
\beqa
\pa_{\cpsi}\pa_{R_{i}} I_{ C_{\ell}}=\hde'_{C_{\ell}}     
 \hal(n, \ldots, 4)R_{3}R_{1}\La^1 R_{1}\La_{1}R\La\cpsi.
\eeqa
This is estimated using (\ref{last-last-consequence}) {\color{\red} and $1\leq |\unk|_{2}^{-1}$}.\\
\nin\textbf{Case 2:} $i\in  C^{2\ell}$ (thus we can assume $i \in C_{\ell}$). 
\beqa
\pa_{\cpsi}\pa_{R_{i}} I_{ C_{\ell}}=\hde'_{C_{\ell}}  (R_{i}\La^iR_{i}\hpa_{i}\hpa_{i-1})   
 \hal(n, \ldots, 1)\pa_{P}\cpsi.
\eeqa
Here we use $\pa_{P}\cpsi=O(\ir)$ and $|\unk|_{i}^{-1}\leq |\unk|_{i-1}^{-1}$.\\

\nin Finally we prove (\ref{La-vector-derivative}). As in the proof of (\ref{La-derivative}) we can assume that 
$i\notin C^{2\ell}$. We have
\beqa
\pa_{\cpsi}\pa_{\La_{i}} I_{C_{\ell}}\y&=&\y\hde_{ C_{\ell} }     
\pa_{\La_{i}} \hal(n,\ldots, 1)\pa_{P}\cpsi\non\\
\y&=&\y O(1)\hde_{ C_{\ell} }    
 \hal(n, \ldots, i+1)  R_{i} \hal(i-1,\ldots, 1)\pa_{P}\cpsi \non\\ 
\y&=&\y O(c^n \prod_{m'=1}^n |\unk|^{-1}_{m'}), \label{La-first-derivative-bound-new}
\eeqa
where we made use of $\pa_{P}\cpsi=O(\ir)$ and standard resolvent bounds. \qed
\begin{appendix}

\section{Proof of estimate (\ref{simple-spectral-bound}), (\ref{infrared-spectral-bound}) with $|\be|=1$ and (\ref{mixed-spectral-bound}) }\label{standard-proofs}
\label{Appendix-trivial}
\setcounter{equation}{0}

In this appendix we establish estimates on $\pa_{k_l^j}f_{P,\si}^{\mm}$, 
$ \pa_{k_l^j} \pa_{k_l^{j'}}f_{P,\si}^{\mm}$, $\pa_{P^j}f_{P,\si}^{\mm}$ and
$\pa_{k_l^{j}}\pa_{P^{j'}}f_{P,\si}^{\mm}$. Since derivatives w.r.t. $k_{l}$ do
not preserve the symmetry of $f_{P,\si}^{\mm}$ under permutations of photon
variables, we will make the dependence on the permutation $\pi\in S_n$ explicit
in our notation. In particular, for a given $\pi$, we will define
\beqa
\unk_{\pi,i}, \quad |\unk|_{\pi,i},\quad r(\unk)_{\pi,i}, \quad R_{\pi,i}, \quad R_{\pi}^i,\quad \La_{\pi,i}, \quad \La_{\pi}^i \label{new-objects}
\eeqa
by replacing $k_{i'}$ with $k_{\pi(i')}$ in  the corresponding expressions from Subsection~\ref{spectral-ingredients-sub}. The properties established
in Subsection~\ref{spectral-ingredients-sub} hold, after obvious modifications, 
for the new objects (\ref{new-objects}).

\subsection{Proof of estimate (\ref{simple-spectral-bound})} \label{k-subsection}

\nin Proof of this estimate is elementary, in particular it does not require the sophisticated bounds from Theorem~\ref{spectr-theorem}   but only standard resolvent estimates from Lemma~\ref{elem-lemma}.  We include it here  for completeness.

We recall  a formula from \cite{Fr73} stated already in (\ref{Froehlich-form})  above,
\beqa
 f_{P,\si}^{\mm}(k_1,\ldots, k_n)=\fr{1}{\sqrt{n!}} \sum_{\pi\in S_n}(-1)^n \lan \Om, 
\prod_{i=n}^1\fr{1}{H_{P-\unk_{\pi,i},\si}-E_{P,\si}+|\unk|_{\pi,i}} \vv^{\si}(k_{\pi(i)}) \cpsi_{P,\si}\ran, \label{Froehlich-formula-one}
\eeqa
where  $S_{\mm}$ is the set of permutations of an $\mm$-element set and for any 
$\pi\in S_{\mm}$ we write $\unk_{\pi,i}:=\sum_{i'=1}^{i} k_{\pi(i')}$ and 
$|\unk|_{\pi,i}:=\sum_{i'=1}^{i} |k_{\pi(i')}|$. We define
\beqa
R_{\pi,i}:=\fr{1}{H_{P-\unk_{\pi,i},\si}-E_{P,\si}+|\unk|_{\pi,i}},\quad
I_\pi:=\big(\vv^{\si}(k_{1})\ldots\vv^{\si}(k_{\mm}) \big)\lan\vac,\prod_{i=n}^{1}   R_{\pi,i} \cpsi_{P,\si}\ran,
\label{I-pi-def}
\eeqa 
and keep in mind that both $I_{\pi}$ and $R_{\pi,i}$ depend on $k_1,\ldots, k_n$.
The first derivative  of $I_{\pi}$  w.r.t. $k_l$ has the form  
\beqa
\pa_{k_l^j}I_\pi\y&=&\y \big( \vv^{\si}(k_{1})\ldots \pa_{k_l^j} \vv^{\si}(k_{l})\ldots   \vv^{\si}(k_{\mm}) \big)
\lan\vac,\prod_{i=\mm}^{1}  R_{\pi,i}  \cpsi_{P,\si}\ran\non\\
& &+\big( \vv^{\si}(k_{1})\ldots  \vv^{\si}(k_{\mm}) \big)    \pa_{k_l^j}\lan\vac,\prod_{i=\mm}^{1} R_{\pi,i}  
\cpsi_{P,\si}\ran.
\label{derivative-terms-one}
\eeqa
The second derivative of $I_\pi$ is given by
\begin{eqnarray}
\pa_{k_l^j} \pa_{k_l^{j'}  }I_\pi
\y&=&\y \big( \vv^{\si}(k_{1})\ldots \pa_{k_l^j} \pa_{k_l^{j'} }\vv^{\si}(k_{l})\ldots   \vv^{\si}(k_{\mm}) \big)
\lan\vac,\prod_{i=\mm}^{1}  R_{\pi,i} \cpsi_{P,\si}\ran\non\\
& &+\bigg(\big( \vv^{\si}(k_{1})\ldots \pa_{k_l^j}\vv^{\si}(k_{l})\ldots   \vv^{\si}(k_{\mm}) \big)   
 \pa_{k_l^{j'}}\lan\vac,\prod_{i=\mm}^{1} R_{\pi,i}  \cpsi_{P,\si}\ran+\{j\leftrightarrow j'\}\bigg)\non\\
& &+\big( \vv^{\si}(k_{1})\ldots  \vv^{\si}(k_{\mm}) \big)    \pa_{k_l^j}\pa_{k_l^{j'}}\lan\vac,\prod_{i=\mm}^{1} R_{\pi,i}
\cpsi_{P,\si}\ran. \label{second-derivative-formula}
\end{eqnarray}
Since for $k_l\in \mcA_{\sigma,\kappa}$ {\color{\red} (see the definition in (\ref{A-set}))} we have
$|\pa_{k_l^j}\chi_{[\si,\ka)}(k_l)|\leq c\chi_{[\si,\kas)}(k_l)$ and also $|\pa_{k_l^j}\pa_{k_l^{j'}}\chi_{[\si,\ka)}(k_l)|\leq c\chi_{[\si,\kas)}(k_l)$, 
 we obtain that
\beqa
& &|\pa_{k_l^j}\vv^{\si}(k_{l})|\leq \fr{c}{|k_l|} \g\fr{\chi_{[\si,\kas)}(k_l)|k_l|^{\alf}   }   {(2|k_l|)^{1/2} },\label{form-factor-first-der}\\
& &|\pa_{k_l^j} \pa_{k_l^{j'}}\vv^{\si}(k_{l})|\leq \fr{c}{|k_l|^2} \g\fr{\chi_{[\si,\kas)}(k_l)|k_l|^{\alf}   }   {(2|k_l|)^{1/2} }. \label{form-factor-second-dar}
\eeqa
(We stress that $c$ may change its value from line to line). Denoting by $(\pa_{k^j_l}I_\pi)^{(1)}$ the first term on the r.h.s. of (\ref{derivative-terms-one}),
and using  (\ref{form-factor-first-der}), (\ref{elem-resolv}),  we   get
\beqa
|(\pa_{k_l^j}I_\pi)^{(1)}|\leq \fr{{ c }^{\mm}}{|k_l|}  \prod_{i=1}^{n} \bigg(\g\fr{\chi_{[\si,\kas)}(k_i) |k_i|^{\alf}  }   {(2|k_i|)^{1/2} }\bigg)\prod_{i=1}^{\mm}\fr{1}{|\unk|_{\pi,i}}\,.\label{I-bound-one}
\eeqa

Now let $(\pa_{k_l^j}I_\pi)^{(2)}$ be the second term on the r.h.s. of (\ref{derivative-terms-one}). Let $\pi, i$ be s.t. $l\in \{\pi(1), \pi(2),\ldots, \pi(i)\}$.
Then
\beqa
\pa_{k_{l}^j} R_{\pi,i}= R_{\pi,i}  \big( (P- \unk_{\pi,i}-P_{\pho})- k_l/|k_l| \big)^jR_{\pi,i}.
 \label{resolvent-first-derivative}
\eeqa
(For $l\notin \{\pi(1), \pi(2),\ldots, \pi(i)\}$ the above derivative is zero). Consequently, making use of standard resolvent
bounds  (\ref{elem-resolv}), we obtain 
\beqa
|\vv^{\si}  (k_{\pi(i)})| \  \|\pa_{k_{l}^j}  R_{\pi,i} \|\leq 
\g\fr{\chi_{[\si,\ka)}(k_{\pi(i)})|k_{\pi(i)}|^{\alf}  }{(2|k_{\pi(i)}|)^{1/2}} \fr{c }{|\unk|_{\pi,i}^2}
\leq \fr{c}{ |k_l|}\g\fr{\chi_{[\si,\kas)}(k_{\pi(i)})|k_{\pi(i)}|^{\alf}  }{(2|k_{\pi(i)}|)^{1/2}} \fr{1}{|\unk|_{\pi,i}}, \label{first-der-estimate}
\eeqa
where in the last step we exploited the fact that $l\in \{\pi(1), \pi(2),\ldots, \pi(i)\}$ which gives $|k_l|\leq |\unk|_{\pi,i}$.
Thus we get that $(\pa_{k^j_l}I_\pi)^{(2)}$ also satisfies a bound of the form (\ref{I-bound-one}). Hence,
\beqa
|(\pa_{k_l^j}I_\pi)|\leq \fr{{c}^{\mm}}{ |k_l| }\prod_{i=1}^n\bigg(\g\fr{\chi_{[\si,\kas)}(k_i)|k_i|^{\alf}   }   {(2|k_i|)^{1/2} }\bigg)\prod_{i=1}^{\mm}\fr{1}{|\unk|_{\pi,i}}.
\eeqa
Now making use of the second identity in (\ref{froehlich-combinatorics}),  we conclude the proof of (\ref{simple-spectral-bound}) in the case of $|\be|=1$. 

To cover the case $|\be|=2$, we still have to estimate the second derivative of the product of resolvents, appearing in 
the last term of (\ref{second-derivative-formula}). We obtain from (\ref{resolvent-first-derivative}) that for $l\in \{\pi(1), \pi(2),\ldots, \pi(i)\}$
\beqa
 \pa_{k_{l}^{j'}}\pa_{k_{l}^j} R_{\pi,i} 
\y&=&\y\{\pa_{k_{l}^{j'} }   R_{\pi,i} \}  \big( (P- \unk_{\pi,i}-P_{\pho})- k_l/|k_l| \big)^j R_{\pi,i} \non\\
& &+   R_{\pi,i} \big( (P- \unk_{\pi,i}-P_{\pho})- k_l/|k_l| \big)^j
\{\pa_{k_{l}^{j'} }  R_{\pi,i}  \} \non\\
& &+  R_{\pi,i}  (k_l^j k_l^{j'}|k_l|^{-3}-\de_{j,j'}(1+|k_l|^{-1}   )) R_{\pi,i}.
 \eeqa
From this expression, Lemma~\ref{elem-lemma} and formula~(\ref{first-der-estimate}), we obtain that
\beqa
|\vv^{\si}  (k_{\pi(i)})| \ \|\pa_{k_{l}^{j'}}\pa_{k_{l}^j}  R_{\pi,i} \| \leq 
\fr{c}{ |k_l|^2}\g\fr{\chi_{[\si,\kas)}(k_{\pi(i)})|k_{\pi(i)}|^{\alf}  }{(2|k_{\pi(i)}|)^{1/2}} \fr{1}{|\unk|_{\pi,i}}. \label{second-der-first}
\eeqa
Now let $\pi$ and $i_1$, $i_2$ be s.t.  $l\in \{\pi(1), \pi(2),\ldots, \pi(i_1)\}$ and $l\in \{\pi(1), \pi(2),\ldots, \pi(i_2)\}$.
Then, making use of (\ref{first-der-estimate}), we get
\beqa
& &|\vv^{\si}  (k_{\pi(i_1)})||\vv^{\si}  (k_{\pi(i_2)})|\, \|\pa_{k_{l}^{j'}} R_{\pi,i_1}  \|
\|\pa_{k_{l}^j} R_{\pi,i_2} \|\non\\ 
& &\ph{4444444444444} \leq\fr{c}{ |k_l|^2}\g\fr{\chi_{[\si,\kas)}(k_{\pi(i_1)})|k_{\pi(i_1)}|^{\alf}  }{(2|k_{\pi(i_1)}|)^{1/2}} \fr{1 }{|\unk|_{\pi,i_1}}
  \g\fr{\chi_{[\si,\kas)}(k_{\pi(i_2)})|k_{\pi(i_2)}|^{\alf}  }{(2|k_{\pi(i_2)}|)^{1/2}} \fr{1 }{|\unk|_{\pi,i_2}}. \label{second-der-second}
\eeqa
Now we use formulas~(\ref{second-der-first})  and (\ref{second-der-second}) to estimate the last term on the r.h.s. of (\ref{second-derivative-formula}).
The remaining terms {\color{\red}in (\ref{second-derivative-formula})} are estimated with the help of (\ref{form-factor-first-der}), (\ref{form-factor-second-dar}), (\ref{first-der-estimate}) and Lemma~\ref{elem-lemma}. 
Altogether we get
\beqa
|\pa_{k_l^j} \pa_{k_l^{j'}}I_\pi|\leq \fr{{c}^{\mm}}{ |k_l|^2 }\prod_{i=1}^{\mm}\bigg(\g\fr{\chi_{[\si,\kas)}(k_{i})|k_{i}|^{\alf}   }   {(2|k_{i}|)^{1/2} }\bigg)\prod_{i=1}^{\mm}\fr{1}{|\unk|_{\pi,i}}.
\eeqa
Making use of   the second identity in (\ref{froehlich-combinatorics}) we complete the proof of (\ref{simple-spectral-bound}) for $|\be|=2$. 

\subsection{Proof of estimate~(\ref{infrared-spectral-bound}) with $|\be|=1$} \label{first-derivative-sub}

We can rewrite  formula~(\ref{Froehlich-formula-one}) as follows
\beqa
\sqrt{n!}(-1)^n f_{P,\si}^{\mm}=v\sum_{\pi\in S_n}\lan \Om, R_{\pi,n}\ldots R_{\pi,1}\cpsi_{P,\si}\ran,
\eeqa
where $v(k_1,\ldots,k_n):=\vv^{\si}(k_{1})\ldots\vv^{\si}(k_{\mm})$.
Consequently, making use of (\ref{R-derivative}), we have
\beqa
\sqrt{n!}(-1)^n \pa_{P^j}f_{P,\si}^{\mm}=vI_{1,n}+v I_{2,n}, \textrm{ where } 
I_{1,n}\y&:=&\y\sum_{\pi\in S_n}\lan \Om, R_{\pi,n}\ldots R_{\pi,1}\pa_{P^j}
\cpsi_{P,\si}\ran, \label{where-formula} \\
I_{2,n}\y&:=&\y\sum_{\pi\in S_n}\sum_{\ell=1}^n\lan \Om, R_{\pi,n}\ldots R_{\pi,\ell}\La_{\pi}^{\ell}R_{\pi,\ell} \ldots R_{\pi,1}\cpsi_{P,\si}\ran \quad\quad  \quad\quad
\eeqa
with $\La_{\pi}^{\ell}:=\La+\unk_{\pi,\ell}$ {\color{\red}(see the definition of $\Lambda$ in (\ref{def-La}))}.
By estimate (\ref{state-bound}), Lemma~\ref{elem-lemma} and the second identity in (\ref{froehlich-combinatorics}) we get
\beqa
|v I_{1,n}|\leq    \frac{1}{\sigma^{\delta_{\lambda_0}}}   g^n_{\sigma} 
\eeqa
which gives one contribution to estimate~(\ref{infrared-spectral-bound}).

To handle $I_{2,n}$, we apply (\ref{R-shift-bound}) i.e. $R_{\pi,i}:=R_{\pi}^i-R_{\pi,i} (\unk_{\pi,i}\cdot \La) R_{\pi}^i$, where
$R_{\pi}^i:=(H_{P,\si}-E_{P,\si}+r(\unk)_{\pi,i})^{-1}$. {\color{\red} In the definition of $R_{\pi,i}$ the reader can recognize a truncated resolvent expansion. In the scalar product defining $I_{2,n}$ below, we use this expansion on each $R_{\pi,i}$, $1\leq i\leq \ell$, starting from $R_{\pi,1}$. Namely,} we write
\beqa
I_{2,n}\y&=&\y \sum_{\pi\in S_n}\sum_{\ell=1}^n\lan \Om, R_{\pi,n}\ldots R_{\pi,\ell}\La_{\pi}^{\ell}R_{\pi,\ell} \ldots R_{\pi,1}\cpsi_{P,\si}\ran\non\\
\y&=&\y \sum_{\pi\in S_n}\sum_{\ell=1}^n\sum_{i=1}^{\ell}(-)\lan \Om, R_{\pi,n}\ldots R_{\pi,\ell}\La_{\pi}^{\ell}R_{\pi,\ell}\ldots R_{\pi,i}(k_{\pi,i}\cdot \La) R_{\pi}^{i} \ldots R_{\pi}^1\cpsi_{P,\si}\ran \label{I-2-n-first-bound} \\
& &+\sum_{\pi\in S_n}\sum_{\ell=1}^n\lan \Om, R_{\pi,n}\ldots R_{\pi,\ell}\La_{\pi}^{\ell}R_{\pi}^{\ell} \ldots R_{\pi}^1\cpsi_{P,\si}\ran. \label{I-2-n-second-bound}
\eeqa
Now making use of Lemma~\ref{elem-lemma}, the relations 
\beqa
R_{\pi}^{i}\cpsi_{P,\si}=r(\unk)_{\pi,i}^{-1}\cpsi_{P,\si}, \quad  
R_{\pi,i}\La\cpsi_{P,\si}=O(\ir), \quad R_{\pi,\ell}\La_{\pi}^{\ell}\cpsi_{P,\si}=O(\ir)
\eeqa
(cf. Corollary~\ref{key-corollary}), the estimate $|\unk|_{\pi,\ell}^{-1}\leq |\unk|_{\pi,i}^{-1}$ for $\ell\geq i$,
and  the second identity in (\ref{froehlich-combinatorics}), we obtain
\beqa
|v I_{2,n}|\leq    \frac{1}{\sigma^{\delta_{\lambda_0}}}   g^n_{\sigma},
\eeqa
which concludes the proof of estimate~(\ref{infrared-spectral-bound}) with $|\be|=1$.

\subsection{Proof of estimate~(\ref{mixed-spectral-bound})    }

Given the results from the last two subsections it is easy to estimate the mixed derivative $\pa_{k_l^{j}}\pa_{P^{j'}}f_{P,\si}^{\mm}$. Starting from 
formula~(\ref{where-formula}) we write
\beqa
\sqrt{n!}(-1)^n \pa_{k_l^{j}}\pa_{P^{j'}}f_{P,\si}^{\mm}=\pa_{k_l^{j}}(vI_{1,n})+(\pa_{k_l^{j}}v) I_{2,n}+
v\pa_{k_l^{j}}I_{2,n}. \label{mixed-derivative-computation}
\eeqa
We note that the considerations in Subsection~\ref{k-subsection} are unchanged if $\cpsi_{P,\si}$ is replaced
with $\pa_{P^{j'}}\cpsi_{P,\si}$, except for an additional factor $\ir$ in the estimates coming from (\ref{state-bound}). Thus we get immediately
\beqa
|\pa_{k_l^{j}}(vI_{1,n})|\leq \frac{1}{\sigma^{\delta_{\lambda_0}}} |k_l|^{-1}  g^n_{\sigma}.
\eeqa
Next, combining the analysis of $I_{2,n}$ from Subsection~\ref{first-derivative-sub} with the bound~(\ref{form-factor-first-der}) we obtain
\beqa
|(\pa_{k_l^{j}}v) I_{2,n}|\leq \frac{1}{\sigma^{\delta_{\lambda_0}}} |k_l|^{-1}  g^n_{\sigma}.
\eeqa
As for the last term on the r.h.s. of (\ref{mixed-derivative-computation}),
by inspection of (\ref{I-2-n-first-bound}) and (\ref{I-2-n-second-bound})
we see that the action of $\pa_{k_{l}^{j}}$ on $I_{2,n}$ leads to an additional
factor $|k_l|^{-1}$ in the estimates. In fact, the relevant ingredients  
 are
\beqa
R^{i}_{\pi}\cpsi_{P,\si}=r(\unk)_{\pi,i}^{-1}\cpsi_{P,\si}, \quad 
\pa_{k_{l}^{j}}r(\unk)_{\pi,i}^{-1}=\mco(|k_l|^{-1} |\unk|_{\pi,i}^{-1}), \quad
\pa_{k_{l}^{j}} \unk_{\pi,i}=\mco(1)=\mco(|\unk|_{\pi,i}/|k_l|)
\eeqa
and the fact that $\pa_{k_l^j} R_{\pi,i}=R_{\pi,i}  \big( (P- \unk_{\pi,i}-P_{\pho})- k_l/|k_l| \big)^jR_{\pi,i}=\mco(|k_l|^{-1}|\unk|^{-1}_{\pi,i})$, cf. (\ref{resolvent-first-derivative}).  Using this and the second identity in (\ref{froehlich-combinatorics}) we conclude the proof.

\section{Proof of Proposition~\ref{pull-through-new}  } \label{pull-through-appendix}
\setcounter{equation}{0}

Our discussion here is similar to the first part of Appendix D of \cite{DP17.1}.
In Subsection~\ref{pull-through-subsection} we defined $\real^3\ni k\mapsto b_W(k)\psi$ for $\psi\in C^{\infty}(H_{P,\free})$ as  vector-valued distributions. For some vectors $\psi\in C^{\infty}(H_{P,\free})$ we can also define $b(k)\psi$ pointwise in $k$ as follows: Let $\eta\in C_0^{\infty}(\real^3)$ be s.t. $\eta\geq 0$, $\eta(0)=1$, $\int \eta(k') d^3k'=1$, $\eta(k)=\eta(-k)$ so that $\eta_k^{\eps}(k'):=\eps^{-3}\eta((k'-k)/\eps^{-1})$ is an approximating sequence of $k'\mapsto \de(k'-k)$ as $\eps\to  0$.   We say that $\psi\in C^{\infty}(H_{P,\free})$ is in $D(b_W(k))$  if the following limit
exists in norm
\beqa
b_W(k)\psi:=\lim_{\eps\to 0}b_W(\eta_k^{\eps})\psi \label{domain-b(k)}
\eeqa
and gives an element of $C^{\infty}(H_{P,\free})$. We   also require that $\lim_{\eps\to 0}b_W(g\eta_k^{\eps})\psi$ exists and equals $g(k)b_W(k)\psi$ for any continuous function $g$.  The domain of the product
$D(b_W(k_1)\ldots b_W(k_m))$ is defined iteratively. As we showed in Lemma D.1 of \cite{DP17.1},    relation~(\ref{domain-b(k)})  is consistent with the  definition as a distribution. 

To prove (\ref{recurrence-relation-W}), we first recall relation (\ref{n-part-pull-through})
\beqa
b_W(k_m)\ldots b_W(k_1)H_{P,\si}\psi\y&=&\y H_{P,\si; k_m,\ldots,k_1}b_{W}(k_m)\ldots b_{W}(k_1)\psi\non\\
& &+\sum_{i=1}^mF_{P,\si}(k_i,k_1+\cdots \check{i} \cdots +k_m)b_{W}(k_m)\cdots \check{i} \cdots b_{W}(k_1)\psi\non\\
& &+\sum_{1\leq i<i'\leq m} G_{P,\si}(k_i,k_{i'})b_{W}(k_m)\cdots \check{i}'\cdots \check{i} \cdots b_{W}(k_1)\psi, \label{pull-through-W-x}
\eeqa
which holds in the sense of distributions. 
Now we set $\psi=\psi_{P,\si}$ and proceed by induction. 
We suppose that  $\psi_{P,\si}\in D(b_W(k_1)\ldots b_W(k_{m-1}))$ for any $k_1, \ldots k_{m-1}\in \real^3\backslash \{0\}$ and we 
will conclude from this that $\psi_{P,\si}\in D(b_W(k_1)\ldots b_W(k_{m-1})b(k_m))$.  (For $m=1$ the former condition is understood to be empty and thus trivially
satisfied). To  this end, we   set 
\beqa
& &A(k'_m):=(E_{P,n}-|\un k|_{m-1}- |k'_m|- H_{P-\un k_{m-1}- k'_m ,n}), \\
&  &F_m^i:=F_{P,\si}(k_i,k_1+\cdots \check{i} \cdots +k_m),\\
& &B_{m-1}:=b_W(k_{m-1})\ldots b_W(k_1), \\ 
& &B_{m-1}^i:= b_W(k_{m-1})\ldots \check{i} \ldots b_W(k_1),\\
& &B_{m-1}^{i',i}:=b_{W}(k_{m-1})\cdots \check{i}'\cdots \check{i} \cdots b_{W}(k_1), 
\eeqa
where we use the notation  $\un k_m:=k_1+\cdots+k_m$
and $|\un k|_m:=|k_1|+\cdots+|k_m|$.   Next, for some $k_m\neq 0$, we rewrite  (\ref{pull-through-W-x}) as follows
\beqa
b_W( f)B_{m-1}\psi_{P,\si}\y&=&F_m^m(f)\,  A(k_m)^{-1} B_{m-1}\psi_{P,\si}+A(k_m)^{-1}\sum_{i=1}^{m-1}(F_m^ib_W)(f)B_{m-1}^i\psi_{P,\si} \label{B-f-formula}\\
& &+A(k_m)^{-1}\sum_{i=1}^{m-1}G(k_i, f)B^i_{m-1}\psi_{P,\si}+ A(k_m)^{-1}\!\!\!\!\!\sum_{1\leq i<i'\leq m-1} G_{P,\si}(k_i,k_{i'}) b_W(f)B_{m-1}^{i',i}\psi_{P,\si} 
\quad\quad \ \label{B-f-formula-one}  \\
& &+\int d^3k_m'\, f(k_m') A(k_m)^{-1}(A(k_m)-A(k'_m))b_W(k'_m)B_{m-1}\psi_{P,\si}, \label{pull-through-error-term}
\eeqa
where we smeared both sides of (\ref{pull-through-W-x}) in one variable with a test-function $f$ vanishing near zero and $F_m^m(f), G(k_i,f), (F_m^ib_W)(f)$ denote smearing of the respective expressions with $f$.
Now we analyze (\ref{pull-through-error-term}).  For simplicity of notation we set in the following $k:=k_m, k':=k'_m$. First, we note that
\beqa
A(k)-A(k')
\y&=&\y  L_{k}(k')+  (P'-P_{\pho})M_{k}(k'),
\eeqa
where $P':=P- \un{k}_{m-1}$ and we defined the functions
\beqa
L_{k}(k'):=(|k'|-|k|)-  \h(k-k')(k'+k), \quad\quad M_{k}(k'):=(k-k').
\eeqa
With these definitions we can write
\beqa
(\ref{pull-through-error-term})=A(k)^{-1}b(L_{k} f)B_{m-1}\psi_{P,\si}+A(k)^{-1}(P'-P_{\pho}) b(M_{k} f)B_{m-1}\psi_{P,\si}. \label{before-iteration}
\eeqa
To conclude the argument, we substitute this back to (\ref{B-f-formula})-(\ref{pull-through-error-term}),  set $f= g \eta^{\eps}_{k}$, note that (\ref{B-f-formula}), 
(\ref{B-f-formula-one}) converge as $\eps\to 0$ to the desired quantities by the inductive hypothesis and (\ref{before-iteration}) tends to zero in this limit since $M_k(k)=L_k(k)=0$ and $\eta^{\eps}_{k}(k')\to \de(k-k')$. To show vanishing of (\ref{before-iteration}),
one combines an iteration argument with energy bounds as in the last part of the proof of Lemma~D.3 of \cite{DP17.1}.

\end{appendix}


\end{document}